\documentclass[aps,pre,reprint,noshowkeys,noshowpacs,nofootinbib,superscriptaddress]{revtex4-1}

\usepackage{hyperref}
\usepackage{amsmath,amssymb}
\usepackage{graphicx,color}
\usepackage{hyperref,comment}
\usepackage{float}
\usepackage{soul}
\usepackage{natbib}
\setcitestyle{numbers,sort&compress}

\graphicspath{{figures/}}

\newcommand{\D}{\mathcal{D}}
\newcommand{\J}{\mathcal{J}}
\newcommand{\K}{\mathcal{K}}
\renewcommand{\H}{\mathcal{H}}
\renewcommand{\P}{\mathcal{P}}
\renewcommand{\S}{\mathcal{S}}
\newcommand{\Z}{\mathcal{Z}}
\newcommand{\F}{\mathcal{F}}
\newcommand{\Dv}{\D{v}}
\newcommand{\Ob}{\mathcal{O}}
\newcommand{\Gnu}{\mathcal{G}^{(\nu)}}
\renewcommand{\Re}{{\rm Re}}

\newcommand{\real}{\operatorname{\mathbb{R}e}}
\newcommand{\imag}{\operatorname{\mathbb{I}m}}

\begin{document}

\title{A Hybrid Monte Carlo algorithm for sampling rare events\\ in space-time histories of stochastic fields}

\author{G.\ Margazoglou}
\affiliation{Department of Physics, University of Rome Tor Vergata, 00133 Rome, Italy}
\affiliation{Computation-based Science and Technology Research Center, Cyprus Institute, 2121 Nicosia, Cyprus}
\author{L.\ Biferale}
\affiliation{Department of Physics, University of Rome Tor Vergata, 00133 Rome, Italy}
\author{R.\ Grauer}
\affiliation{Institute f\"ur Theoretische Physik I,  Ruhr-University Bochum, 44780 Bochum, Germany}
\author{K.\ Jansen}
\affiliation{NIC, DESY, 15738 Zeuthen, Germany}
\author{D.\ Mesterh\'azy}
\email[Corresponding author: ]{mesterh@itp.unibe.ch}
\affiliation{Institute for Theoretical Physics, University of Bern, 3012 Bern, Switzerland}
\author{T.\ Rosenow}
\affiliation{Brandenburg University of Technology, 03046 Cottbus, Germany}
\author{R.\ Tripiccione}
\affiliation{Department of Physics, University of Ferrara, 44121 Ferrara, Italy}
\affiliation{INFN Sezione di Ferrara, 44122 Ferrara, Italy}

\begin{abstract}

We introduce a variant of the Hybrid Monte Carlo (HMC) algorithm to address large-deviation statistics in stochastic hydrodynamics. Based on the path-integral approach to stochastic (partial) differential equations, our HMC algorithm samples space-time histories of the dynamical degrees of freedom under the influence of random noise. First, we validate and benchmark the HMC algorithm by reproducing multiscale properties of the one-dimensional Burgers equation driven by Gaussian and white-in-time noise. Second, we show how to implement an importance sampling protocol to significantly enhance, by orders of magnitudes, the probability to sample extreme and rare events, making it possible to estimate moments of field variables of extremely high order (up to 30 and more). By employing reweighting techniques, we map the biased configurations back to the original probability measure in order to probe their statistical importance. Finally, we show that by biasing the system towards very intense negative gradients, the HMC algorithm is able to explore the statistical fluctuations around instanton configurations. Our results will also be interesting and relevant in lattice gauge theory since they provide insight into reweighting techniques.

\end{abstract}


\maketitle

\section{Introduction}
\label{Sec:Introduction}

Intermittency and anomalous scaling are two key features of turbulent flows important for fundamental questions of both out-of-equilibrium systems and applied flow configurations \cite{Sreenivasan:1997,Frisch:1995}. Although these phenomena have been subject of research for decades it is fair to say that we are still far from understanding their origin and controlling their statistical properties from first principles. Intermittency is connected to the strong non-Gaussian nature of turbulent energy dissipation, which is dominated by localized quasisingular structures. Anomalous scaling is connected to intermittency via the inertial-range turbulent energy cascade, which proceeds from large to small scales, breaking self-similarity, with power-law correlation functions that do not follow dimensional scaling. The two phenomena are correlated, with the small-scale energy dissipation being the result of the inertial-range energy transfer \cite{Frisch:1995}. The problem is therefore how to characterize the statistical properties of intense, but rare hydrodynamical fluctuations, an issue that is difficult to attack with brute force forward-in-time evolution of the underlying partial differential equations due to the unpredictability and sparsity of such events. This sobering state of affairs prompted repeated speculations whether techniques developed for quantum field theory (QFT) might eventually turn out to be useful to attack the existence of these (quasi)singular structures in a nonperturbative way, free from any modeling assumptions \citep{Kraichnan:1958,Rosen:1960a,Rosen:1960b,Wyld:1961,Forster:1976zz,Yakhot1986,Polyakov:1995mn,Antonov:2003}.

The way to proceed is to use the Janssen--de Dominicis \cite{Janssen1976,dedominicis:1976} path integral approach based on the seminal work by Martin \textit{et al.} \cite{Martin:1973zz,Rosen:1960b,Hosokawa:1966,Hosokawa:1968,Rosen:1983,Thacker:1997} to describe the space-time flow configuration when stirred by a random external forcing. This formalism is based on the introduction of an action that depends on the flow configuration and constructs the measure as a weighted sum of all possible flow realizations. This opens up the possibility to address Navier-Stokes equations using Markov chain Monte Carlo methods well known from lattice QFT and/or statistical mechanics by sampling full space-time histories. Although computationally challenging, this provides a unique perspective on the problem of turbulence in the sense that it allows us to consider systematic improvements of the importance sampling in regions of the phase space where standard (forward-in-time) numerical integration faces difficulties, e.g., due to insufficient statistics. In particular, it allows us to address questions regarding the probability of rare events associated with exceptionally large fluctuations, which are at the focus of turbulence research and often attacked by semi-analytical tools based on instanton calculus and large-deviation theory.

Instantons were introduced in turbulence theory in \cite{Gurarie:1995qc} where the probability densities of positive velocity gradients and increments (smooth ramps) were calculated analytically. The calculation of the probability densities of negative velocity gradients and increments (shocks) were performed in \cite{Balkovsky:1997zz} where the asymptotic behavior could be determined utilizing the Cole-Hopf transformation \cite{Cole:1951,Hopf:1950}. Using instantons in the calculation of rare, irregular transitions between different attractors in fluid flows was presented in \cite{Bouchet2009,Bouchet2011}.

The development of numerical methods for the investigation of rare events is a long-standing effort that has been pursued in many disciplines. Important examples are the adaptive multilevel splitting techniques (see \cite{brehier2016,Rolland2016,Ferre-Touchette:2018} and references therein), transition path sampling, \cite{Dellago1998,Bolhuis2002}, and the cloning algorithm \cite{Giardina2011}. Significant advances have also been established in the field of Molecular Dynamics \cite{TORRIE1977,Shirts2008}, where interesting rare events, such as protein folding, occur on disparate timescales \cite{karplus2002molecular}. See Ref.\ \cite{Ragone2018} for a recent study of extreme heat waves in climate models. Comparison of these methods with our path integral based approach is envisaged for future studies. 

The objective of this work is to implement, test, and employ a Hybrid Monte Carlo algorithm \cite{Duane:1987de,Sexton:1992nu} for hydrodynamic turbulence. The HMC algorithm was developed to tackle outstanding problems in the theory of strong interactions \cite{Gottlieb1987} and is advantageous for problems where the classical action involves nonlocal terms. We address the case of the one-dimensional random-noise-driven Burgers equation \cite{Burgers:1974}, which is widely considered the perfect testbed for new ideas in turbulence \cite{Bec:2007}. A previous attempt based on the path integral for hydrodynamical systems was explored in \cite{Duben:2008iw,Mesterhazy:2011kr,Mesterhazy:2013oaa}, based on a local successive over-relaxation algorithm \cite{Adler:1981sn,Adler:1987ce}.

From a methodological point of view, our first important result is the validation of the HMC against pseudospectral (PS) forward-time-integration techniques that are widely used in simulations of the random-noise-driven Burgers equation. We clearly stress that while the HMC is certainly not competitive with standard PS methods whenever the interest is confined to low order flow moments, e.g., the total mean energy and total mean energy dissipation, it becomes unavoidable if the focus is on very large fluctuations, e.g., either high order moments of velocity increments or extreme events for the space-time distribution of the energy dissipation. Indeed, the main quantitative result about the properties of the Burgers equation is the implementation of an importance sampling technique to steer the HMC algorithm to explore the phase-space region where rare and extreme fluctuations happen. We show later that due to several technical improvements of the basic HMC algorithm we are able to probe fluctuations 30 (and more) standard deviations away from the mean for the velocity gradient probability distribution function (PDF), something that would be simply impossible to achieve with standard time-advancing algorithms.

The outline of this article is as follows. In Sec.\ \ref{Sec:Stochastic Burgers equation: A simple model for hydrodynamic turbulence} we briefly discuss the phenomenology of the random-noise-driven Burgers equation and in Sec.\ \ref{Sec:Path integral for stochastic dynamics} we introduce the path integral for stochastic dynamics. Section \ref{Sec:Hybrid Monte Carlo algorithm} introduces the HMC algorithm and details the individual steps of our implementation. In Sec.\ \ref{Sec:Benchmarking the HMC against a first-order Euler-Maruyama explicit solver} we show that the HMC algorithm successfully reproduces the results of a standard PS forward-time-integration method [hereafter also referred to as direct numerical simulation (DNS)] at the example of the stochastic Burgers equation. In Sec.\ \ref{Sec:Constrained space-time evolution using HMC}, we investigate different boundary conditions and constraints in space and time. In Sec.\ \ref{Sec:Time-periodic boundary conditions} we impose periodic boundary conditions in time, while in Sec.\ \ref{Sec:Enhanced sampling of extreme and rare events} we show how the HMC is capable of consistently enhancing the sampling of extreme and rare events by imposing field-force constraints to systematically support the occurrence of strong negative velocity gradients. Here, we will also discuss the significant performance improvements by the HMC compared to a standard DNS method in regard to the sampling of the tails of the probability distribution function of observables. In Sec.\ \ref{Sec:The relevance of instantons in extreme events} we emphasize the significance of instantons for the theory of turbulence and derive the instanton configuration the Burgers equation. We also show numerical results associated with the methods developed in Sec.\ \ref{Sec:Enhanced sampling of extreme and rare events} to support the relevance of instantons in extreme events. We summarize in Sec. \ref{Sec:Conclusions}.

\section{Stochastic Burgers equation: A simple model for hydrodynamic turbulence}
\label{Sec:Stochastic Burgers equation: A simple model for hydrodynamic turbulence}

\begin{figure*}[!t]
  \centering
  \includegraphics[width=0.48\textwidth]{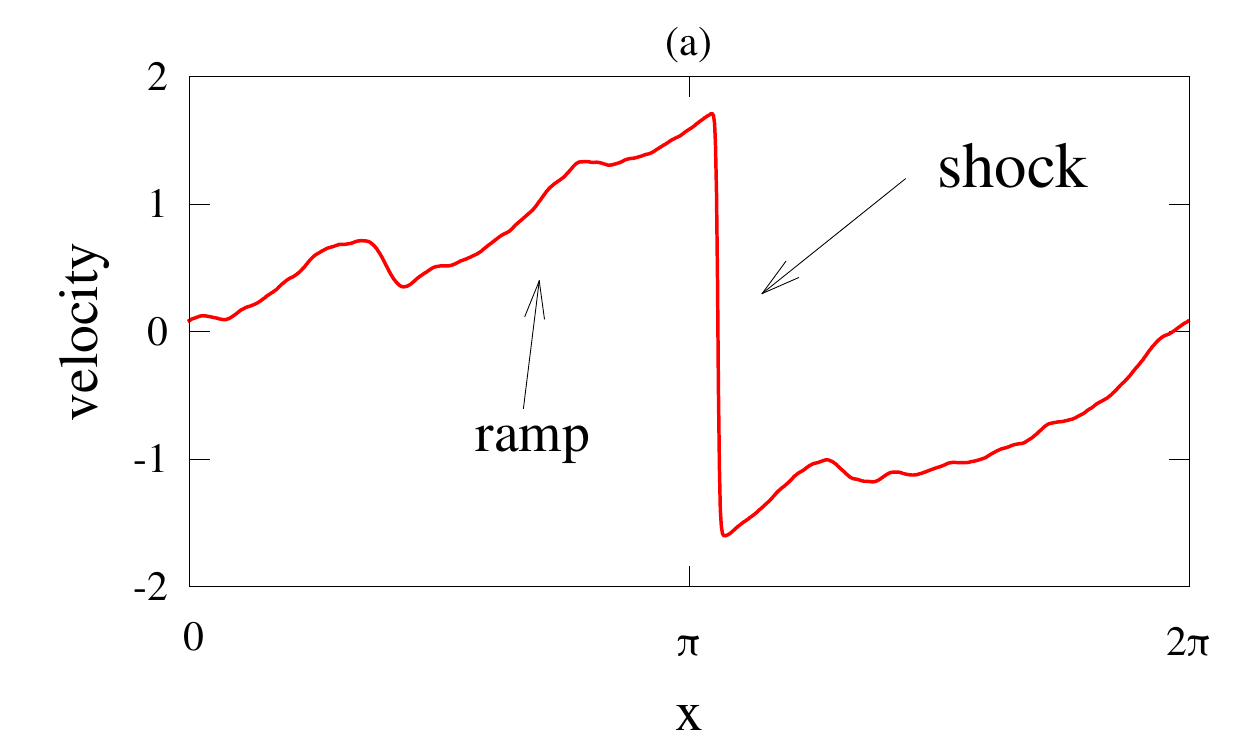} \quad
  \includegraphics[width=0.48\textwidth]{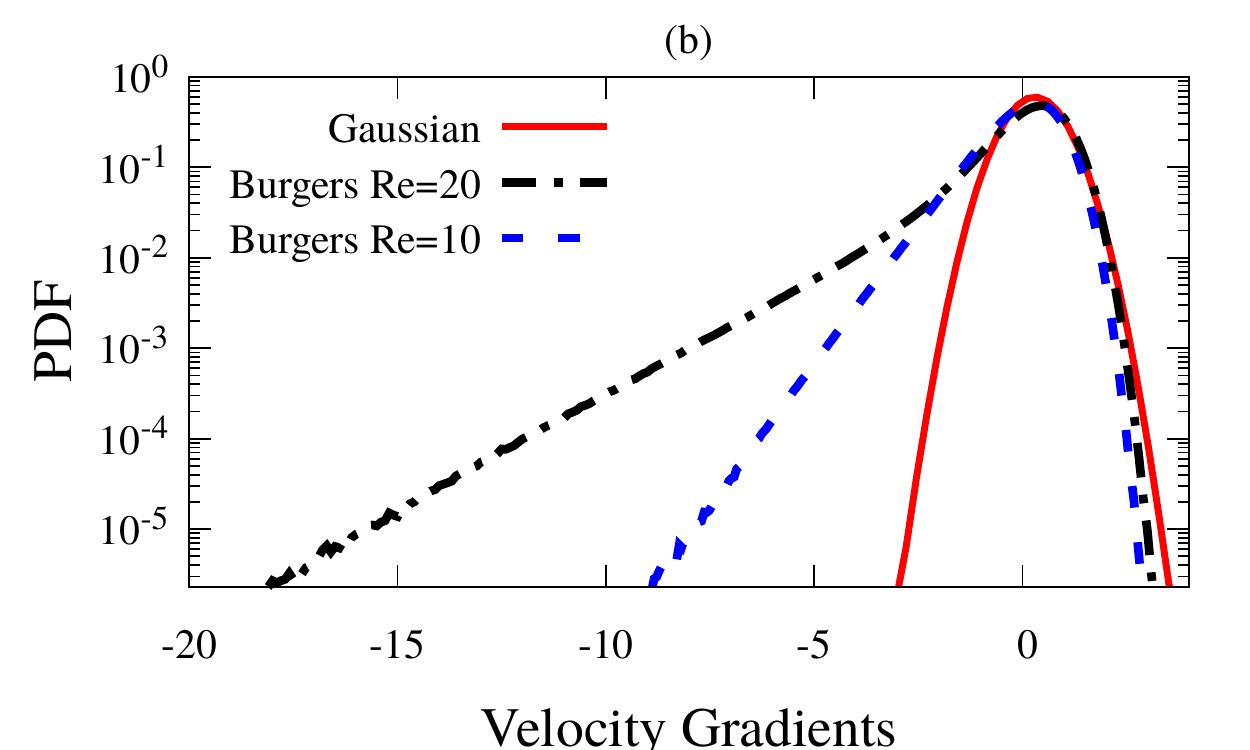}
  \caption{(a) Shock wave as a particular realization of Burgers equation. Its two main features are the ramps and jumps of finite width. (b) PDF of velocity gradients of Burgers equation for two different Reynolds numbers compared with a Gaussian distribution. The finite jumps in the shock wave profile are responsible for the heavy left tail, while the right tail is due to the ramps.}
\label{fig:burgers}
\end{figure*}

In this work we are concerned with the one-dimensional, random-noise-driven Burgers equation \cite{Burgers:1974}, which can be seen as a prototype system for compressible hydrodynamic turbulence and is given by
\begin{equation}
 \label{Eq:burgers}
 \partial_t v + v\partial_x v - \nu \partial_x^2 v = \eta .
\end{equation}
Specifically, we consider the time evolution of the scalar velocity field $v$ in a periodic spatial domain $x \in [-L/2,L/2]$ on a given time interval $t \in [t_0,t_f]$ of length $T = t_f-t_0$; $\nu$ is the kinematic viscosity and the random noise $\eta = \eta(x,t)$ is assumed to be centered and Gaussian distributed. Thus, the random noise can be fully characterized in terms of two-point correlations
\begin{equation}
\label{eq:pdfeta}
\langle \eta(x,t) \eta(x', t') \rangle \equiv \int\D\eta\,\P_\eta\, \eta(x,t) \eta(x',t') ,
\end{equation}
where $\P_\eta\equiv\P[\eta]$ is the random-noise probability distribution functional and the integration $\int\D\eta ~ \cdots$ is taken over all field configurations $\eta = \eta(x,t)$.

Generally, in the following, $\D\phi$ will denote a functional measure associated with the field $\phi$. Path integrals $\int\D\phi\,\, \cdots$ will always be supplied with ``boundary conditions'' in field space. Furthermore, where appropriate, ensemble averages will be denoted by angular brackets, $\langle \,\cdots\,\rangle$. For example, for an observable $\Ob_\phi\equiv \Ob[\phi]$, we have $\langle\Ob_\phi\rangle = \int\D\phi\, \P_\phi\, \Ob_\phi$ and $\int \D\phi\, \P_\phi = 1$. Depending on context, we might drop the index indicating the field degrees of freedom to be averaged over.

In this paper we restrict our attention to the case where the random noise is  self-similar in space and $\delta$ correlated in time, with the corresponding two-point Fourier correlation given by
\begin{equation}
  \langle \eta(k,t) \eta(k',t') \rangle = \Gamma(k)  \delta_{k+k',0} \delta (t-t') ,
  \label{eq:forcing_cor}
\end{equation}
where $k,k' \in \mathbb{Z}$ and $\Gamma(k) = \Gamma_0 |k|^\beta$ with a negative power-law exponent $\beta$ that controls the scale-by-scale energy injection.\footnote{In practice, when $\beta < -1$, the correlator will be regularized by an infrared cutoff $k_{\rm IR} \sim 1/L$ and $k_{\rm IR} > 0$.}

From Eq.\ \eqref{Eq:burgers} it is easy to derive the evolution equation for the energy spectrum $E(k,t) = |v(k,t)|^2$,
\begin{equation}
  \label{Eq:burgersF}
  \partial_t E(k,t) = T(k,t) - 2 \nu k^2 E(k,t) + 2 \real \{ v(k,t)^\ast \eta(k,t) \} ,
\end{equation}
where $T(k,t) = (k/L) \sum_{k'} \imag \{v(k,t)^\ast v(k',t) v(k-k',t) \}$ is the energy transfer \cite{Frisch:1995}. Equation \eqref{Eq:burgersF} can be further simplified if we average over noise realizations and assume stationarity:
\begin{equation}
  \label{Eq:burgersF2}
  \langle T(k)\rangle - 2 \nu k^2 \langle E(k) \rangle + 2 \Gamma(k) = 0 .
\end{equation}
The ensemble-average cumulative energy injection due to the stochastic forcing $\eta$ is given by $\langle \varepsilon_{{\rm in}}(k) \rangle = (2/L) \sum_{|k'|\leq k} \Gamma (k')$ and is dominated by the infrared regime only if $\beta < -1$. Thus, in order to mimic the standard large-scale injection we will always keep $\beta = -3$ in this paper (see \cite{Medina:1989,Chekhlov:1995a,Hayot:1996,Verma:2000ue} for a detailed investigation of the statistical properties at changing the forcing slope).

It is well known that the evolution to the Burgers equation is characterized by the formation of quasisingular shocks, i.e., localized events with a steep negative velocity gradient where all the dissipation is concentrated. In the small-viscosity limit the typical width of the shock becomes smaller, but the ensemble-average mean energy dissipation $\langle \varepsilon_{{\rm diss}} \rangle = (2\nu / L) \sum_k k^2 \langle E(k) \rangle$ remains nonvanishing. Since $T(k)$ only transfers the energy between different modes but does not contribute to the total energy, the total energy injection matches the energy dissipation:
\begin{equation}
  \label{Eq:balance}
  \lim\limits_{k\to\infty} \langle \varepsilon_{{\rm in}}(k) \rangle = \langle \varepsilon_{{\rm diss}} \rangle .
\end{equation}

Writing Eq.\ \eqref{Eq:burgers} in a dimensionless way reveals that the problem has only one control parameter, the Reynolds number, $\Re$. This is made manifest by introducing characteristic scales of length $L_0$, and velocity $V_0$, and a time scale $T_0 = L_0 / V_0$. We change $x$, $t$, $v$, and $\eta$ according to
\begin{eqnarray}
&& x \mapsto x \, L_0 , \quad v \mapsto v \, V_0, \quad t \mapsto t \, T_0 = t \, L_0 / V_0, \nonumber\\ && \eta \mapsto \eta \, V_0 / T_0 = \eta \, V_0^2/L_0,
\end{eqnarray}
to obtain the dimensionless stochastic Burgers equation
\begin{equation}
  \label{Eq:burgers_nondim}
  \partial_t v + v\partial_x v - \frac{1}{\Re} \partial_x^2 v = \eta ,
\end{equation}
with $\Re \equiv L_0 V_0 / \nu$. Consequently, in the remainder of this paper we will speak about the large-Reynolds-number and small-viscosity limits interchangeably.

In Figure\ \ref{fig:burgers} two of the most characteristic elements of Burgers turbulence are shown. Fig.\ \ref{fig:burgers}(a) depicts the shock formation as a solution of the Burgers equation, which is described by the finite-width jumps and approximately linear ramps. In Fig.\ \ref{fig:burgers}(b) we show the probability distribution function of the velocity gradients, defined as
\begin{equation}
  P(w) = \langle \delta (\partial_x v(x,t) - w) \rangle .
\end{equation}
The localized jumps at the shock are the source of intermittency in this model, and contribute to the heavy left tail of the PDF, while the ramps are related to the right tail (we refer the reader to \cite{Bec:2007} for an in-depth review of Burgers turbulence). From the previous discussion the question we ultimately want to address becomes clear: Is it possible to develop algorithms which are able to focus specifically on the phenomenon of shock formation by exploring only the far left tail of the PDF shown in Fig.\ \ref{fig:burgers}(b)? This will be the aim of the HMC approach we propose.

\section{Path integral for stochastic dynamics}
\label{Sec:Path integral for stochastic dynamics}

The path integral for stochastic dynamics was first introduced in Refs.\ \cite{Phythian:1977,Langouche:1979,Jensen:1981}. To make our exposition self-consistent, however, we briefly repeat the main steps of its derivation. While we employ the same notation as in Eq.\ \eqref{Eq:burgers} we emphasize that the following reasoning is in principle applicable to any stochastic (partial) differential equation (SPDE) driven by Gaussian random noise, $\delta$ correlated in time. We will denote these SPDEs by the following short-hand notation
\begin{equation}
  \label{Eq:stochastic_eq}
   F(x, t, v, \partial_x^m v, \partial_t^n v) = \eta ,
\end{equation}
with $m,n\in \mathbb{N}_0$. Here $F\equiv F(x, t, v, \partial_x^m v, \partial_t^n v)$ should be interpreted as a (nonlinear) differential operator, which acts on the dynamical field $v = v(x, t)$. We will only make some minimal assumptions regarding its form, namely, that it should yield a well-posed initial value problem. By well-posed we mean that for any given random noise realization $\eta$, there exists one and only one solution $v$ to Eq.\ \eqref{Eq:stochastic_eq} in the domain $-L/2 \leq x\leq L/2$ and for finite times $0\leq t\leq T$.

To derive the path integral associated with Eq.\ \eqref{Eq:stochastic_eq} we define the partition sum $\Z$ by integrating $\P_\eta$ over all noise realizations. Since $\eta$ is Gaussian and white in time, we have
\begin{equation}
  \label{eq:prob_distr_eta}
  \P_\eta \propto e^{-\frac{1}{2} \int dt\, \int dx\, \eta(x,t) \int dx' \,\Gamma^{-1}(x-x') \eta(x',t)} ,
\end{equation}
where $\Gamma^{-1}$ is the inverse of the correlation function of the noise defined in Eq.\ \eqref{eq:forcing_cor}. Accordingly, we define the partition sum as
\begin{equation}
  \Z = \int\D\eta\, e^{-\frac{1}{2} \int dt \, (\eta, \,\Gamma^{-1} \ast \eta )} ,
  \label{eq:functional_int_noise}
\end{equation}
where the binary operator $\ast$ denotes the convolution, i.e., $(f\ast g)(x) = \int dx'\, f(x') g(x-x')$ and by $(\cdot\,, \cdot)$ we designate the integral over the (bounded) spatial domain $[-L/2, L/2]$, i.e., $(f\,,g) \equiv \int dx\, f(x)\, g(x)$, with $||f||^2 \equiv (f,f)$. Changing the integration in Eq.\ \eqref{eq:functional_int_noise} from $\eta$ to $v$ modifies the functional measure as
\begin{equation}
  \D\eta = \Dv \left|\det\left( \delta F /\delta v \right)\right| ,
  \label{Jacobian}
\end{equation}
where $\J = \left|\det\left( \delta F /\delta v \right)\right|$ is the Jacobian associated with the map $v\mapsto \eta$. The latter is assumed to be nonsingular and therefore $\J > 0$.

Putting everything together, we may write the partition sum in the form of a path integral over $v$,
\begin{eqnarray}
  \Z &=& \int\Dv\, \J e^{-\frac{1}{2} \int dt \, (F, \Gamma^{-1} \ast F)} \equiv \int\Dv \, e^{-\S},
  \label{eq:functionalint_mean}
\end{eqnarray}
with the action
\begin{equation}
  \S = \frac{1}{2} \int dt \, (F, \,\Gamma^{-1} \ast F) - \ln\J ,
  \label{eq:action_S}
\end{equation}
associated with the stochastic partial differential equation (PDE) \eqref{Eq:stochastic_eq}. Note that the action bears resemblance to the well-known Onsager-Machlup functional \cite{Onsager:1953}. More generally, one arrives at the following action:
\begin{equation}
  \label{eq:general_action}
  \S = -\ln \{\Z \P_\eta[\eta = F] \} - \ln\J ,
\end{equation}
The probability distribution functional $\P_v$ for the dynamical field $v$ is given by $\P_v = \Z^{-1} e^{-\S}$ and satisfies the normalization condition $\int\Dv\, \P_v = 1$. Specifically, for the Burgers equation, $F = \partial_t v + v\partial_x v - \nu \partial_x^2 v$ and $\J = \textrm{const.}$ (which holds for causal forward-time propagation, see, e.g., \cite{Nakazato:1990kk} and Sec.\ \ref{Sec:Benchmarking the HMC against a first-order Euler-Maruyama explicit solver} of this paper), the action \eqref{eq:action_S} takes the form
\begin{eqnarray}
  \S &=& \frac{1}{2} \int dt \int dx\, \left(\partial_t v + v\partial_x v - \nu \partial_x^2 v \right) \nonumber\\ && \hspace{10pt} \times \int dx' \, \Gamma^{-1}(x-x') \left( \partial_{t} v + v\partial_{x'} v - \nu \partial_{x'}^2 v \right) , \nonumber\\[-8pt]
  && 
    \label{eq:action_S_burgers}
\end{eqnarray}
where we dropped the constant contribution from the Jacobian.

\section{Hybrid Monte Carlo algorithm}
\label{Sec:Hybrid Monte Carlo algorithm}

The Hybrid Monte Carlo algorithm, originally introduced in \cite{Duane:1987de}, has become a standard computational tool to tackle demanding numerical simulations of quantum field theories in the path-integral formulation (see \cite{Kennedy:2006ax,Luscher2010ae} for reviews). It belongs to the broad class of Markov Chain Monte Carlo methods, and uses artificial Hamiltonian dynamics, frequently termed Molecular Dynamics (MD), to advance the dynamical degrees of freedom in Monte Carlo time to generate unbiased field samples. A main feature of the HMC is that dynamical fields and their conjugate momenta can be evolved in parallel in a given time step of the evolution, if, e.g., a leap-frog type integrator is used. This makes the HMC most suitable for problems where the classical action of the theory features nonlocal terms. They may arise from the stochastic equation itself (as the pressure term in the Navier-Stokes equation), or, in addition, from the convolution with the inverse force correlator (as in the present case).

In this work we apply the HMC algorithm for a stochastically driven PDE to the example of the Burgers equation. In order to be self-contained, we will first briefly review its basic elements. Then we proceed to discuss important improvements to the HMC algorithm, which allow for a significant enhancement of performance to sample various statistical estimators in a stable and consistent way.

In the HMC algorithm a set of momenta is introduced which are conjugate to the, in our case, velocity fields. Adding these momenta to the partition sum of Eq.\ \eqref{eq:functionalint_mean} leads to an (artificial) Hamiltonian, which governs the dynamics in a fictitious Molecular Dynamics time $s$ via Hamilton's equations of motion. In practice, the numerical solution of Hamilton's equations starting from some initial  MD time, say $s = 0$ to a final time $s = \tau$ has to be performed in discrete steps, which leads to the fact that the energy of the artificial Hamiltonian system is not conserved. This can be repaired by adding a global reject or accept step which makes the algorithm exact and guarantees the convergence to the desired probability distribution. See \cite{Neal2012} for a general review of the HMC algorithm.

To be more concrete, the HMC algorithm starts by generating a set of Gaussian-distributed momenta $\pi = \pi(x,t)$ such that the partition sum is modified as
\begin{equation}
  \Z \propto \int\D\pi\, e^{-\frac{1}{2} \int dt\, ||\pi(t)|| ^2} \int\Dv\, e^{-\S} .
  \label{eq:hami}
\end{equation}
Identifying $\K = \frac{1}{2} \int dt\, ||\pi(t)||^2$ as the ``kinetic term'' and $\S$ as the ``potential'' we may interpret $\H = \K + \S$ as the Hamiltonian of the system, with the probability distribution functional: $\P_{v,\pi} \propto e^{-\H}$. Since $\int \D\pi\, \P_{(v,\pi)} = \P_v$, the ensemble average of any velocity-dependent observable $\Ob_v$ remains unaltered. The so constructed Hamiltonian system can now be evolved using Hamilton's equations of motion. In this evolution, the role of ``time'' is played by $s$. In order to make the dependence on  $s$ explicit, we will introduce $v_s(x,t)$ and $\pi_s(x,t)$ where the subscript indicates the  MD time. Hamilton's equations for the Hamiltonian $\H$, with the action as in Eq.\ \eqref{eq:action_S_burgers}, are then given by
\begin{subequations}
\label{eq:MD}
  \begin{align}
  \frac{dv_s}{ds} &= \frac{\delta \H}{\delta \pi_s(x,t)} = \pi_s(x,t) , \\
  \frac{d\pi_s}{ds} &= -\frac{\delta \H}{\delta v_s(x,t)} = -\frac{\delta \S}{\delta v_s(x,t)} .
\end{align}
\end{subequations}
In the case of the one-dimensional Burgers equation, the Molecular Dynamics forces $\Phi_\pi\equiv -\frac{\delta \S}{\delta v_s(x,t)}$ acting on the conjugate momenta are given by
\begin{eqnarray}
  \Phi_\pi(x,t) &=& \left(\partial_t + v\partial_x + \nu \partial_x^2 \right) \nonumber\\ && \hspace{10pt} \times \int dx' \, \Gamma^{-1}(x-x') \left( \partial_{t} v + v\partial_{x'} v - \nu \partial_{x'}^2 v \right) . \nonumber\\[-8pt]
  \label{eq:MD_forces}
\end{eqnarray}

\subsection{HMC implementation}
\label{Sec:HMC implementation}

The equations of motion \eqref{eq:MD} are solved for $(v_s, \pi_s)$, $0\leq s\leq \tau$, starting at  MD time $s = 0$ and integrating up to $s = \tau$; $\tau$ defines the trajectory length. We apply a symmetric symplectic integrator (leapfrog scheme) with stepsize $\Delta \tau = \tau / N_{\Delta \tau}$, with $N_{\Delta \tau}$ a parameter that is the discrete number of steps of the Hamiltonian evolution for a single HMC iteration, so that the trajectory of length $\tau$ is completed.

Due to the finite integration stepsize error, the Hamiltonian reached at $s=\tau$ will be different from the initial Hamiltonian. To correct for this deficiency, we apply a global Metropolis accept or reject step of the proposed new momentum and velocity field configuration: The new field configuration is accepted with probability $p = \min \left( 1, e^{-\Delta \H} \right)$, where $\Delta \H = \H[v_{\tau}, \pi_{\tau}] - \H[v_0, \pi_0]$, i.e., the difference of the Hamiltonian at the beginning and the end of the trajectory. If the proposal is rejected, we resample the conjugate momenta and restart from the old set $v = v_0$. After each completed HMC iteration, we resample the momenta, regardless the outcome of the Metropolis step. This is necessary to satisfy ergodicity. Other important requirements for the HMC algorithm to be exact are the preservation of the phase-space volume and the reversibility in the fictitious time $s$. These are inherent properties of the Hamiltonian dynamics.

Reversibility, which is a necessary condition of detailed balance, is in practice measured by first performing a Hamiltonian evolution $(v_{0}, \pi_{0}) \mapsto (v_{\tau}, \pi_{\tau})$, and then negating the momenta $\pi_{\tau} \mapsto -\pi_{\tau}$. Now, starting from $(v_{\tau}, -\pi_{\tau})$ and performing another Hamiltonian evolution, we return to $(v'_{0}, \pi'_{0})$. In any numerical implementation we expect violations of reversibility which are quantified by $\max(|v'_0 - v_0|)/v_{\text rms}$. Large reversibility violations will spoil the invariance of the desired distribution (here $e^{-\S}$) under the HMC updates. Therefore, reversibility violations need to be monitored in the actual simulation. Indeed, we checked that reversibility violations are negligible in our simulations, i.e., of order $10^{-12} - 10^{-14}$.

\begin{figure}[t!]
  \centering
  \includegraphics[width=\linewidth]{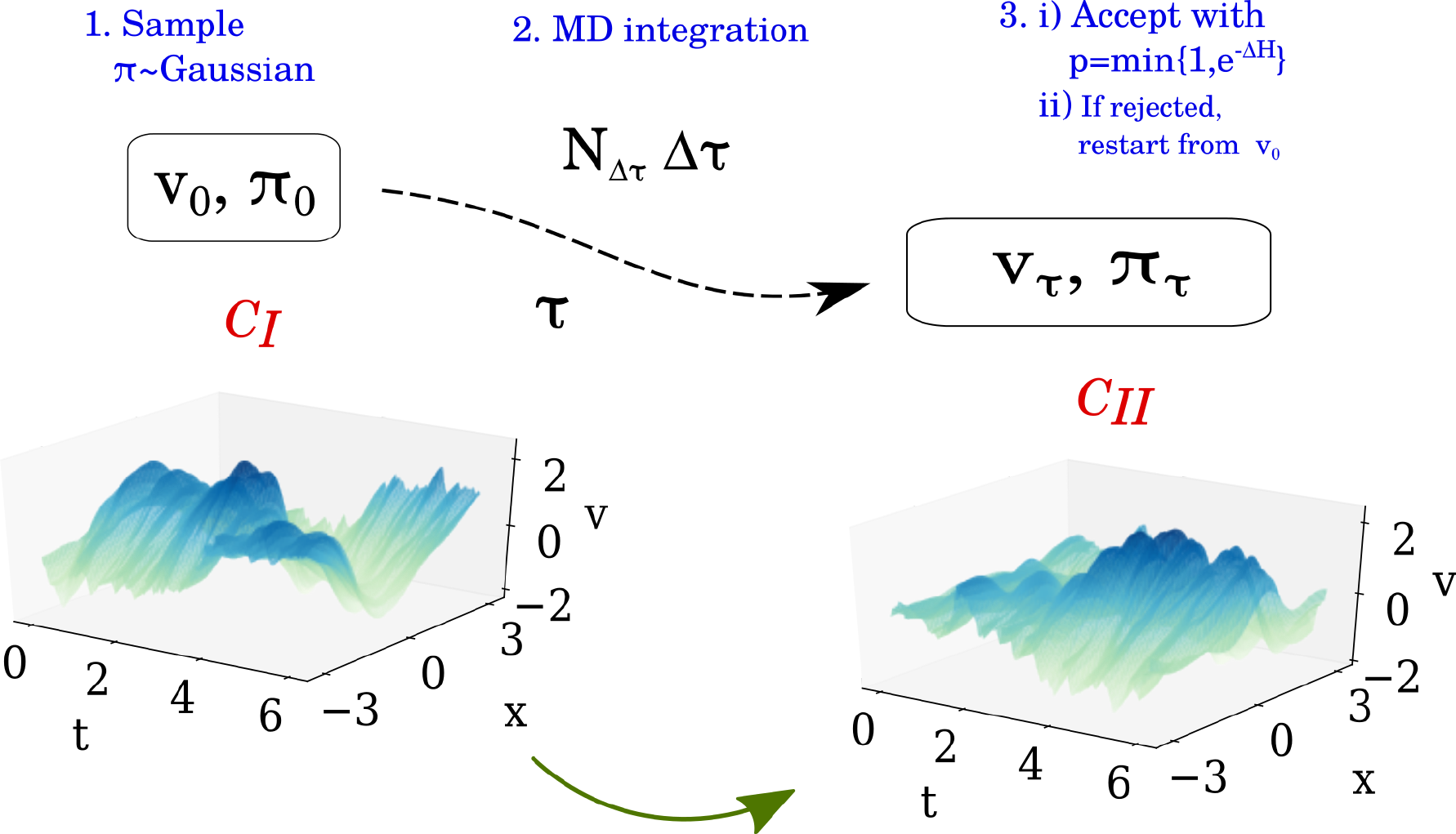}
  \caption{Schematic description of a single HMC iteration of trajectory length $\tau$. The dashes in the top arrow represent the $N_{\Delta \tau}$ number of intermediate steps of size $\Delta\tau$, so that it holds $\tau = N_{_{\Delta \tau}} \, \Delta\tau$. After the numerical integration of Eqs.\ \eqref{eq:MD}, a configuration $v_\tau$ is proposed, and is accepted with probability $p = \min \left(1, e^{-\Delta \H}\right)$. If accepted, we use $v_\tau$ as the initial field for the next iteration, otherwise we restart from $v_0$. In both cases, we discard $\pi_\tau$ and resample the momenta according to the Gaussian distribution for the next HMC iteration. (Note that the conjugate momentum fields are not plotted).}
  \label{fig:conf_phase_space}
\end{figure}

Let us finally briefly summarize, and illustrate in Fig.\ \ref{fig:conf_phase_space}, the three basic steps of the HMC algorithm for a single HMC iteration:
\begin{enumerate}
\item {\it Momentum heat bath.} Sample $\pi_0$ according to the Gaussian distribution 
  \begin{equation}
    \P_\pi \propto e^{-\frac{1}{2} \int\, dt\, ||\pi_0(t)||^2} .
  \end{equation}
\item {\it Hamiltonian evolution.} Use a symplectic integrator to numerically solve the system of equations (\ref{eq:MD}) starting from $(v_0, \pi_0)$ and propose $(v_{\tau}, \pi_{\tau})$.
\item {\it Metropolis step.} Accept the proposed field configuration $(v_{\tau}, \pi_{\tau})$ with probability
  \begin{equation}
    p = \min \left(1, e^{-\Delta \H}\right) ,
  \end{equation}
  where $\Delta \H = \H[v_{\tau}, \pi_{\tau}] - \H[v_0, \pi_0]$.
\end{enumerate}

The generated ensemble is a Markov chain of configurations $v_s(x,t)$ that results from numerous HMC iterations, by repeating steps 1-3. In our simulations the parameters $\tau$ and $N_{\Delta \tau}$ are tuned such that acceptance rate originating from step 3 is close to 90\% and more. This ensures that the autocorrelation time does not become too large and also avoids too many rejected velocity configurations. 

\subsection{Fourier acceleration}
\label{sec:Fourier acceleration}

We observe that the application of the standard HMC, based on Eq.\ \eqref{eq:hami}, leads to very large autocorrelation times. The problem with the large autocorrelation time is essentially due to the multiscale nature of the stochastic forcing, which in turn means that different Fourier modes are forced with different intensity. In order to deal with this problem, we made use of a well-known approach from the area of lattice field theory, i.e., the method of Fourier acceleration \cite{Batrouni1985,Davies:1989vh,Catterall2002,Gerhold:2010wy}. The latter assigns different effective trajectory lengths to the evolution of the Fourier modes. Indeed, this technique proved highly effective in our approach and it improved the performance of the HMC algorithm by considerably decreasing autocorrelation effects.

In practice, we apply the Fourier acceleration by introducing the space-time dependent kernel $\Omega(x,t)$ to multiply the momenta $\pi_s(x,t)$. This gives rise to the following ``effective Hamiltonian''
\begin{eqnarray}
  \H^{\textrm{eff}} &=& \frac{1}{2} \int dt (\pi_s, \Omega \ast \pi_s) + \S .
  \label{eq:eff_hami}
\end{eqnarray}
It is important to note that we consider $\Omega(x,t)$ as being independent of the  MD time $s$. The introduction of the kernel $\Omega(x,t)$ and the redefinition of the Hamiltonian do not affect the physical results, as the redefined kinematic term is still independent of the velocity field, and can be factored out of the path integral \eqref{eq:hami}.

In our HMC implementation, we propose to define $\Omega$ as
\begin{equation}
\Omega(x, t) \propto \frac{1}{\langle \left|\Phi_\pi(x, t)\right| \rangle^{2}} ,
\label{Eq:Omega}
\end{equation}
with $\Phi_\pi(x,t) =-\frac{\delta \S}{\delta v_{s=\tau}(x, t)}$ the MD forces recorded at $s = \tau$, and defined in Eq.\ \eqref{eq:MD_forces}. For a detailed derivation see Appendix \ref{Sec_ap:Implementation of the leapfrog integrator and the Fourier acceleration}. The ensemble average on the right-hand side of Eq.\ \eqref{Eq:Omega} denotes the MC average and implies that the Fourier acceleration scheme needs to be adjusted dynamically during the initial convergence phase of the Markov chain. Initially, we use the ansatz
\begin{equation}
\Omega (k,t) \propto \Gamma^{-1}(k) \delta(t-t').
\label{Eq:Omega_ini}
\end{equation}
Then, after a fixed number of HMC steps, a tuning stage follows in which we measure the forces $\Phi_\pi$. This tuning stage can be composed of several cycles, while at the end of each cycle we set $\Omega$ to the new forces. As soon as $\Omega$ has converged, the tuning stage is completed, and we can start the measurement of physical observables. Typically, we require five to ten reset cycles of about $10^3$--$10^4$ iterations, to achieve an optimal choice of $\Omega$. Finally, since the stochastic forcing is $\delta$ correlated in time, it turns out in practice that the final choice for $\Omega$ is approximately time independent, i.e., $\Omega(k,t) \approx \Omega(k)\delta(t-t')$. The technical details regarding the implementation of the Fourier acceleration will be discussed elsewhere \cite{HMC2019b} and are beyond the scope of the present paper.

\section{Benchmarking the HMC against a first-order Euler-Maruyama explicit solver}
\label{Sec:Benchmarking the HMC against a first-order Euler-Maruyama explicit solver}

\subsection{Fixed/open boundary conditions in time}
\label{Sec:Fixed/open boundary conditions in time}

As this is a different approach for the sampling of stochastic PDEs, we took considerable care to benchmark the HMC with standard numerical methods that are employed in computational fluid dynamics. In the following we will demonstrate that our simulations match results obtained via a first-order Euler-Maruyama explicit solver (DNS) for a wide range of viscosities \cite{Higham2001,Kloeden:1992}.

\begin{table}[!t]
  \centering
  \begin{tabular}{cccccccc}
    &$\nu$ & $\Re$ &$v_{\textrm{rms}}$ &$\ell_d$ &$\ell$  &$\langle\bar\varepsilon_{{\rm diss}} \rangle$ &$T_\ell$  \\ \hline \hline
    &0.08 &90 &1.14  &0.15  &1.5  &1 &1.31    \\
    &0.1 &70 &1.12  &0.18  &1.4  &1 &1.24     \\
    &0.2 &30 &1.03  &0.3  &1.1   &1 &1.03  \\ \hline
  \end{tabular}
  \caption{\label{table1} Parameters and observables of the numerical simulations for fixed and open boundary conditions. Here we employ the following parameters:
    $N_t=1056$ number of grid points in time,
    $N_x=128$ number of grid points in space,
    $T=6$ and $L=2\pi$.
    The Reynolds number is defined as $\Re = \frac{v_{\textrm{rms}} \, L}{\nu}$ with root-mean-square velocity
    $v_{\textrm{rms}}=\langle \sqrt{||v||^2/L} \rangle$. Here
    $\ell_d =(\nu^3/\langle \bar\varepsilon_{{\rm diss}}  \rangle)^{1/4}$ defines the Kolmogorov dissipation length scale and 
    $\ell = \frac{\bar\varepsilon_{\rm kin}^{3/2}}{\langle \bar\varepsilon_{{\rm diss}}\rangle}$ is the integral length scale 
    with $\bar\varepsilon_{\rm kin} = v_{\rm rms}^2$. In addition, $\langle \bar\varepsilon_{{\rm diss}}\rangle$ denotes the ensemble-average mean energy dissipation, i.e., $\langle \bar\varepsilon_{{\rm diss}} \rangle= 2 \nu \left\langle \left|\left|\partial_x v \right|\right|^2 \right\rangle /L$ and $T_\ell = \ell/v_\mathrm{rms}$ is the large-eddy turnover time.}
\end{table}

\begin{figure}[!t]
	\centering
	\includegraphics[width=0.48\textwidth]{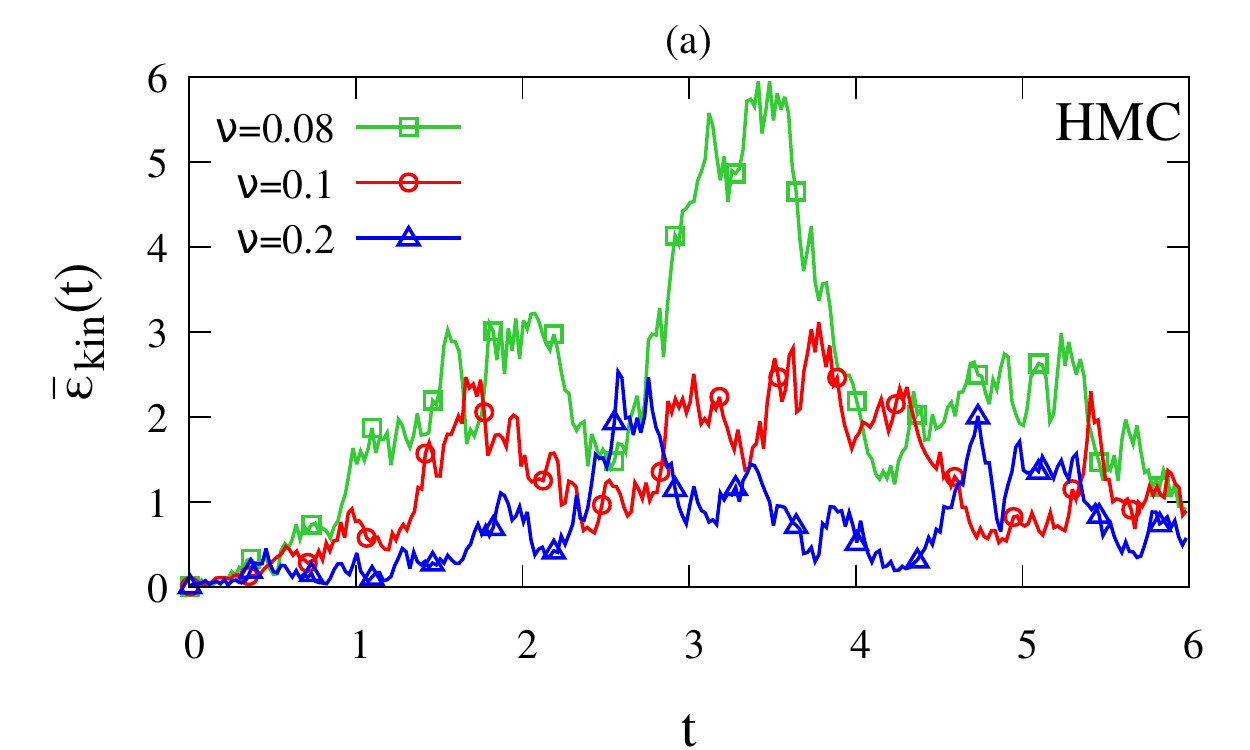} \quad
	\includegraphics[width=0.48\textwidth]{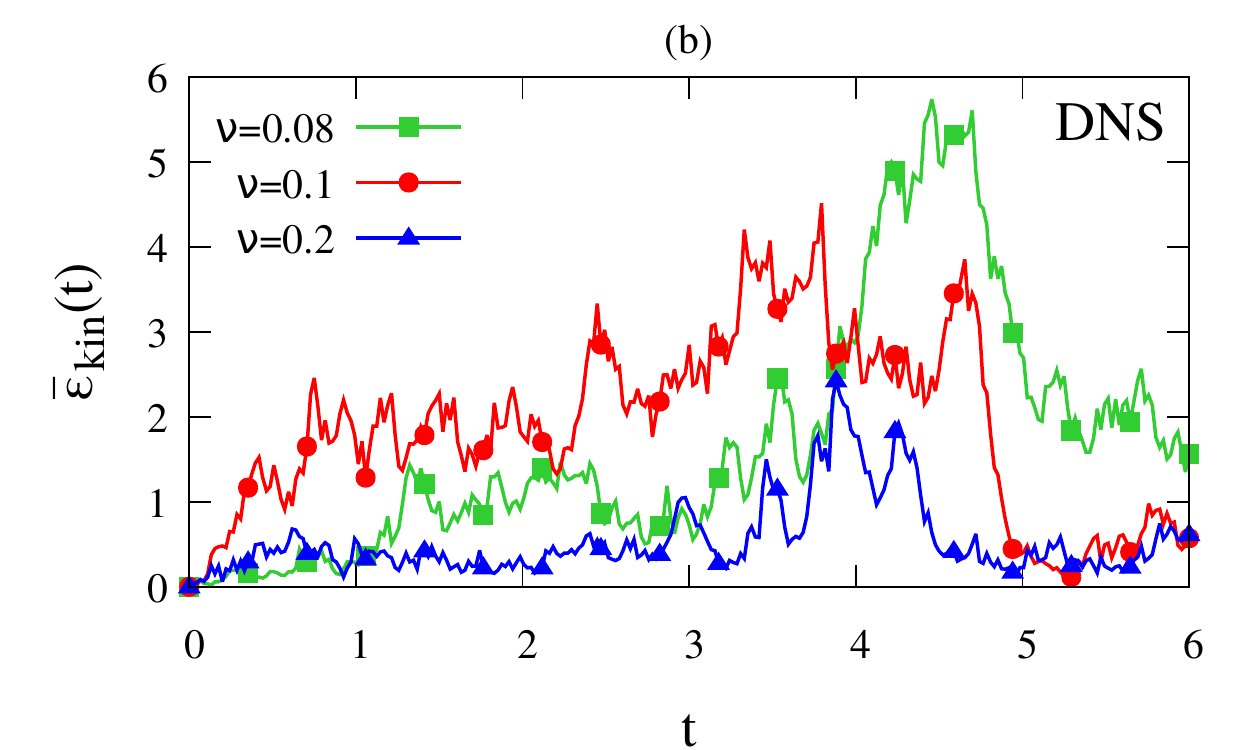} 
	\caption{Kinetic energy as a function of time for three samples corresponding to three different viscosities for the (a) HMC and (b) DNS.}
	\label{fig:kin_energy}
\end{figure}

\begin{figure}[!t]
	\centering
	\includegraphics[width=0.48\textwidth]{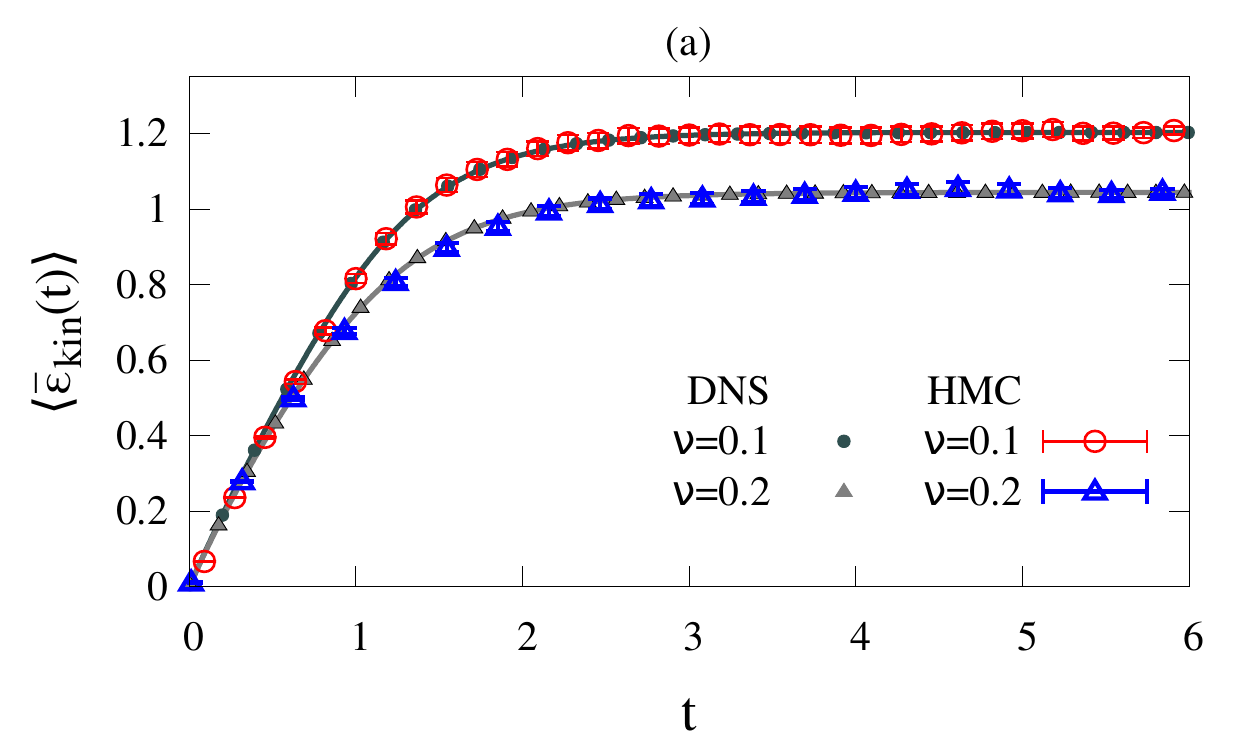} \quad
	\includegraphics[width=0.48\textwidth]{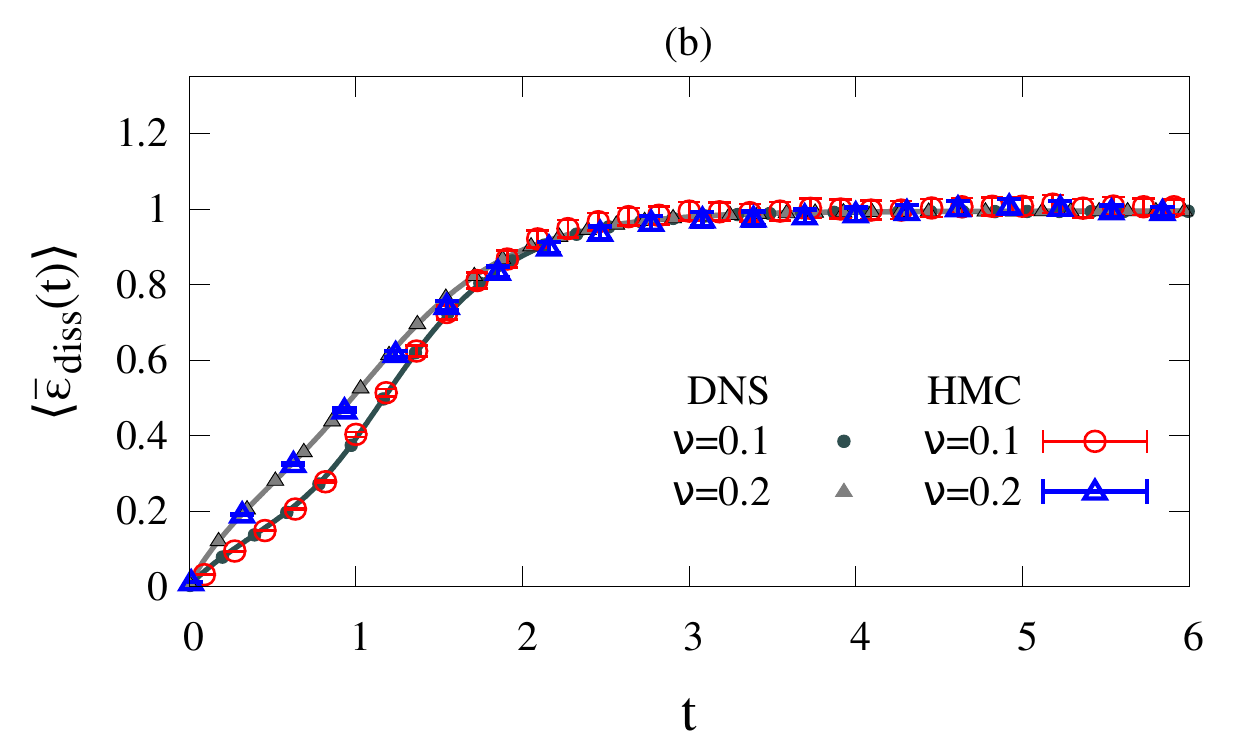} 
	\caption{(a) Temporal evolution of the ensemble-average mean kinetic energy, $\langle \bar\varepsilon_{\rm kin}(t) \rangle $. (b) Temporal evolution of the ensemble-average mean energy dissipation $\langle \bar\varepsilon_{{\rm diss}} (t) \rangle$. Note that $\langle \bar\varepsilon_{{\rm diss}} (t>t_s) \rangle = 1$ because the energy injection is fixed at $\lim\limits_{k\to\infty}\langle \bar\varepsilon_{{\rm in}}(k) \rangle = 1$; Closed gray symbols correspond to the DNS results (lines denote their interpolation) and  open colored symbols correspond to the HMC results.}
	\label{fig:dissip}
\end{figure}

In both implementations, the spatio-temporal domain is discretized uniformly and the Burgers equation is expressed in Fourier space, with the nonlinear term being written in a flux-conservative form, i.e., $v\partial_x v = \frac{1}{2}\partial_x(v^2)$. We apply the pseudospectral method, i.e., first $v^2$ is measured in real space and afterward transformed to Fourier space so that the partial derivative can be conveniently treated as $\partial_x \mapsto ik$. Therefore, the nonlinear term is calculated as $\tfrac{ik}{2} \F(v^2)$, where by $\F$ we denote the (forward) Fourier transform. To further ensure stability we apply two further steps in the numerics: First, we transform $v(k,t) \to \exp ( -\nu \, k^2 \,\Delta t ) v(k,t)$, which corresponds to an exact integration of the viscous term. It relaxes the restriction on the time step $\Delta t$ by the diffusive term and significantly improves the convergence for large wave numbers. Second, we effectively remove the aliasing error by setting $v(2 \pi |k| / L\ge N_x/3, t) = 0$ (see \cite{Canuto1988}).

Here we present three different runs, with the parameters summarized in Tables.\ \ref{table1} and \ref{table4}. Both the DNS and the HMC share the same setup, i.e., the same forcing correlation function, the same discretization, and the same periodic boundary conditions in space. As for the HMC, we choose fixed and open boundary conditions in time, corresponding to a standard initial-value problem. Note, that this choice yields a Jacobian $\J$ that is field-independent \cite{Nakazato:1990kk}, which therefore can be neglected for the purposes of importance sampling.

\begin{figure*}[!t]
  \centering
  \includegraphics[width=0.48\textwidth]{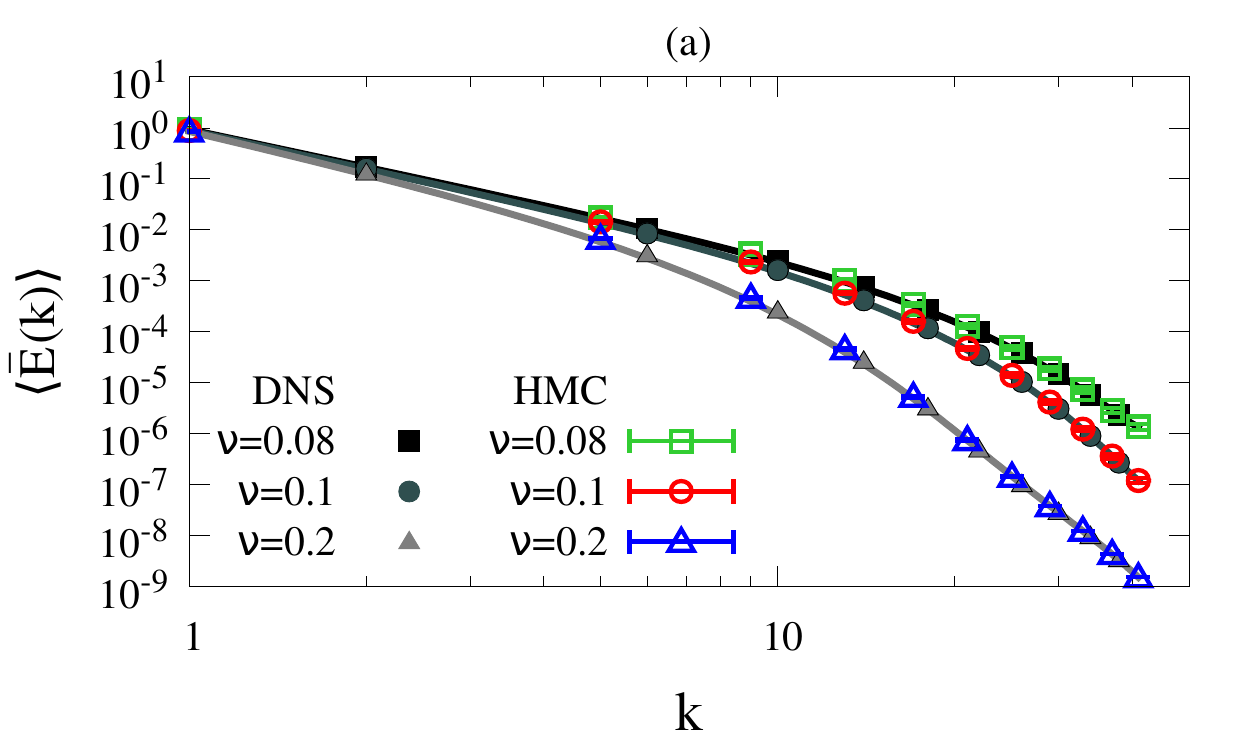}  \quad
  \includegraphics[width=0.48\textwidth]{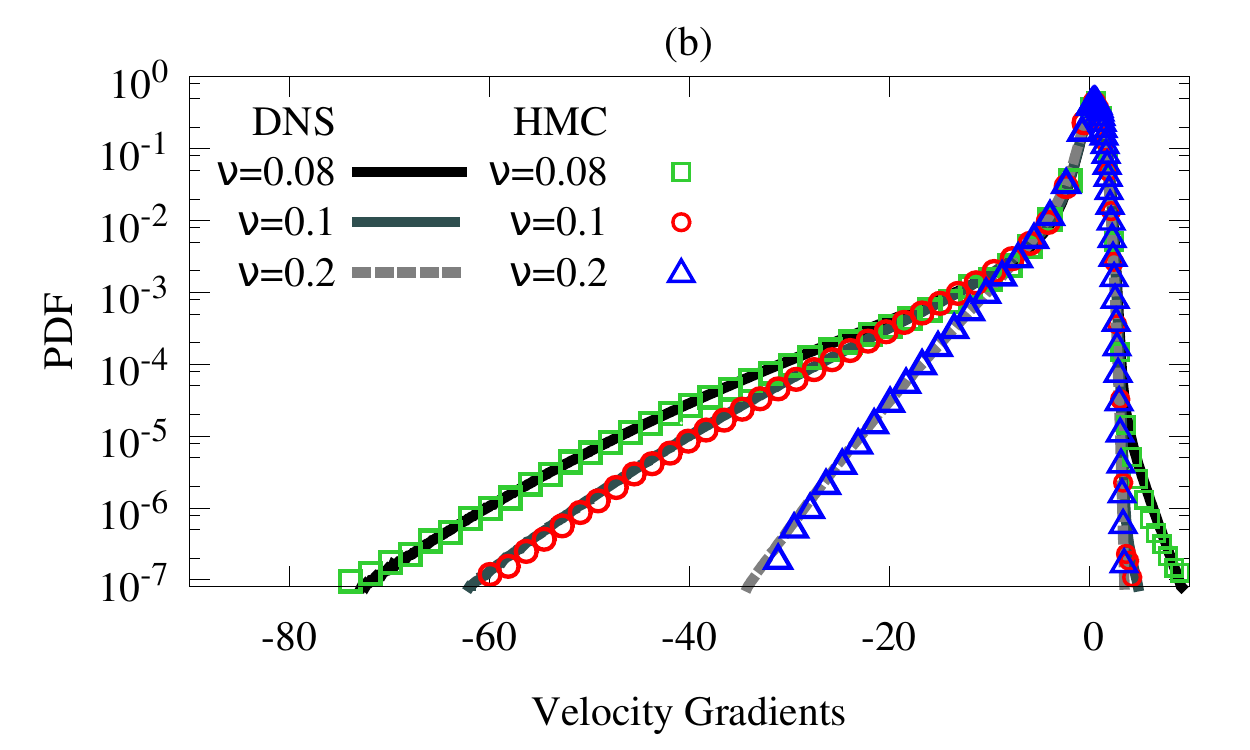}  
  \caption{(a) ensemble-average energy spectra, $\langle \bar E(k)\rangle$. Closed gray symbols correspond to the DNS results (lines denote their interpolation) and  open colored symbols correspond to the HMC results. (b) Probability distribution functions of velocity gradients. The  gray dashed line correspond to the DNS results, while open colored  symbols correspond to the HMC results.}
  \label{fig:dv_pdf}
\end{figure*}

In Fig.\ \ref{fig:kin_energy}(a) we compare the HMC and DNS temporal evolution of the mean kinetic energy, $\bar\varepsilon_{\rm kin}(t) = ||v(t)||^2/L$, for configurations corresponding to three different viscosities. As one can see the overall intensities of fluctuations are very similar. More quantitatively, in Fig.\ \ref{fig:dissip}(a) we show the temporal evolution of the ensemble average of the mean kinetic energy, i.e., $\langle \bar\varepsilon_{\rm kin}(t) \rangle$, starting from $v(x,t_0)=0$. Around time $t_s \approx 3$ the system reaches stationarity, meaning that the dissipative and injection forces are balanced and the system is driven to a nonequilibrium steady state -- beyond $t_s$, the $\langle  \bar\varepsilon_{\rm kin}(t) \rangle$ is constant in time. Figure \ref{fig:dissip}(b) shows the temporal evolution of the ensemble-average mean energy dissipation $\langle \bar\varepsilon_{{\rm diss}}(t) \rangle$, where $\bar\varepsilon_{{\rm diss}}(t) = 2 \nu ||\partial_x v(t)||^2/L$, while in Fig.\ \ref{fig:dv_pdf}(a) we consider the ensemble average of the energy spectrum $\langle \bar E(k) \rangle$, which is averaged in time $t$, i.e., $\bar E(k) = \frac{1}{T'} \int_{t_s}^{t_f} dt \, E(k,t)$, $T'=t_f-t_s$. 

In Fig.\ \ref{fig:dv_pdf}(b) we show the probability distribution function of the velocity gradients. In practice, the PDF is approximated by determining the counts of a fixed number of bins $(w_{{\rm min}},w_{{\rm max}})$ with equal width $\delta w$. The velocity gradients measured on the generated ensemble are counted only if $t>t_s$. The resulting histogram is normalized by dividing with the total number of counts, i.e.,
\begin{equation}
  \sum_i \int_{w_i - \delta w/2}^{w_i + \delta w/2} dw\, P(w) = 1 .
\end{equation}

From Figs.\ \ref{fig:kin_energy}--\ref{fig:dv_pdf}, we conclude that the HMC produces the same results as the DNS. Furthermore, we identify the same discretization effects in both implementations, which can be removed by taking the continuum limit. This has been thoroughly checked, but we skip this discussion here.

\begin{table}[!t]
  \centering
  \begin{tabular}{cccccc}
    &$\nu$ &$\tau$ &$N_{\Delta \tau}$ &Ensemble size &$\tau_{\text{int}}$ \\ \hline \hline
    &0.08 &1024 &20480 &$2\times10^5$ &$70$   \\
    &0.1 &1024 &20480 &$2\times10^5$ &$40$   \\
    &0.2 &1024 &20480 &$10^5$ &$10$ \\ \hline
  \end{tabular}
  \caption{\label{table4} Technical parameters and observables of the HMC simulations for the three different sets of runs, using fixed and open boundary conditions. Here $\tau$ is the trajectory length, and $N_{\Delta \tau}$ is the number of steps of the symplectic integrator for a single HMC iteration. 
    The fourth column gives the ensemble size per viscosity or, in other words, the number of HMC iterations performed. Finally,  $\tau_{\text{int}}$ is the integrated autocorrelation time evaluated here for the kinetic energy and measured in units of $\tau$ \cite{WOLFF2004143}. The ratio of ensemble size to $\tau_{\text{int}}$ estimates the effective statistically uncorrelated ensemble size.}
\end{table}

There are two interesting remarks regarding the behavior of the HMC and in connection with Table \ref{table4}. First, we notice that for fixed resolution and trajectory length $\tau$, the integrated  autocorrelation time $\tau_{\rm int}$ increases with decreasing viscosity. Second, we manage to perform highly efficient simulations simply by increasing the trajectory length $\tau$ while keeping $\Delta \tau$ fixed. Contrary to common practice in lattice QCD, where $\tau$ is kept of $O(1)$, to avoid energy and reversibility violations \cite{MEYER200791}, in our case it proved a safe and beneficial choice to set $\tau$ of order $\tau \approx 10^2$ or $\tau \approx 10^3$ without introducing significant effects of reversibility violations or loss of acceptance rate. This allowed us to significantly decrease autocorrelation times, and avoided the disposal of many generated configurations between measurements. In principle we can increase the trajectory length to higher values that will allow us to generate statistically independent configurations at each HMC iteration, as it is done for the runs in Sec.\ \ref{Sec:Constrained space-time evolution using HMC}, but we did not check this systematically for the present section.

Finally, a key element of a Monte Carlo-based approach is a rigorous error analysis, for which there are well-established methodologies \cite{MullerKrumbhaar1973,Madras1988,Ferrenberg1991}. Since Markov chain Monte Carlo simulations, are known to be prone to autocorrelation effects, we had to go through a thorough investigation of the integrated autocorrelation times $\tau_{\rm int}$ for each observable. Therefore, as a post production step, we used the data analysis package provided in \cite{WOLFF2004143} as a tool to estimate the errors of the observables, which takes into account the corresponding autocorrelation effects. This recipe for the error calculation will be followed throughout this article.

\begin{figure*}[!t]
  \centering
  \includegraphics[width=0.48\textwidth]{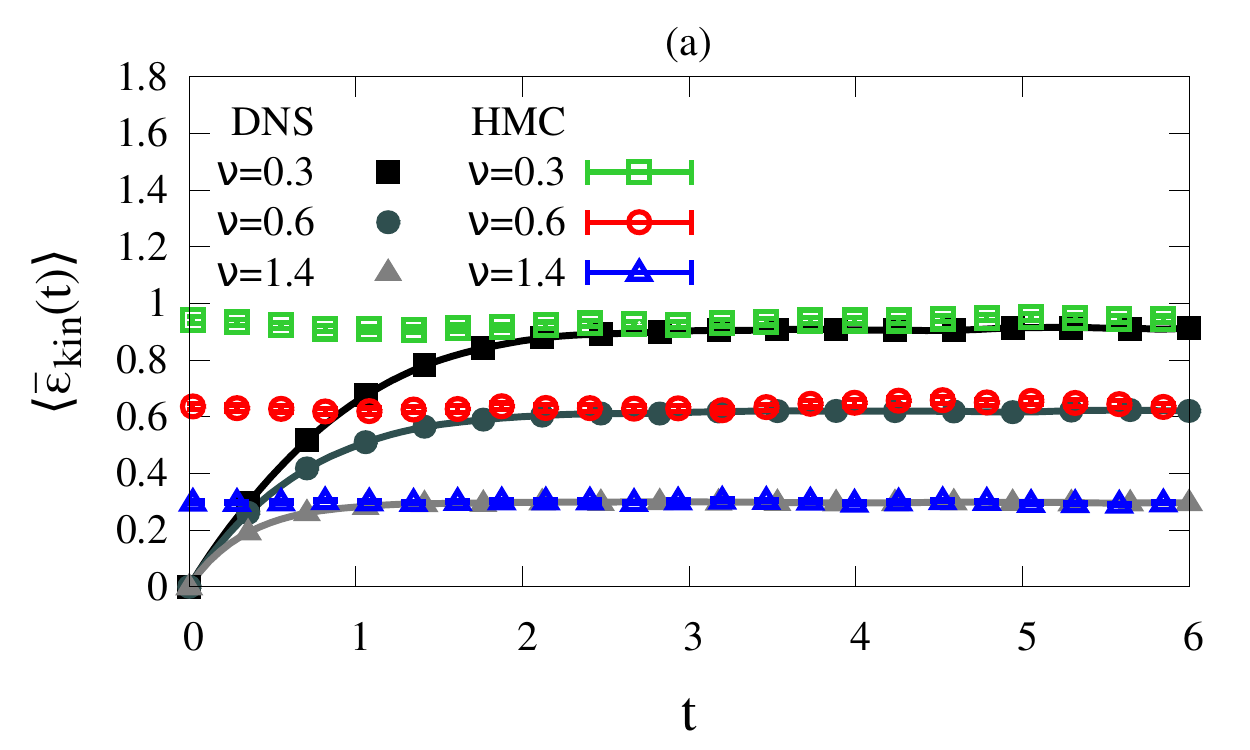}  \quad 
  \includegraphics[width=0.48\textwidth]{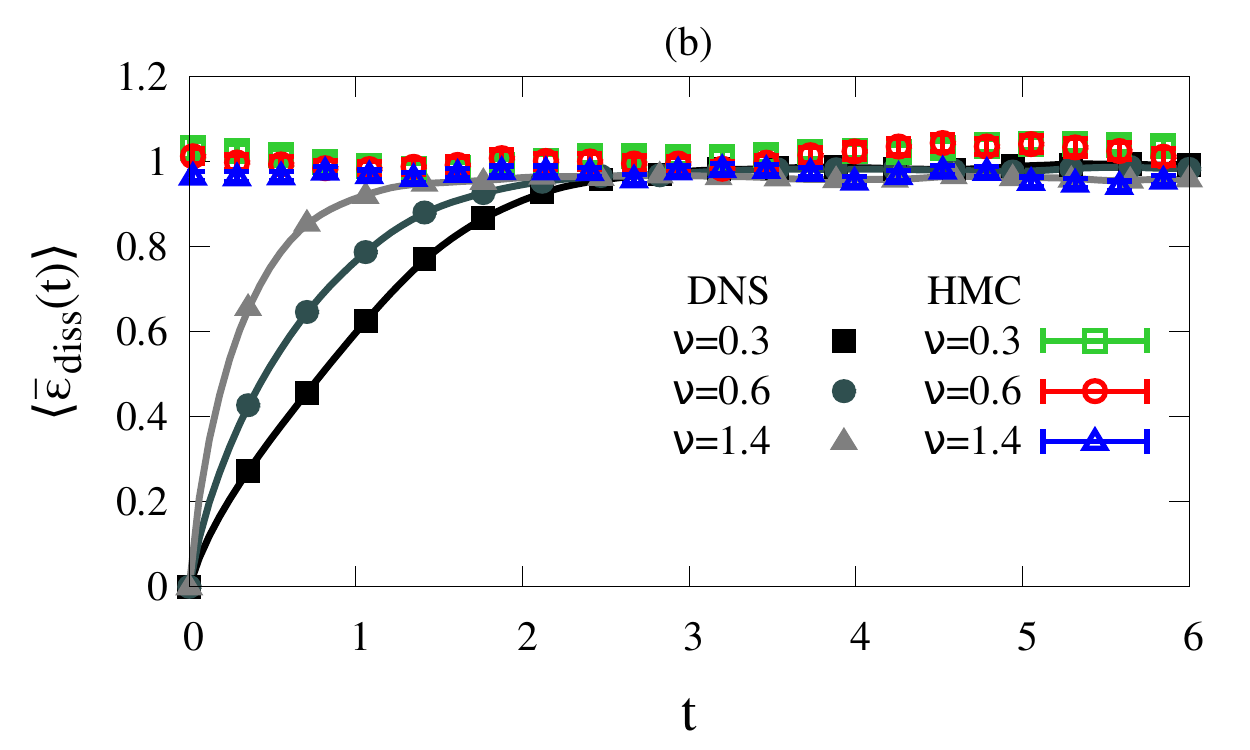}
  \caption{Results for the HMC using periodic boundary conditions in time. (a) Temporal evolution of the ensemble-average mean kinetic energy $\langle \bar\varepsilon_{\rm kin}(t) \rangle$. (b) Temporal evolution of the ensemble-average mean energy dissipation $\langle\bar\varepsilon_{{\rm diss}} (t) \rangle$. Closed gray symbols correspond to the DNS results (lines denote their interpolation) and  open colored symbols correspond to the HMC results.}
  \label{fig:kin_energy_pbc}
\end{figure*}

\section{Constrained space-time evolution using HMC}
\label{Sec:Constrained space-time evolution using HMC}

Now that we have benchmarked the HMC against a standard DNS algorithm, we will present the features and advantages that this path-integral-based approach can bring to the numerical studies of turbulent models, as well as stochastic PDEs in general. First, since the HMC considers the full temporal evolution of the field, this provides an additional flexibility towards the choice of boundary conditions in time. Therefore, in Sec.\ \ref{Sec:Time-periodic boundary conditions} we show, for instance, that one can apply periodic boundary conditions in time, i.e., $v(x,t) \equiv v(x,t+T)$. Then in Sec.\ \ref{Sec:Enhanced sampling of extreme and rare events} we turn towards the motivation for this article. That is to introduce field constraints, which will affect the Monte Carlo sampling in a controlled way, in order to favor the generation of specific configurations that will comply with the imposed constraint. More specifically, as a first application, we apply a protocol to systematically generate configurations where a large negative velocity gradient is produced at a prescribed space-time point. This also provides with some insight into the underlying dynamics of how the system evolved in time $t$ to reach this extreme condition.

\subsection{Time-periodic boundary conditions}
\label{Sec:Time-periodic boundary conditions}

As a first application, we discuss the use of periodic boundary conditions in time. Under this scenario, we observe that after the system has equilibrated (to the desired target distribution), the ensemble consists of configurations that have reached stationarity at any time $t\in[t_0,t_f]$. This can be better understood by looking at Fig.\ \ref{fig:kin_energy_pbc}, where the ensemble average of the mean kinetic energy [Fig.\ \ref{fig:kin_energy_pbc}(a)] and of the mean energy dissipation [Fig.\ \ref{fig:kin_energy_pbc}(b)] are constant in time in the example of the HMC (colored symbols). We also show the results of the DNS (gray lines and symbols) using zero initial conditions as a further comparison. The parameters used for the three different runs are summarized in Table  \ref{table2}.

The use of periodic boundary conditions in time leads to a field-dependent Jacobian $\J$ \cite{Nakazato:1990kk} and therefore we must expect it to affect the importance sampling. Nevertheless, in this work, we have consistently neglected the evaluation of the Jacobian (which, in the lattice field theory literature, is often referred to as the quenched limit). To get a better impression of the systematic error associated with this approximation, we have chosen to compare our results with periodic boundary conditions to the case of fixed and open boundary conditions. As can be seen from Fig.\ \ref{fig:kin_energy_pbc} our results overlap with the stationary regime attained by using fixed and open boundary conditions with reasonable accuracy. In fact, it is possible to show that the error by neglecting the Jacobian vanishes in the limit $T \rightarrow \infty$. Thus, for those cases considered here, the systematic error is likely negligible. We defer the evaluation of the Jacobian determinant to future work.

\begin{table}[!t]
  \centering
  \begin{tabular}{cccccccc}
    &$\nu$ &$\Re$ &$v_{\textrm{rms}}$ &$\ell_d$ &$\ell$   &$\langle\bar\varepsilon_{{\rm diss}} \rangle$ &$T_\ell$ \\ \hline \hline
    &0.3 &20 &0.93  &0.40  &0.81 &1 &0.87    \\
    &0.6 &7 &0.64  &0.68  &0.26  &0.99 &0.41     \\
    &1.4 &1 &0.31 &1.29  &0.03   &0.97 &0.09  \\
    \hline
  \end{tabular}
  \caption{\label{table2} Parameters and observables of the numerical simulations for periodic boundary conditions of the HMC. Here the fixed parameters for both implementations are $N_t=544$, $N_x=64$,  $T=6$, and $L=2\pi$. Also for the HMC $\tau=128$, and $N_{\Delta \tau}=2560$. See also Tab.\ \ref{table1} for definitions.}
\end{table}

\subsection{Enhanced sampling of extreme and rare events}
\label{Sec:Enhanced sampling of extreme and rare events}

\begin{figure*}[!t]
  \centering
  \includegraphics[width=0.95\textwidth]{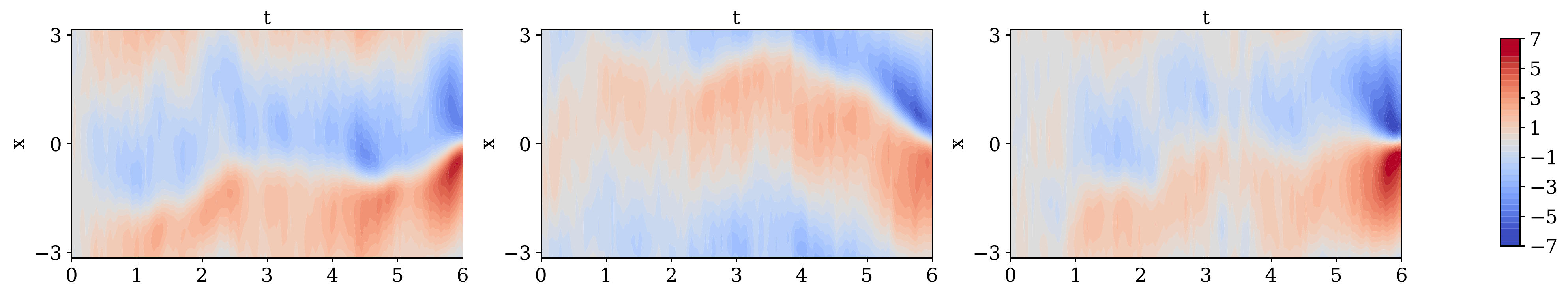}
  \caption{Sample configurations from HMC simulation using the constraint $\Delta \S _1$ with $c_1=1.9$, enforcing a negative velocity gradient maximization at $x=0$, $t=t_f$. The color spectrum corresponds to the intensity of the velocity field.}
  \label{fig:HMC_samples}
\end{figure*}

We will now describe the important steps towards constraining the sampling of the HMC to generate a large negative velocity gradient at a specified space-time point. Also, we will explain how to directly compare the observables obtained from the constrained ensemble with the ones related to an unconstrained ensemble, by using reweighting techniques, and therefore estimate their relative importance with respect to the typical statistics of the system. We note two important points. First, we will use the same boundary conditions as in Sec.\ \ref{Sec:Benchmarking the HMC against a first-order Euler-Maruyama explicit solver}, i.e., periodic in space and fixed and open in time. Second, the statistics of the DNS will be referred to as the ones related to the unconstrained system. We could use the corresponding ones from the HMC with unconstrained sampling, but another purpose of ours is to demonstrate the benefits of employing this method for the purpose of systematically sampling extreme and rare events, compared to a standard DNS implementation, where such instances are a matter of chance.

\paragraph{Reweighting}
\label{Sec:Reweighting}

Reweighting is a standard technique introduced in \cite{Ferrenberg1988} that has proved very helpful in the study of phase transitions and critical phenomena. In short, it allows one to exploit the information of a generated ensemble of a single Monte Carlo simulation performed at a certain parameter (e.g., at fixed inverse temperature $\beta$) and obtain results for a range of nearby parameters (e.g., $\beta_i$). Reweighting can also provide a way to modify the sampling in a Monte Carlo simulation, which is how we use it here by constraining the sampling of the HMC to enhance the generation of strong negative gradients. What is common in both cases is that we include a reweighting factor in the ensemble averages to obtain the desired ensemble (see \cite{Pelissetto2014} for a review on the topic).

In our application, reweighting is employed at the post-production stage as a means to relate the observable $\langle \Ob \rangle'$, measured using the ensemble which is generated by sampling with respect to the action $\S'$, to $\langle \Ob \rangle$, measured on the ensemble sampled with the action $\S$. We briefly revisit here the standard steps of the derivation. Note that any probability density functional $\P'_v = e^{-\S'} / \Z'$ can be related to another $\P_v$ for the same field configuration $v = v(x,t)$ via
\begin{equation}
  \P_v = \frac{1}{\Z} e^{-\S} = \frac{\Z'}{\Z} e^{-(\S-\S')} \, \P'_v .
\end{equation}
The expectation value of an observable  $\langle\Ob\rangle$ using $\S$ is given by
\begin{eqnarray}
  \langle\Ob\rangle &=& \int\Dv\, \P_v\, \Ob \nonumber\\ &=& \frac{\Z'}{\Z} \int\Dv\,  \P'_v\, e^{-(\S-\S')} \Ob \nonumber\\ &=& \frac{\Z'}{\Z} \langle e^{-(\S-\S')} \Ob \rangle' ,
\end{eqnarray}
where the notation $\langle\,\cdots\,\rangle'$ implies that the expectation value is evaluated with the action $\S'$. From the identity $\langle 1 \rangle = 1$ we may derive the relation $\Z / \Z' = \langle e^{\Delta \S} \rangle'$, with $\Delta \S = \S' - \S$, whereby
\begin{equation}
  \langle\Ob\rangle = \frac{\langle e^{\Delta \S} \Ob \rangle'}{\langle e^{\Delta \S} \rangle '} .
  \label{eq:reweighting}
\end{equation}

As the next step we determine the error of the estimator $\langle\Ob\rangle$. Note that in Eq.\ \eqref{eq:reweighting} both the numerator and denominator have fluctuations. Furthermore, as they are calculated from the same ensemble we expect that both errors are correlated. As explained in \cite{Ferrenberg95}, to estimate the error $\delta\langle\Ob\rangle$ of $\langle\Ob\rangle $ we employ the propagation of error of two dependent variables including the covariance and the cross  covariance of the nominator and the denominator. Simplifying Eq.\ \eqref{eq:reweighting} to $\langle\Ob\rangle = A/B$, the final expression is
\begin{equation}
  \delta\langle\Ob\rangle = \langle\Ob\rangle \sqrt{\left(\frac{\delta A}{A} \right)^2 + \left(\frac{\delta B}{B}  \right)^2 - 2\left( \frac{\delta(A B)}{A B} \right)^2} ,
  \label{eq:error_propagation}
\end{equation}
where $\delta(AB) = \langle A B\rangle - \langle A\rangle \langle B\rangle + 2\sum_{i,j>i} (A_i-\langle A\rangle)(B_j-\langle B\rangle)$.

\paragraph{Implementation of sampling constraints}
\label{Sec:Implementation of sampling constraints}

The idea is to define a different action $\S'$ to sample via the HMC, which consists of the original $\S$ [Eq.\ \eqref{eq:action_S}] in addition to a constraint functional $\Delta \S$:
\begin{equation}
  \S' = \S + \Delta \S .
  \label{eq:contraint_action}
\end{equation}
The choice of $\Delta \S$ cannot be arbitrary. If there is a tiny overlap of the distributions $e^{-\S}$ and $e^{-\S'}$, the reweighting procedure will most likely not work. Therefore, it is not clear from the beginning, for which parameter values a successful reweighting can be performed. We remark that in cases where reweighting fails, one could attempt to insert intermediate reweighting steps as explained in \cite{Ferrenberg1989}. We also need to stress that any constraint functional $\Delta \S$ will contribute to the MD forces $\Phi_\pi$ through the functional derivative $\delta \S' / \delta v = \delta \S / \delta v + \delta \Delta \S / \delta v \,$ and this contribution needs to be evaluated exactly.

Nevertheless, the observables of the HMC are not directly comparable to the DNS. In the following, we will explain how to directly compare the statistics of the HMC using the action $\S'$, with the typical unconstrained statistics using the action $\S$, by utilizing reweighting techniques.

In order to demonstrate the application of Eq.\ \eqref{eq:reweighting}, we first discuss the example of the ensemble-average mean kinetic energy before and after reweighting. This is also a sufficient step to further ensure the consistency with the unconstrained statistics, meaning that after reweighting the observable measured by the constrained ensemble should collapse, within error bars, with the  corresponding unconstrained one. Following Eq.\ \eqref{eq:reweighting}, the reweighted ensemble-average mean kinetic energy will be
\begin{equation}
  \langle \bar\varepsilon_{\rm kin}(t) \rangle = \frac{\langle e^{\Delta\S} \bar\varepsilon_{\rm kin}(t) \rangle'}{\langle e^{\Delta\S} \rangle'} .
  \label{eq:kinerg_rew}
\end{equation}

As a first attempt we tried a series of local constraint functionals, with a suitable shape, that enhance the probability to produce a large negative velocity gradient at a certain point in the middle of the spatial domain at the last time slice, i.e., $x=0$, $t=t_f$. The parameters that we used for the HMC are summarized in Table \ref{table3}. A general way to define the local functional $\Delta \S$ is
\begin{equation}
  \Delta \S_i = c_i  \int dt\,\int dx\, { g_i(\partial_x v / w_i) \, \delta(x) \delta(t-t_f)} ,
  \label{eq:constr_func}
\end{equation}
where $c_i$ is a prefactor to characterize the strength of the functional and $w_i$ is an imposed velocity gradient value around which we want our simulation to sample at ($x=0$, $t=t_f$). With the index $i$ we label the different choices of $g_i$, for which we have tested
\begin{subequations}
  \label{eq:constr_functionals}
  \begin{eqnarray}
    g_1(z)  &=& z , \label{eq:s1}\\
    g_2(z) &=& \left(z + 1 \right)^2 ,\label{eq:s2}\\
    g_3(z) &=&\left( z^2 - 1 \right)^2 . \label{eq:s3}
  \end{eqnarray}
\end{subequations} 

\begin{table}[!t]
  \footnotesize
  \centering
  \begin{tabular}{ccccccccc}
    & \multicolumn{8}{c}{only HMC} \\
    &$c_1$ &$w_1$ &$\Re_\ell$   &$\Re$  &$v_{\rm rms}$  &$\langle\bar\varepsilon_{{\rm diss}}\rangle$  &$\ell$  &$\kappa$  \\ \hline
    & 1.2 &1 &1.02(3)  &10(2)  &0.8(1)   &0.96(3)  &0.61(2) &1.12   \\
    & 1.6 &1 &1.0(2)  &10(4)  &0.8(3)   &0.9(1) &0.6(1) &1.21    \\
    & 1.9 &1 &0.7(4)  &9(5)  &0.7(4)   &0.7(2)  &0.5(2)  &1.93 \\ \hline
    &$c_2$ &$w_2$ &$\Re_\ell$  &$\Re$  &$v_{\rm rms}$  &$\langle\bar\varepsilon_{{\rm diss}}\rangle$  &$\ell$   &$\kappa$ \\ \hline
    & 80 &12 &0.83(5)  &10(2)  &0.8(2)   &0.79(4)  &0.55(3) &418    \\
    & 80 &18 &0.8(2)  &9(4)  &0.8(4)   &0.8(2) &0.6(1) &$2.6\times10^5$    \\
    & 160 &24 &1.4(6)  &12(6)  &1.0(5)   &1.2(3)   &0.7(3)  &$5\times10^9$  \\
    & 160 &30 &1.7(6)  &14(7)  &1.1(5)  &1.6(4)   &0.8(2)  &$2.6\times10^{11}$  \\ \hline
    &$c_3$ &$w_3$ &$\Re_\ell$  &$\Re$ &$v_{\rm rms}$  &$\langle\bar\varepsilon_{{\rm diss}}\rangle$  &$\ell$   &$\kappa$  \\ \hline
    & 80 &12 &1.5(4)  &12(6)  &1.0(5)   &1.3(3)  &0.8(2) &$3.9\times10^{5}$    \\
    & 80 &18 &1.0(3)  &11(5)  &0.8(4)   &1.0(2) &0.6(1) &$5.4\times10^{8}$     \\
    & 80 &24 &1.2(4)  &11(5)  &0.9(4)   &1.1(3)   &0.7(2)  &$5.2\times10^{11}$   \\
    & 120 &30 &1.2(3)  &11(5) &0.9(4)  &1.0(2)   &0.7(2)  &$4.5\times10^{16}$  \\ \hline
    & \multicolumn{8}{c}{only DNS} \\
    & -- &-- &1.01 &10.5 &0.83  &0.95 &0.61 &-- 
  \end{tabular}
  \caption{\label{table3} Parameters for HMC simulations with constrained sampling. The integral length scale Reynolds number is defined as $\Re_\ell=\frac{v_{\rm rms}\, \ell}{\nu}$ and the large-scale Reynolds as $\Re=\frac{v_{\rm rms}\, L}{\nu}$. The results for the HMC have been reweighted and the temporal interval for averaging corresponds to the stationary regime. The number in parentheses gives the error of the last digit of the mean. Here the fixed parameters for both DNS and HMC are: $N_t=144$, $N_x=64$, $T=6$, $L=2\pi$, $\nu=0.5$, and $\ell_d=0.59$. In addition, $\kappa$ is defined in Eq.\ \eqref{eq:rescaling_factor}. Specifically for the HMC, $\tau= 128 $, and $N_{\Delta \tau} = 2560$.  }
\end{table}

Note that the chosen constraints $\Delta S_i$ result in singular derivatives in the Hamiltonian  \eqref{eq:eff_hami}. In our case the regularization of the $\delta$ function happens through the space-time grid and its finite lattice spacing. In the approximation of the integral, which is a finite sum, the $\delta$ function's approximation appears as a properly normalized Kronecker delta. Our discretization of the $\delta$ function in space, $\delta = \delta(x_i - x_j)$, is $\delta_{i, j} / \Delta x$. Accordingly, in time we discretize $\delta = \delta(t_m - t_n)$ by $\delta_{m, n} / \Delta t$.

The HMC will sample around the region where $e^{-\S'}$ is maximal, i.e., where $\S'$ is minimal, and the constraint functionals $\Delta \S_i$ contribute towards this procedure. In particular, the constraints imposed by $\Delta\S_2$ and $\Delta\S_3$ are of a localization nature in the sense that the generated configurations comply with the constraint by sampling in a narrow region around the imposed gradient $w_i$, where $\Delta\S_2$ and $\Delta\S_3$ are minimal. In the same spirit, as $\Delta\S_1$ is a linear function of $\partial_x v$, then for any negative $\partial_x v$ it will have a negative contribution to the action, which will favor the sampling towards this direction. It therefore allows us to sample across a wider range of negative velocity gradients. Nevertheless, we can redefine $\Delta\S_1$, as in this case $w_1$ can be absorbed by $c_1$. Thus, we set $w_1=1$ and present only values of $c_1$. Accordingly, if we wish to sample positive gradients at ($x=0$, $t=t_f$) in the same manner, it is sufficient to consider negative values of $c_1$. That is also possible for $\Delta\S_2$, where according to Eq.\ \eqref{eq:s2}, if we consider $w_2<0$, then the algorithm will preferably sample positive gradients at ($x=0$, $t=t_f$). In contrast, for $\Delta\S_3$, due to it's symmetry around $\partial_x v=0$, one could expect the sampling of both positive and negative gradients around $w_3$. However, the physics of the Burgers equation favors the generation of strong negative gradients instead of their positive counterparts, i.e., notice the asymmetric PDF of velocity gradients [Fig.\ \ref{fig:burgers}(b)]. Hence, by choosing $\Delta\S_3$, negative gradients will be preferred at ($x=0$, $t=t_f$).

\begin{figure}[!t]
  \centering
  \includegraphics[width=0.45\textwidth]{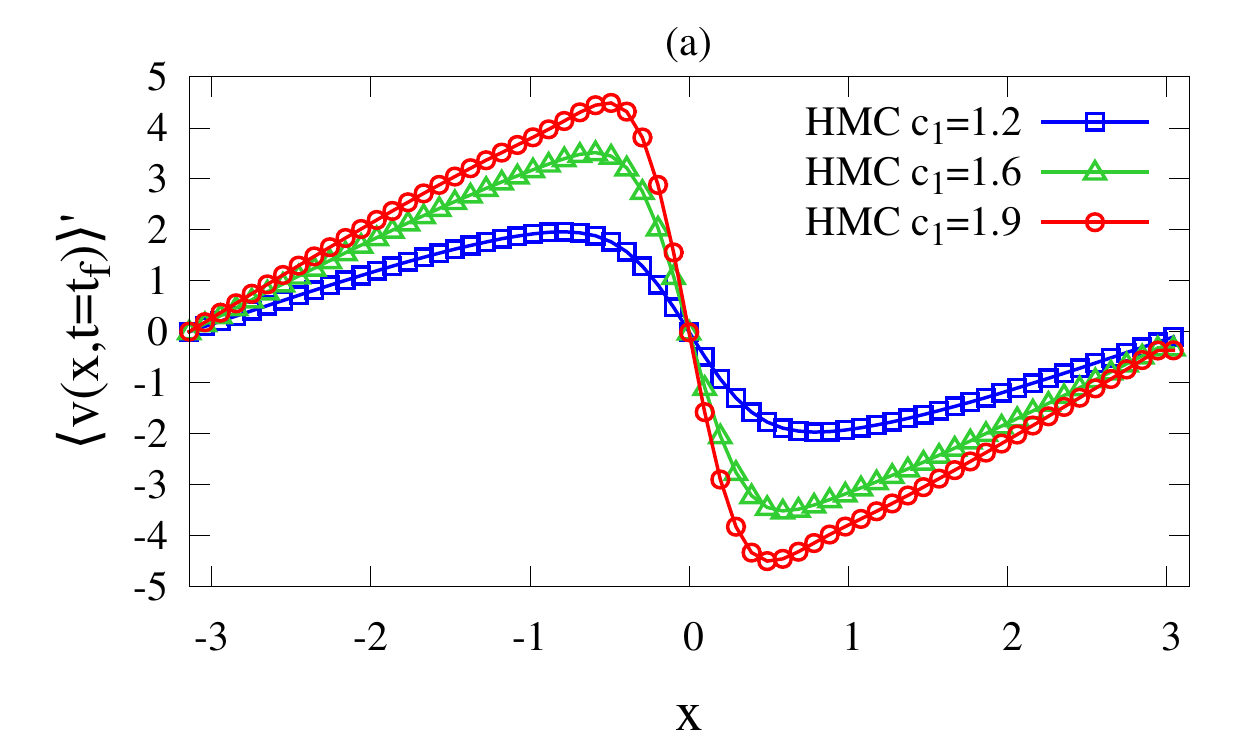} \quad
  \includegraphics[width=0.45\textwidth]{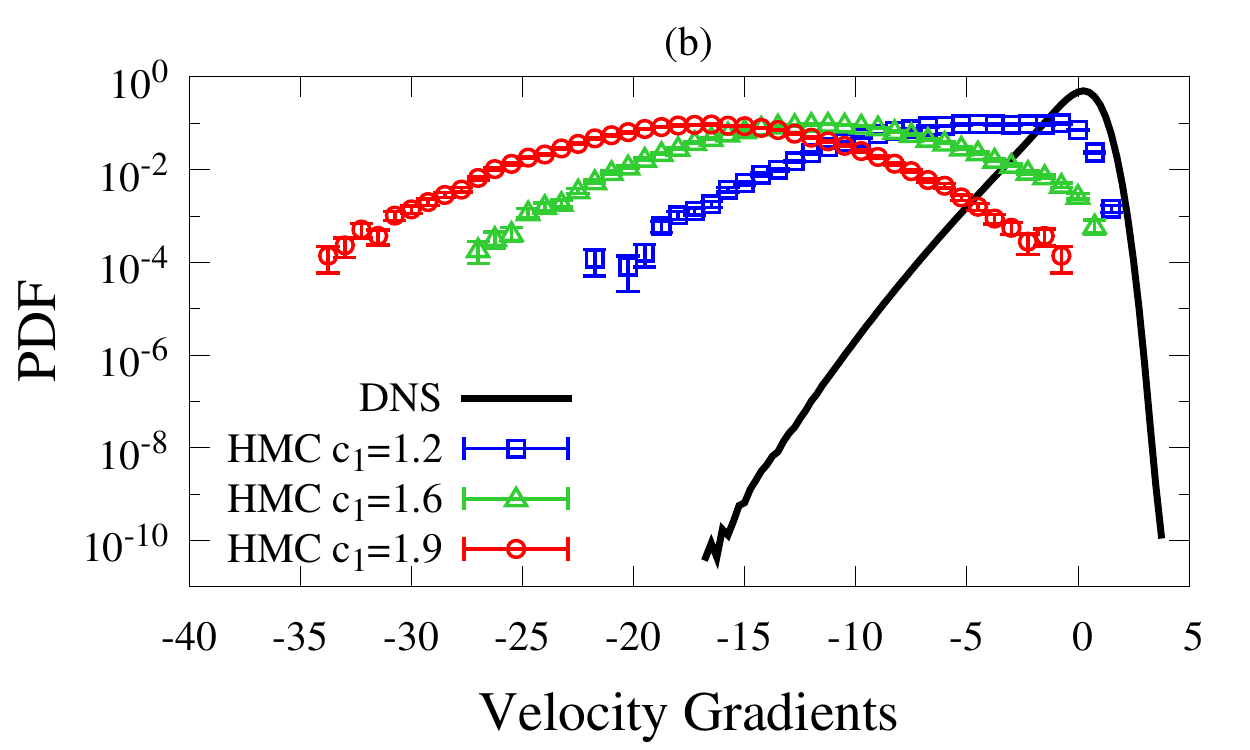} 
  \caption{(a) Ensemble average of the velocity field $v(t=t_f)$ using the HMC with the action $\S' = \S + \Delta \S_1$ for different values of $c_1$. (b) PDF of velocity gradients (DNS versus the HMC). For the HMC we measure $P'(w)$ only at the space-time point where we constrain the ensemble, i.e., at $x=0$, $t=t_f$.}
  \label{fig:HMC_diff_c}
\end{figure}

\begin{figure}[!t]
  \centering
  \includegraphics[width=0.45\textwidth]{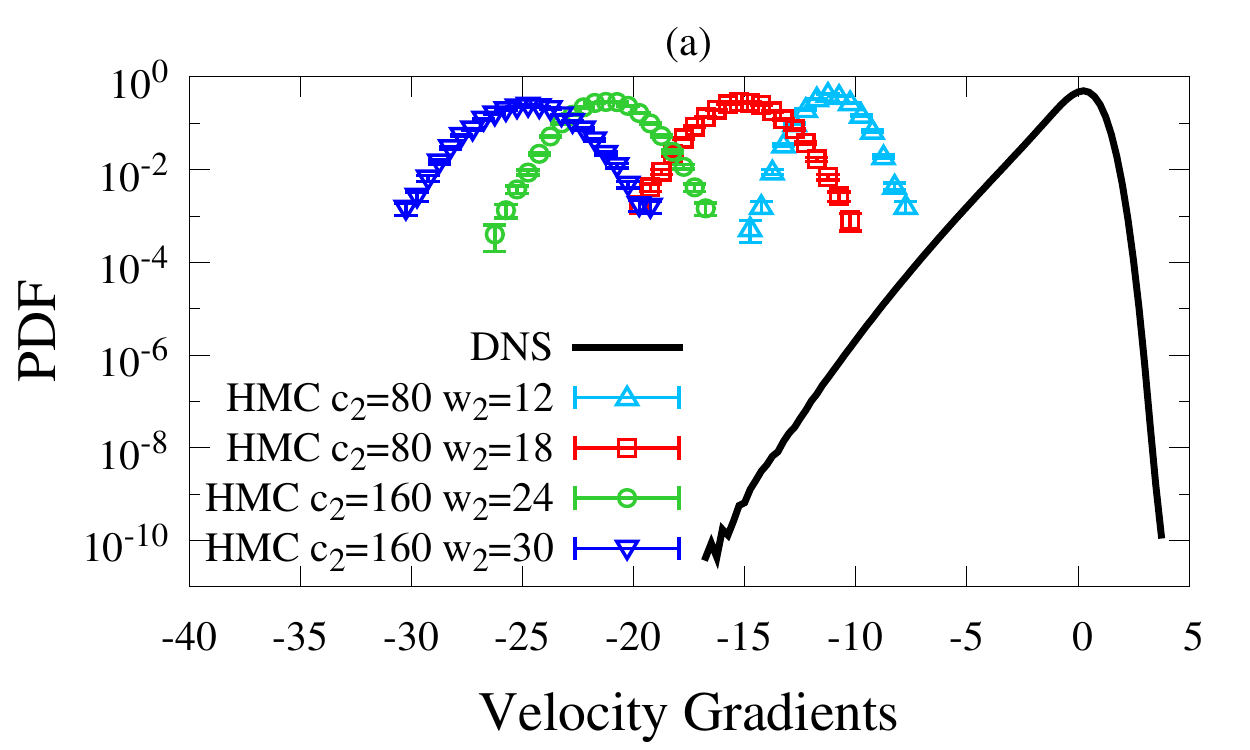} \quad
  \includegraphics[width=0.45\textwidth]{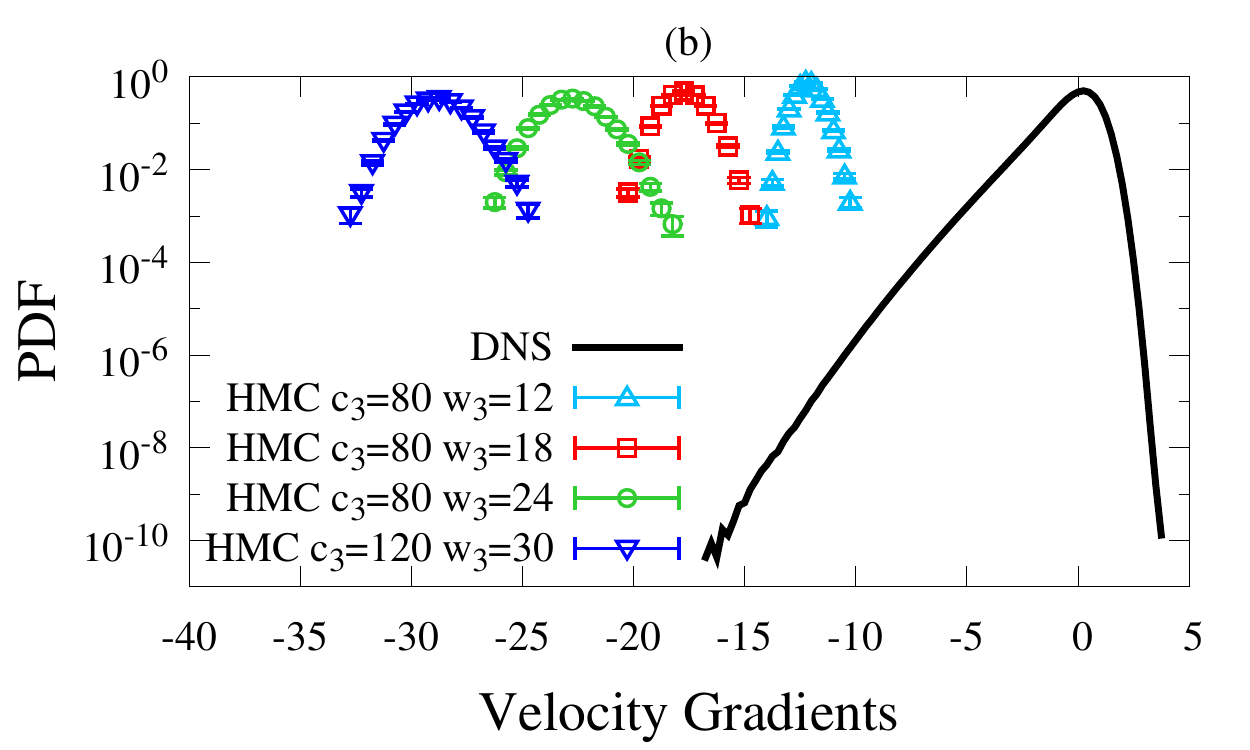}
  \caption{PDF of velocity gradients (DNS versus the HMC), i.e., $P'(w)$, generated using the action ${\S'}$, and measured only at the point that we constrain, i.e., at ($x=0$, $t=t_f$), for different values of $c_i$ and $w_i$, $i=2,3$, using (a) $\Delta \S _2$ and (b) $\Delta \S _3$.}
  \label{fig:HMC_diff_c_s}
\end{figure}

\begin{figure*}[!t]
  \centering
  \includegraphics[width=0.31\textwidth]{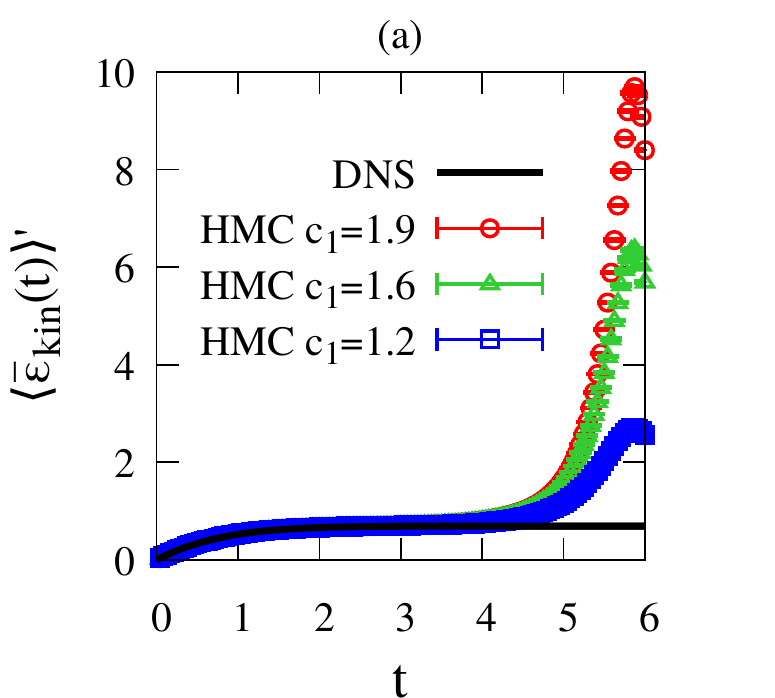} \quad
  \includegraphics[width=0.31\textwidth]{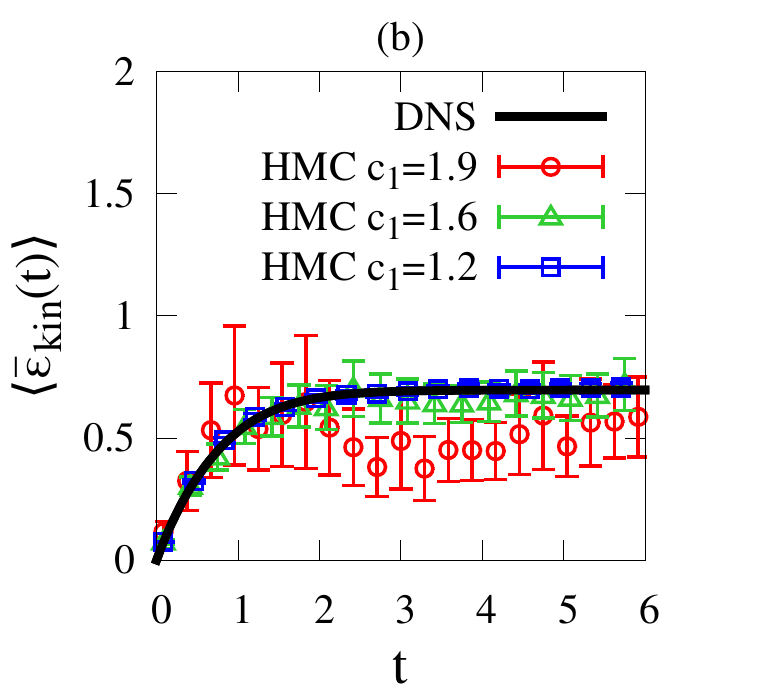} \quad
  \includegraphics[width=0.31\textwidth]{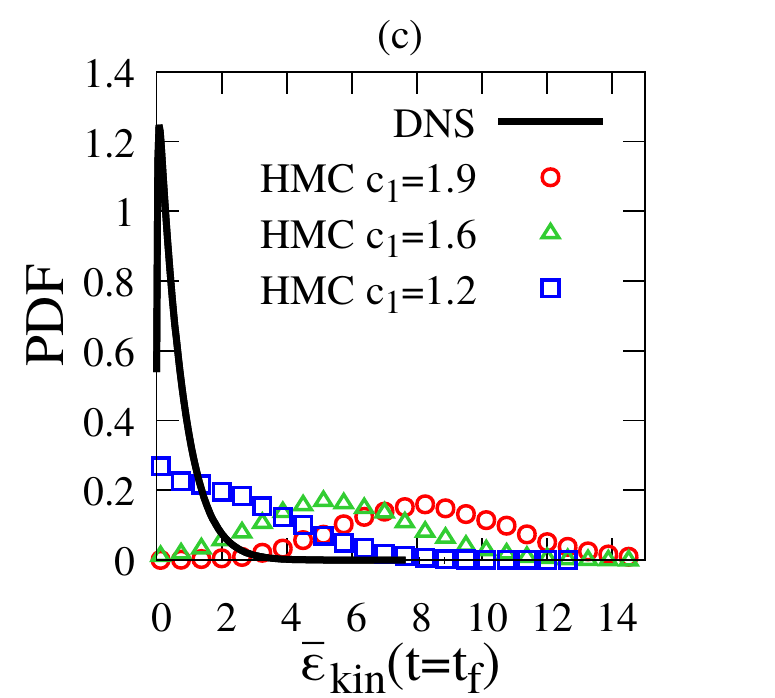}
  \caption{Ensemble-average mean kinetic energy of the HMC vs DNS using $\Delta \S _1$ for different $c_1$ (a) before reweighting and (b) after reweighting. Notice that the error bars for the $c_1=1.9$ case after reweighting are pronounced as for this choice the fluctuations introduced by the reweighting factor become significantly large. (c) PDF of the non reweighted mean kinetic energy $\bar\varepsilon_{\rm kin}(t)$ measured only for the last time slice $t=t_f$, in the case of the HMC, and using $\Delta \S_1$ for different $c_1$. }
  \label{fig:kinerg_rew}
\end{figure*}

As for the numerical stability, we note that the grid resolution should always be sufficient to ``fit'' the strong shock. Therefore, we cannot increase $c_i$ and $w_i$ unconditionally for a fixed resolution. In practice, for a particular discretization, there is a threshold beyond which the HMC is not reliable anymore.

To identify the impact of constraining the sampling of the HMC on the generated configurations, we show three independent samples in Fig.\ \ref{fig:HMC_samples}. A large negative velocity gradient at ($x=0$, $t=t_f$) is achieved in all cases. The general idea here is that we provide the HMC with a certain constraint, local or global, by which the HMC will consider all the possible realizations in the configuration space to fulfill the corresponding condition on the velocity field. In the case of extreme and rare events, for instance, the HMC provides a systematic way to sample the fluctuations around a particular extreme event (e.g., the occurrence of a strong velocity gradient).

Focusing now on the constraint functionals of Eqs.\ \eqref{eq:constr_functionals}, Fig.\ \ref{fig:HMC_diff_c}(a) shows the ensemble average of the velocity field for the final timeslice $\langle v(x,t=t_f) \rangle'$ at changing $c_1$. It further indicates the functionality of $c_1$ and the effect it has on the sampled configurations, i.e., the larger $c_1$ is, the more negative the sampled gradient will be. This can also be justified from Fig.\ \ref{fig:HMC_diff_c}(b), which, for different $c_1$, depicts the PDF of the velocity gradients measured only at the point that we constrain, i.e., at ($x=0$, $t=t_f$). It is defined as
\begin{equation}
  P'(w) = \langle \delta (\partial_x v(0,t_f) - w) \rangle' ,
  \label{eq:pdf_non_rew}
\end{equation}
where $w$ is the value of the bin which is incremented according to the value of the velocity gradient $\partial_x v(0,t_f)$ and is generated using the action $\S'$. In this plot we see that by increasing $c_1$, the peak of the histogram moves to the left towards larger negative velocity gradients.

As for the prefactors $c_2$ and $c_3$, they have a slightly different behavior with respect to $c_1$. In fact, as we increase $c_2$ and $c_3$, the HMC will sample more systematically around the prescribed velocity gradient $w_i$. In Fig.\ \ref{fig:HMC_diff_c_s} we show $P'(w)$ at varying $c_i$ and $w_i$, with $i=2,3$. In Fig.\ \ref{fig:HMC_diff_c_s}(a) the functional $\Delta \S _2$ has been used, and in Fig.\ \ref{fig:HMC_diff_c_s}(b) the functional $\Delta \S _3$. For the same parameters, the quartic functional $\Delta \S _3$ has a slightly better performance towards sampling the prescribed velocity gradient $w$ than the quadratic functional $\Delta \S _2$. Notice that in Fig.\ \ref{fig:HMC_diff_c}(b) and both plots of Fig.\ \ref{fig:HMC_diff_c_s} we also include the PDF of the velocity gradients of the DNS (black line) to give a qualitative description of how the constrained sampling compares with the original statistics. 

To be more specific, we first refer to Fig.\ \ref{fig:kinerg_rew}(a), where we show the non reweighted ensemble-average mean kinetic energy, defined as $\langle \bar\varepsilon_{\rm kin}(t) \rangle'$ at changing $c_1$, using the functional $\Delta \S_1$, and we compare it with the ensemble-average kinetic energy of the DNS (black line, unconstrained statistics). The larger the value of $c_1$ is, the more pronounced the kinetic energy will be closer to the final time $t=t_f$, where the constraint is applied. Figure  \ref{fig:kinerg_rew}(b) depicts the corresponding reweighted data, i.e., $\langle \bar\varepsilon_{\rm kin}(t) \rangle $, by using \eqref{eq:kinerg_rew}, where both the DNS and the reweighted HMC collapse within error bars.

We remark two points. First, through Fig.\ \ref{fig:kinerg_rew}(a), we can also get an estimate of how important the constraint is as a function of time. For instance, on average, at time $t \approx 3$ the effects of $\Delta \S_1$ seem to have decayed. Second, for the particular observable, by increasing here $c_1$ we get increased error bars after reweighting. For instance, in the case of $\langle \bar\varepsilon_{\rm kin}(t) \rangle $ for $c_1=1.2$ we notice small error bars and very good agreement with the DNS, while for $c_1=1.9$ the $\langle \bar\varepsilon_{\rm kin}(t) \rangle $ has much more pronounced error bars. This is related to a previous comment on the applicability of reweighting, for which we stated that the distributions $e^{-\S}$ and $e^{-\S'}$ should have a sufficient overlap. In this example, for $c_1=1.2$, the distribution of $\bar\varepsilon_{\rm kin}(t)$, for $t=t_f$, of the constrained ensemble and the distribution of $\bar\varepsilon_{\rm kin}(t)$, for $t>t_s$, of the unconstrained system do overlap considerably, as can be seen in Fig.\ \ref{fig:kinerg_rew}(c) [blue and black lines accordingly], which leads to the resulting collapse of the data [same colors in Fig.\ \ref{fig:kinerg_rew}(b)]. The difference with $c_1=1.9$ [red line in Fig.\ \ref{fig:kinerg_rew}(a)] is that the corresponding overlap with the DNS is marginal. Also $c_1=1.9$ favors more the sampling of extreme velocity gradients $\partial_x v$, which, together with a (finite) characteristic dissipation scale $\ell_d$, implies large values of $v_d \sim (\partial_x v) \ell_d$ [see Fig.\ \ref{fig:HMC_diff_c}(a)]. The averaged kinetic energy is a global observable, which is mostly related to the bulk of the statistics of $v$ and consequently not sensitive to very strong and rare fluctuations. Therefore, if we want to improve the behavior of $\langle \bar\varepsilon_{\rm kin}(t) \rangle $ for $c_1=1.9$, we should simply increase the statistics of the particular constrained ensemble to capture, by chance, events with smaller $v$ that are more representative of the unconstrained ensemble. This translates to the fact that for the constrained ensemble, a rare event can be an event which, for the unconstrained ensemble, is a typical one.

To sum up, reweighting of the ensemble-average kinetic energy is a sufficient but not a necessary condition to determine whether the particular constrained ensemble is representative of the original system. In fact, here it was a simple demonstration of the reweighting technique \eqref{eq:reweighting} in our application. As we will see in the following, we can achieve a very-well-behaved reweighting for the PDF of the velocity gradients for any $c_i$ and $w_i$, considering that the latter are appropriately chosen, as stated earlier, so that the HMC is numerically stable.

\subsection{Velocity gradient statistics}

To assess the performance of generating extreme and rare events, we compare the HMC, when using sampling constraints, with the DNS by studying the statistics related to the velocity gradients, such as their PDF. We note that, in the following, the observables that we consider are measured only at the single point that we constrain, i.e., at ($x=0$, $t=t_f$). This is related to the introduction of the local constraint $\Delta \S$, which breaks the space-time symmetry of the system. In principle, after applying Eq.\ \eqref{eq:reweighting} we restore the symmetries of the system, in the limit of infinite statistics, but in practice this is not the case. However, for histogram reweighting, by considering only the site on which the local constraint acted, we restore homogeneity and we will show that it is sufficient to obtain a systematic comparison with the unconstrained statistics, regardless the mutual overlap of the non reweighted histogram and the unconstrained histogram (e.g., of the DNS). For instance, in Fig.\ \ref{fig:HMC_diff_c_s}(b) the case of the HMC with $c_3=120$ and $w_3=30$ has no overlap with the DNS (even though there would be if we increased the statistics to infinite), yet in the following we will demonstrate that this particular PDF, together with other similar cases, will be successfully reweighted to the unconstrained statistics. Nevertheless, if we consider other sites, we encounter problems similar to the ones discussed in the preceding section, e.g., for the kinetic energy, where by increasing $c_i$ and $w_i$ we notice increasing error bars. 

\begin{figure}[!t]
	\centering
	\includegraphics[width=0.45\textwidth]{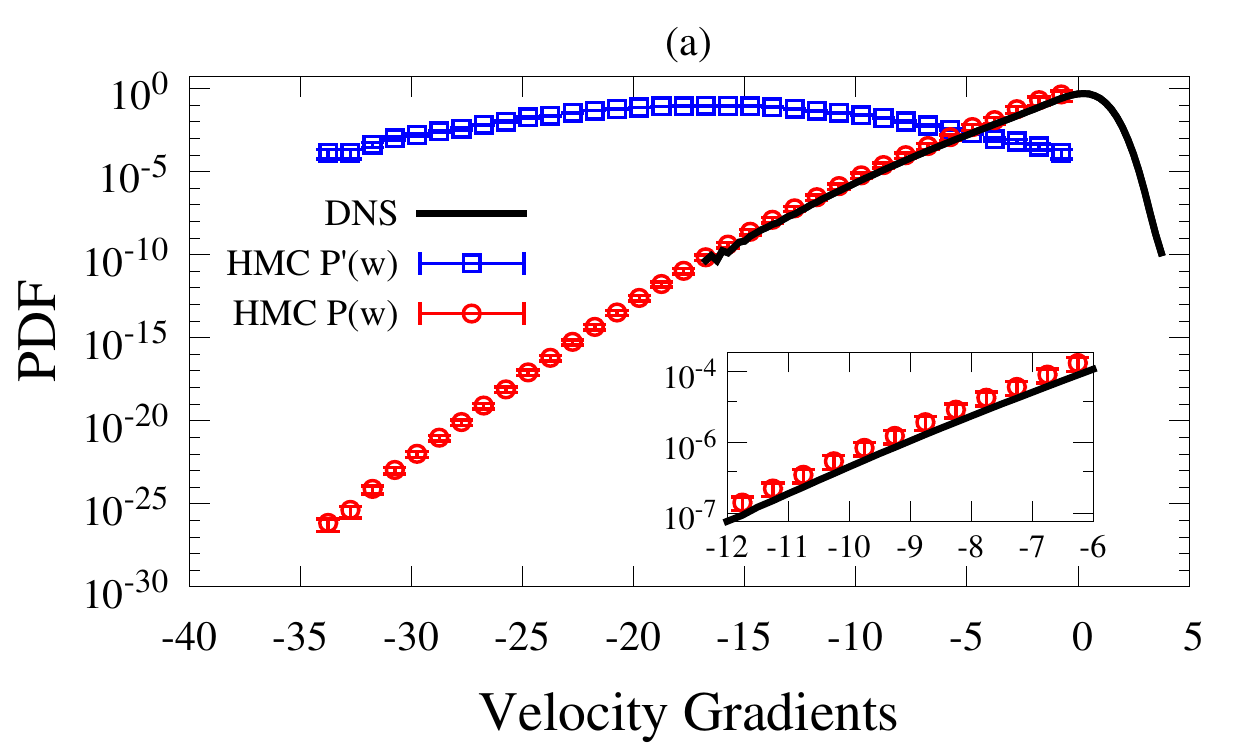} \quad
	\includegraphics[width=0.45\textwidth]{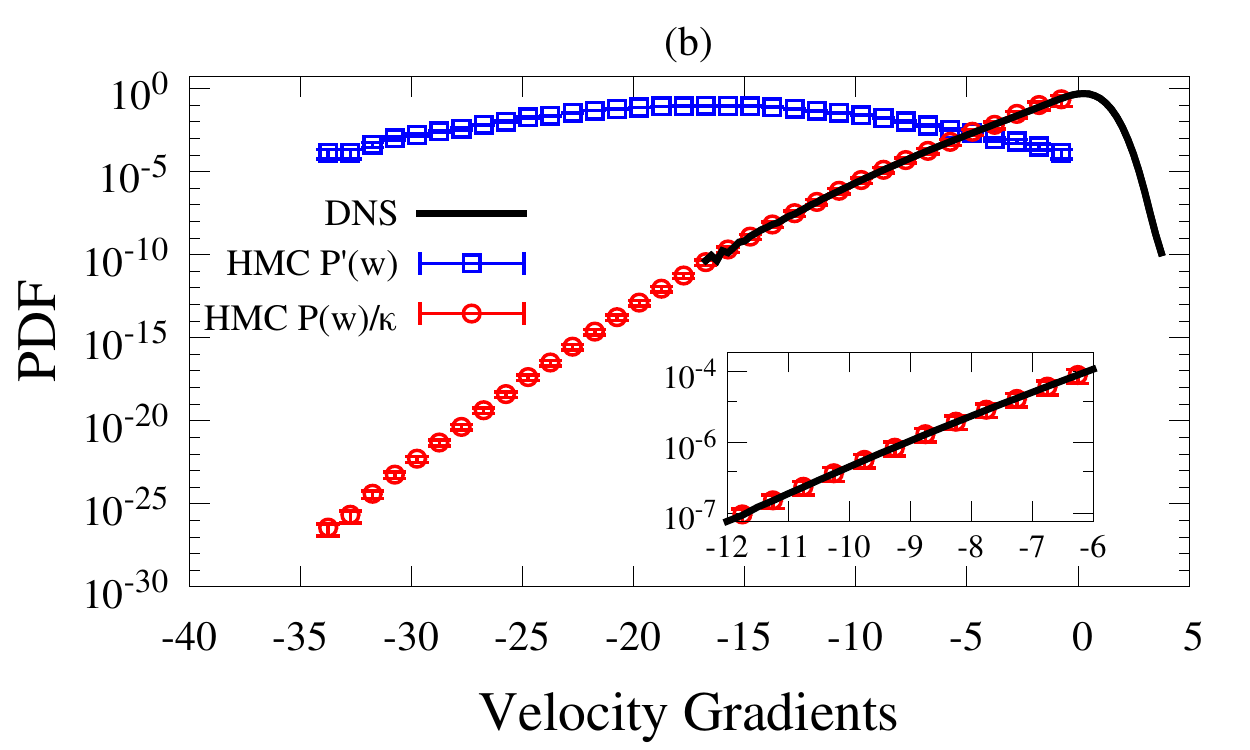} 
	\caption{PDF of the velocity gradients for the HMC (blue and red symbols) versus DNS (black line) (a) without rescaling and (b) rescaling the reweighted HMC histogram by dividing it with $\kappa=1.93$. For the HMC we use the constraint $\Delta \S_1$ with $c_1=1.9$, and $P'(w)$ is measured considering only the site ($x=0$, $t=t_f$) on which the constraint is enforced. The inset plot shows the chosen interval $[-12,-6]$.}
	\label{fig:rescaling}
\end{figure}

\begin{figure}[!t]
	\centering
	\includegraphics[width=0.45\textwidth]{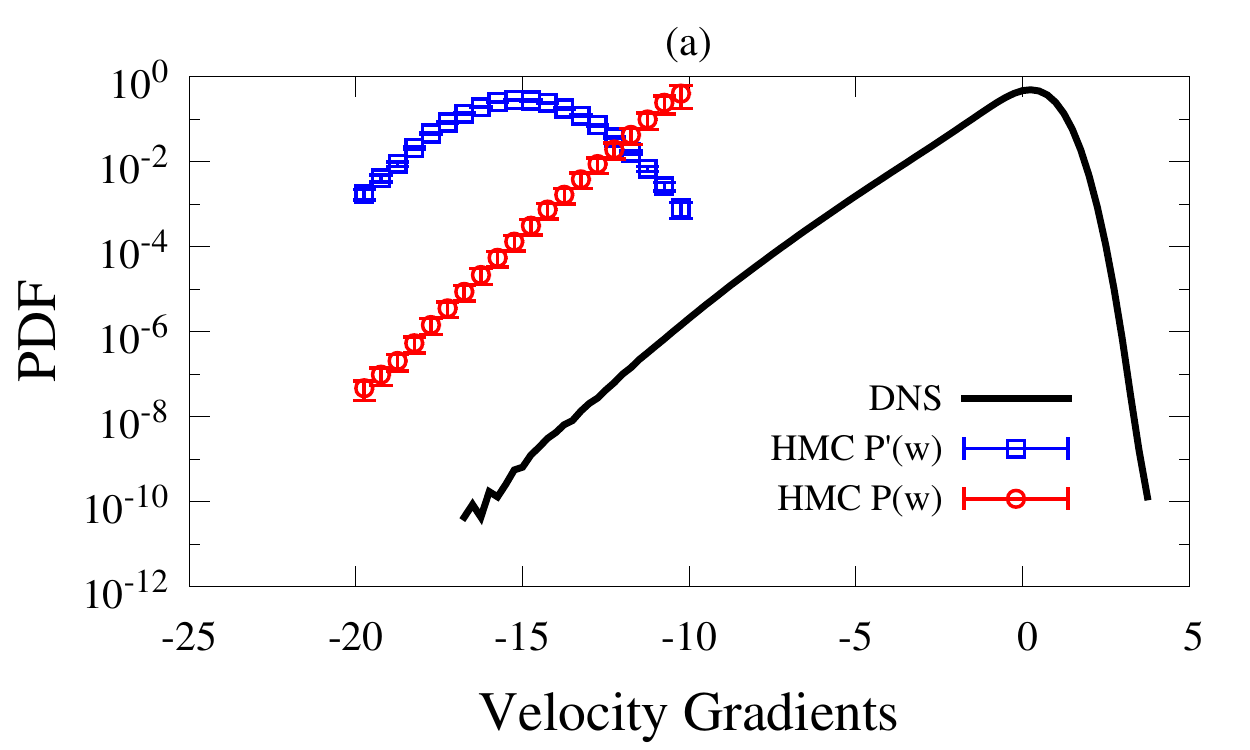} \quad
	\includegraphics[width=0.45\textwidth]{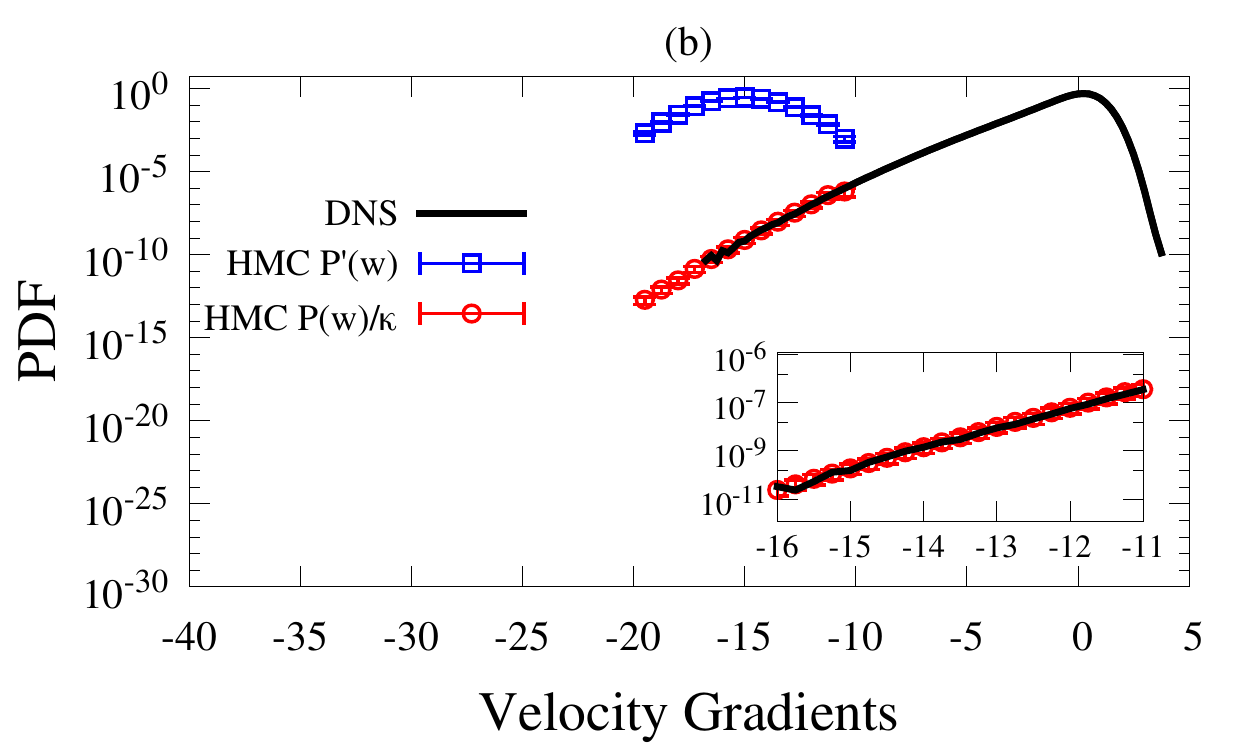}
	\caption{PDF of the velocity gradients for the HMC (blue and red symbols) versus DNS (black line) (a) without rescaling and (b) rescaling the reweighted HMC histogram by dividing it with $\kappa=2.63\times10^5$. We consider only the extracted histogram from the lattice point on which the constraint $\Delta \S_2$ acted, i.e., $x=0$, $t=t_f$, in the case of the HMC, with $c_2=80$ and $w_2=18$. The inset shows the chosen interval [-16,-11] for the rescaling.}
	\label{fig:rescaling_x_2}
\end{figure}

\begin{figure*}[!t]
  \centering
  \includegraphics[width=0.45\textwidth]{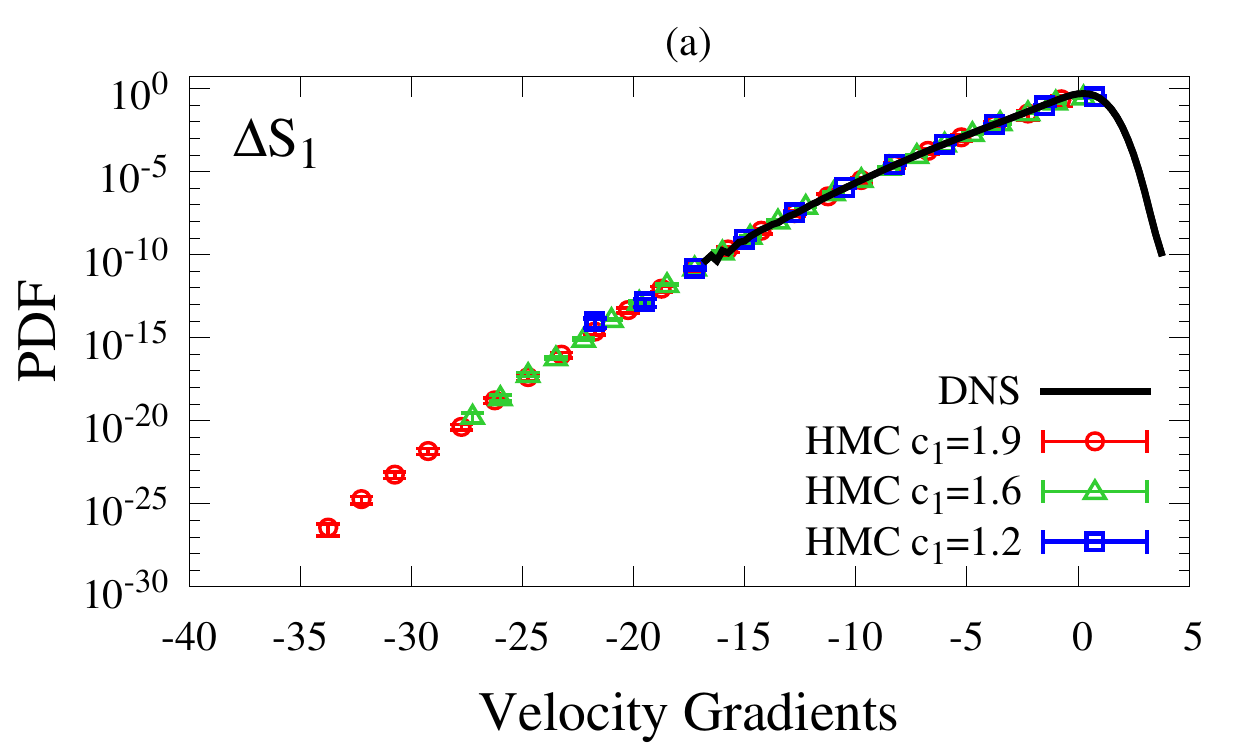} \quad
  \includegraphics[width=0.45\textwidth]{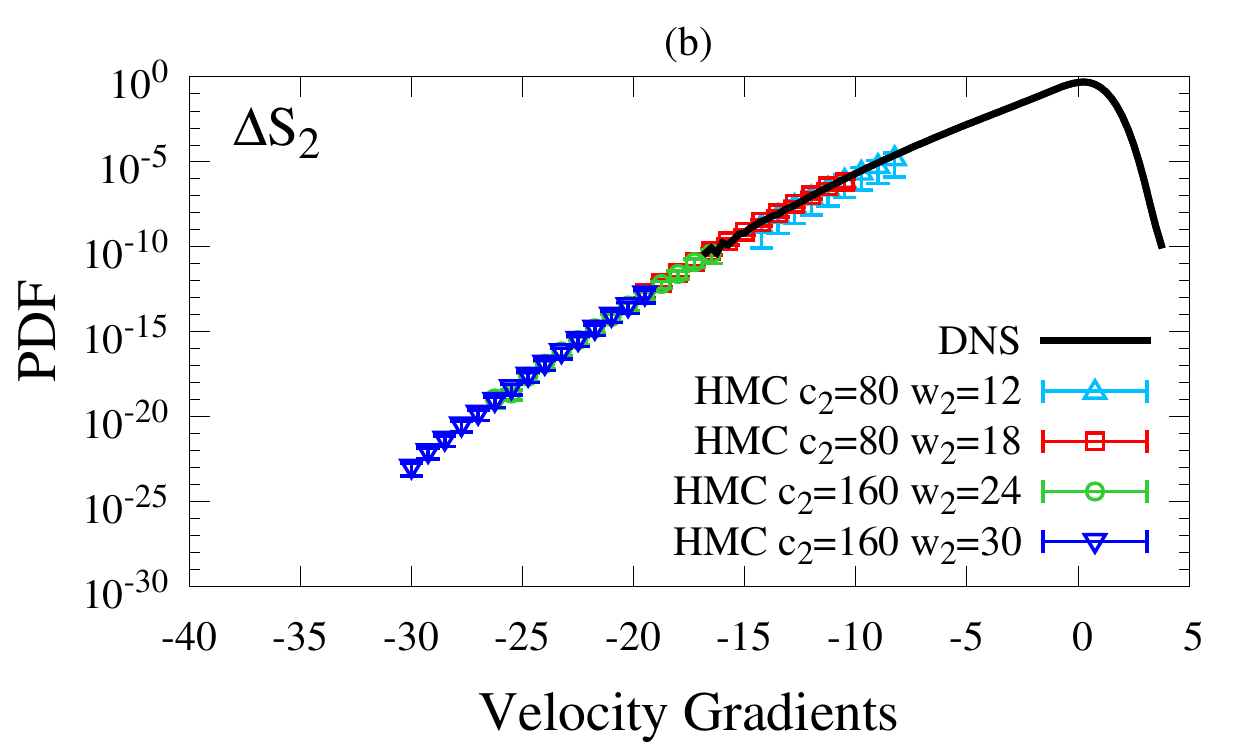}
  \includegraphics[width=0.45\textwidth]{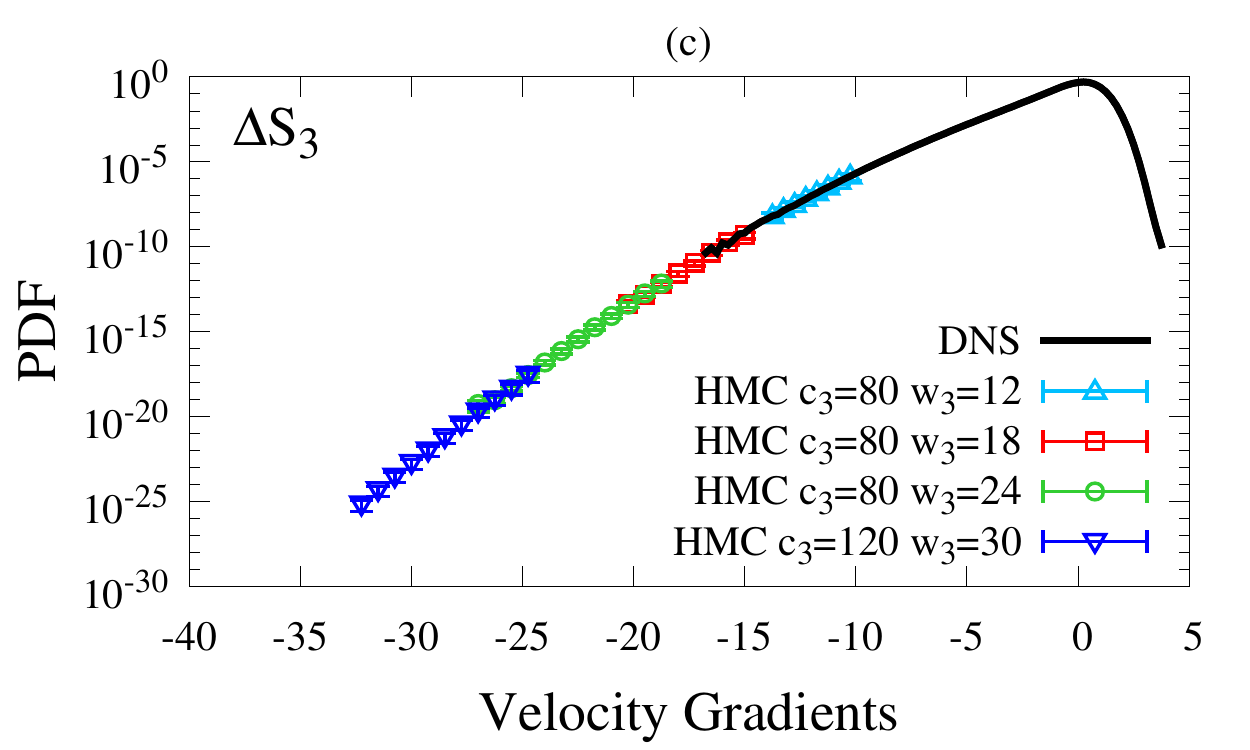} \quad
  \includegraphics[width=0.45\textwidth]{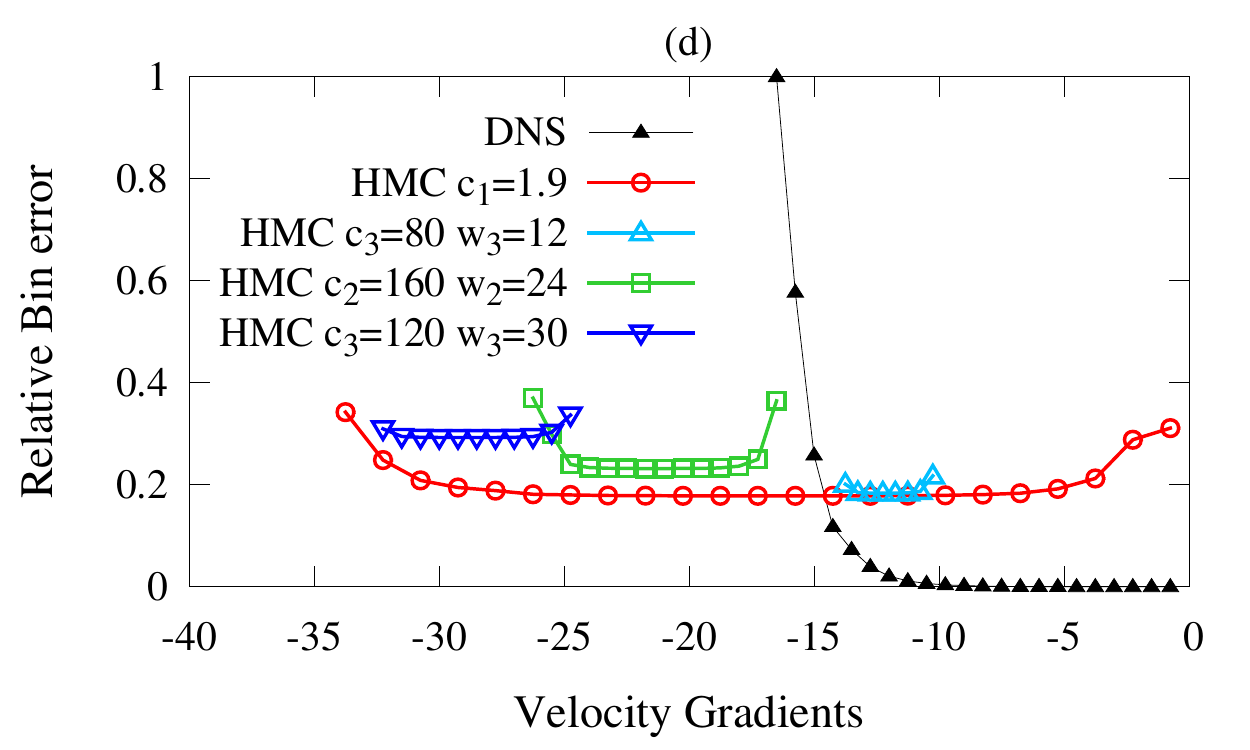}
  \caption{PDF of velocity gradients $P(w)$ for HMC and DNS. We consider here only the extracted histogram from the lattice point on which the constraint $\Delta \S$ acted (i.e.,$x=0$, $t=t_f$) in the case of the HMC. We show the effect of reweighting for different parameters of the constraints. (a) Using $\Delta \S_1$. (b) Using $\Delta \S_2$. (c) Using $\Delta \S_3$. (d) Relative bin error $ \delta P(w) / P(w)$, with the error being evaluated using Eq.\ \eqref{eq:error_propagation}. Regarding rescaling, for those $P(w)$ of which the overlap with the DNS was marginal or nonexistent, the rescaled $P(w)$ for $c_1=1.9$ was used.}
\label{fig:constr_HMC_pdf}
\end{figure*}

To reweight the PDF of the velocity gradients  $P'(w) = {\langle \delta (\partial_x v(0,t_f) - w) \rangle '}$, we use Eq.\ \eqref{eq:reweighting} to get
\begin{equation}
  P(w) = \frac{\langle \delta (\partial_x v(0,t_f) - w) e^{\Delta \S} \rangle '}{\langle e^{\Delta \S} \rangle '} ,
 \label{eq:v_grad_PDF}
\end{equation}
where, in practice, for each measurement $i$ of the ensemble, we increment the bin $w$ by $e^{\Delta \S_i}$. In Fig.\ \ref{fig:rescaling} we apply \eqref{eq:v_grad_PDF} on the ensemble that generated the non reweighted PDF $P'(w)$ (open blue squares) to produce the reweighted histogram $P(w)$ (open red circles) and compare it with the corresponding PDF of the DNS (black line). In Fig.\ \ref{fig:rescaling}(a) we identify a slight discrepancy between the $P(w)$ and the DNS (seen more clearly in the inset), while the trend is similar. This is related to the fact that the HMC is constrained to systematically sample large negative velocity gradients (far left tail) and therefore the support on the right tail is limited. As a result, by strictly applying \eqref{eq:v_grad_PDF}, and since it normalizes the area under  $P(w)$ to 1, the comparison between the HMC and the DNS is not straightforward, as $P(w)$ is actually an excerpt of the original PDF of velocity gradients, which is assumed to be the curve of the DNS here. For the same reason, $P(w)$ cannot be considered as a PDF. What is missing is rescaling of $P(w)$ with an appropriate factor $\kappa$ so that both the DNS and the HMC calculate the same probability $p(a,b)$ to sample in a particular interval $(a,b)$ of velocity gradients. By definition, $p(a,b)=\sum_{a}^{b}P(w)\delta w$, with $\delta w$ the bin width, so $\kappa$ is defined as the ratio of the two probabilities measured by the HMC and the DNS,
\begin{equation}
  \kappa = \frac{p(a,b)_{\textrm{HMC}}}{p(a,b)_{\textrm{DNS}}} ,
  \label{eq:rescaling_factor}
\end{equation}
where we have tested that by increasing the statistics of the HMC, $\kappa \rightarrow 1$. We also assume that the DNS has enough support in both tails to be claimed as a PDF and therefore to be considered as a reliable benchmark for the rescaling of $P(w)$. In Fig.\ \ref{fig:rescaling}(b) we show the rescaled $P(w) / \kappa$, with $\kappa=1.93$. Also here $\kappa$ is measured in the interval $[-12,-6]$ for the rescaling. In this way we achieve a collapse of the HMC and the DNS data. What is striking, in this example, is the unique ability of the HMC to systematically sample intense gradients that are up to $\sim 30\sigma$ and more, with $\sigma=0.99$, far from the mean. For a similar discussion on the resultant statistical efficiency of the chosen constraint functionals we refer the reader to \cite{Ushnish2018,Todd2015}. 

Another example where the need to further treat the reweighted velocity gradients histogram $P(w)$, by rescaling it with an appropriate factor $\kappa$, becomes more evident is when we consider one of $\Delta \S_2$ or $\Delta \S_3$. In Fig.\ \ref{fig:rescaling_x_2} we show $P'(w)$ (open blue squares) and $P(w)$ (red open circles) using the functional $\Delta \S_2$, with $c_2=80$ and $w_2=18$ in the case of the HMC, against the DNS (black line). In Fig.\ \ref{fig:rescaling_x_2}(a), $P(w)$ (red symbols) is derived by applying \eqref{eq:v_grad_PDF} to the PDF of the HMC (blue symbols). As before, the area below $P(w)$ is equal to 1. However, by considering $P(w)$, the probability $p(a,b)_{\textrm{HMC}}$ to sample within an interval $(a,b)$ of velocity gradients does not correspond to the one of the DNS, $p(a,b)_{\textrm{DNS}}$, so we employ again \eqref{eq:rescaling_factor} to get $\kappa=2.63\times10^5$. In Fig.\ \ref{fig:rescaling_x_2}(b) we plot $P(w) / \kappa$ (red symbols) instead, in order to achieve the collapse with the PDF of the DNS.

Now that we have clarified how to derive $P(w)$\footnote{Note that for the rest of article, when referring to $P(w)$, it is implied that $P(w)$ is rescaled with an appropriate $\kappa$.} and explained the need for a further rescaling with a constant, we can do the same procedure for all the different runs using the three different constraint functionals of Eqs.\ \eqref{eq:constr_functionals}. This is done in Fig.\ \ref{fig:constr_HMC_pdf}, where we compare $P(w)$ for different combinations of $\Delta \S_i$, $c_i$, and $w_i$ with the velocity gradients PDF of the DNS (black line).

In Fig.\ \ref{fig:constr_HMC_pdf}(a) we show $P(w)$ for $\Delta \S_1$ and different $c_1$, in Fig.\ \ref{fig:constr_HMC_pdf}(b) the results correspond to $P(w)$ for $\Delta \S_2$ and different $c_2$ and $w_2$, and Fig.\ \ref{fig:constr_HMC_pdf}(c) depicts the $P(w)$ for $\Delta \S_3$ and different $c_3$ and $w_3$. An important remark is that for those cases of the reweighted histogram $P(w)$, where the overlap with the DNS is marginal or absent, we used the rescaled $P(w)$ for $\Delta \S_1$ and $c_1=1.9$ as a guide to rescale them. For instance, this was necessary for $w_i=24,\,30$. Furthermore, the different $\kappa$ that were used for each case are shown in Table \ref{table3}. Finally, Fig.\ \ref{fig:constr_HMC_pdf}(d) shows the relative bin error $\delta P(w) / P(w)$ as a measure of the statistical efficiency of each different constraint $\Delta \S_i$. For the HMC, we used Eq.\ \eqref{eq:error_propagation} to measure $\delta P(w)$, while for the DNS it is simply equal to $\delta P(w) = 1/\sqrt{\text{counts}}$. Interestingly, the HMC has a constant ratio for extreme values of the velocity gradients, while the DNS quickly diverges as soon as the statistics are limited. Note that a typical ensemble size of the HMC runs is of the order of $10^4$, while that of the DNS is of the order of $10^9$.

\begin{figure*}[!t]
	\centering
	\includegraphics[width=0.31\textwidth]{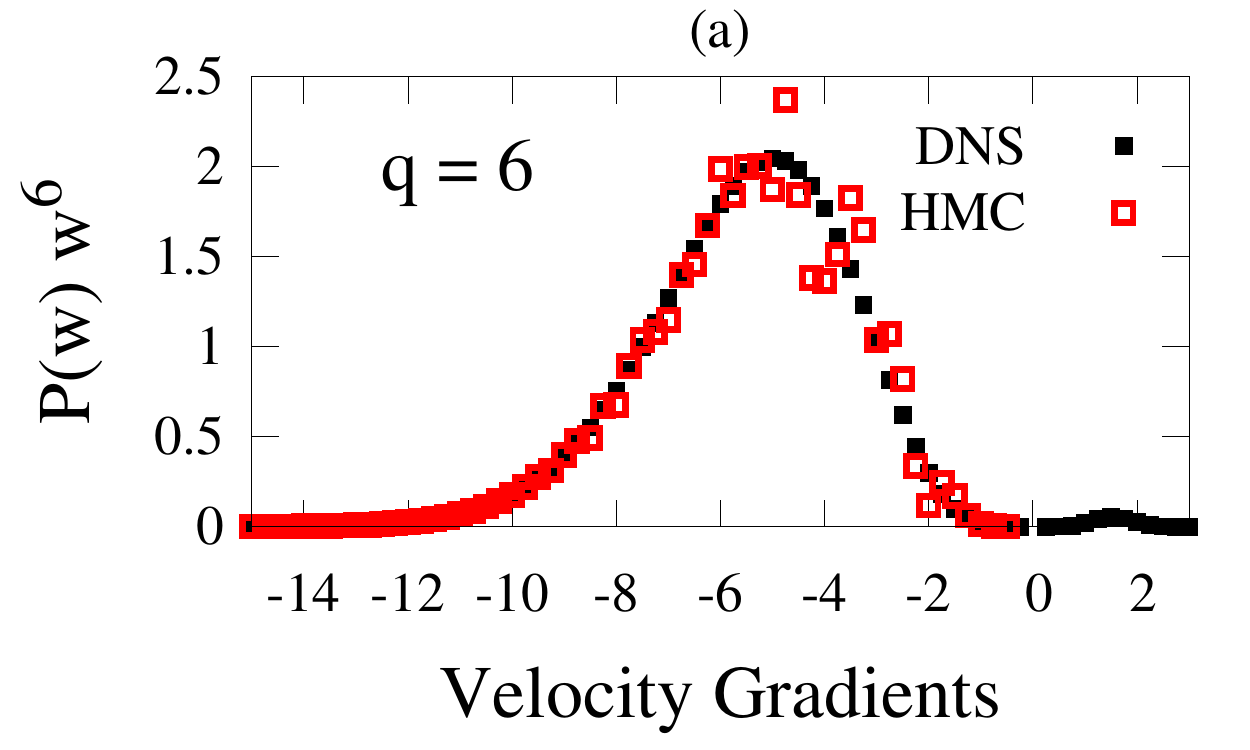} \quad
	\includegraphics[width=0.31\textwidth]{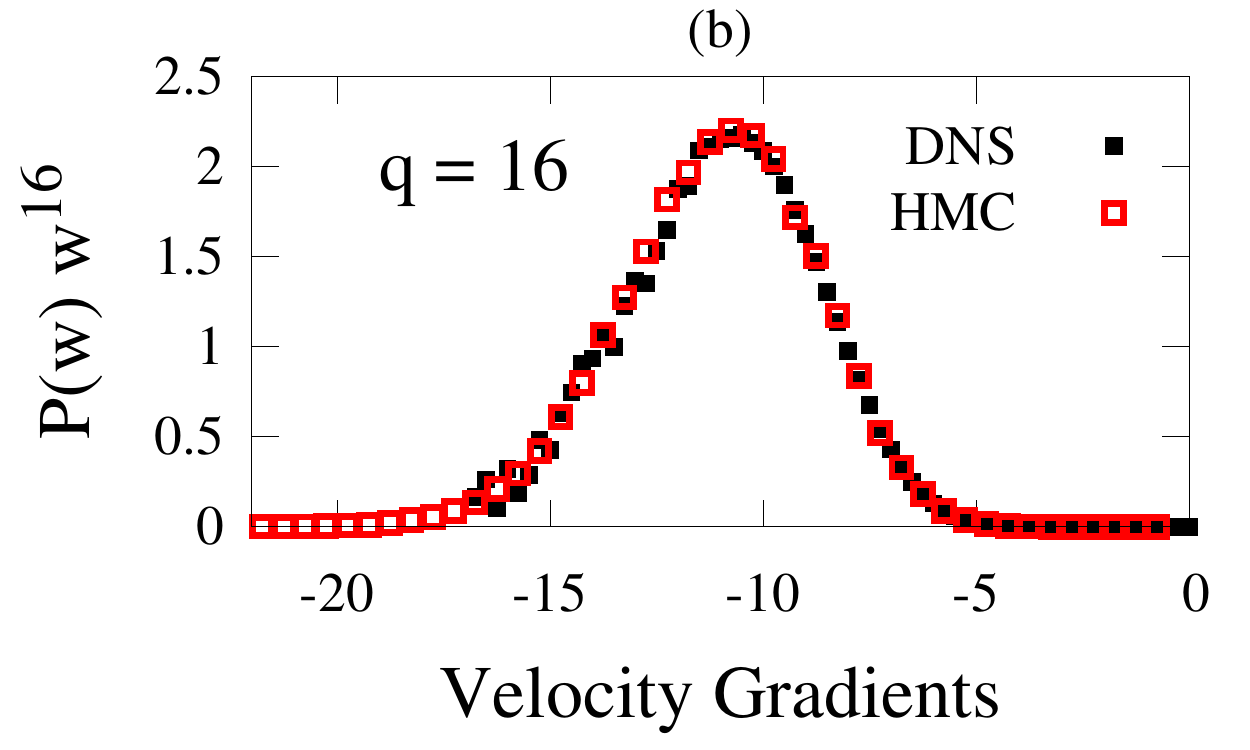} \quad \includegraphics[width=0.31\textwidth]{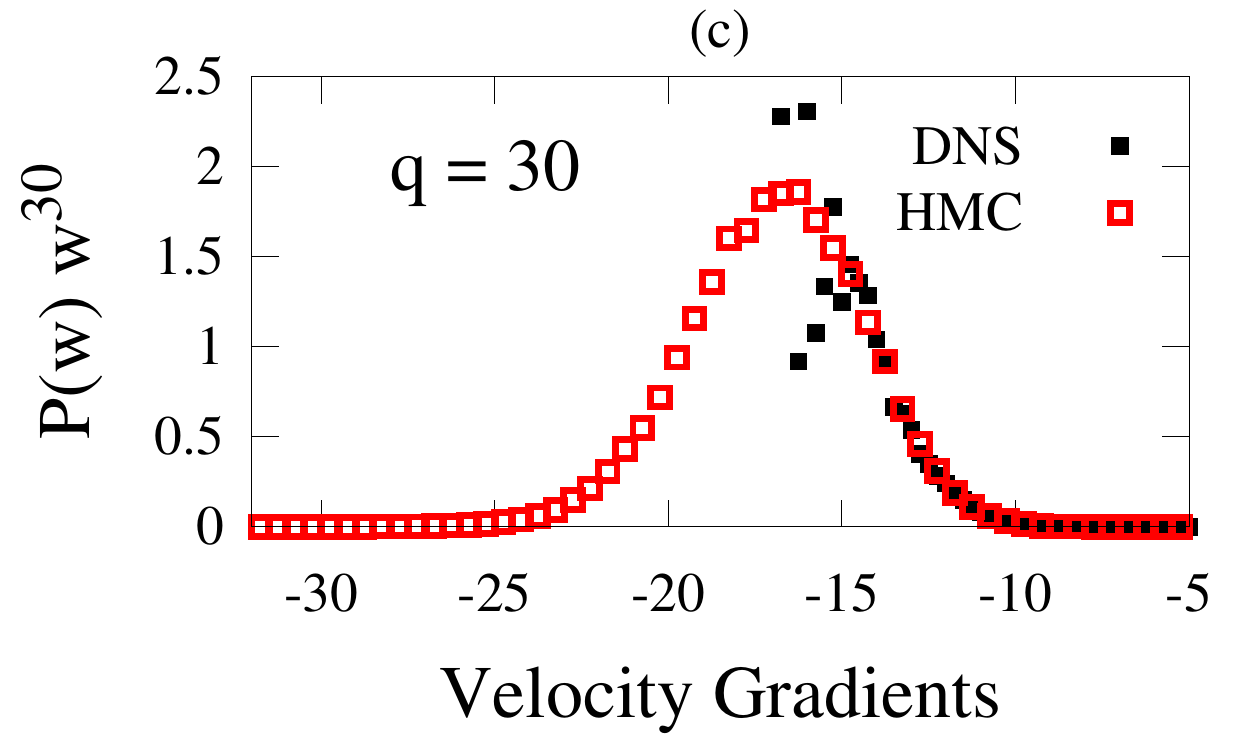} 
	\includegraphics[width=0.31\textwidth]{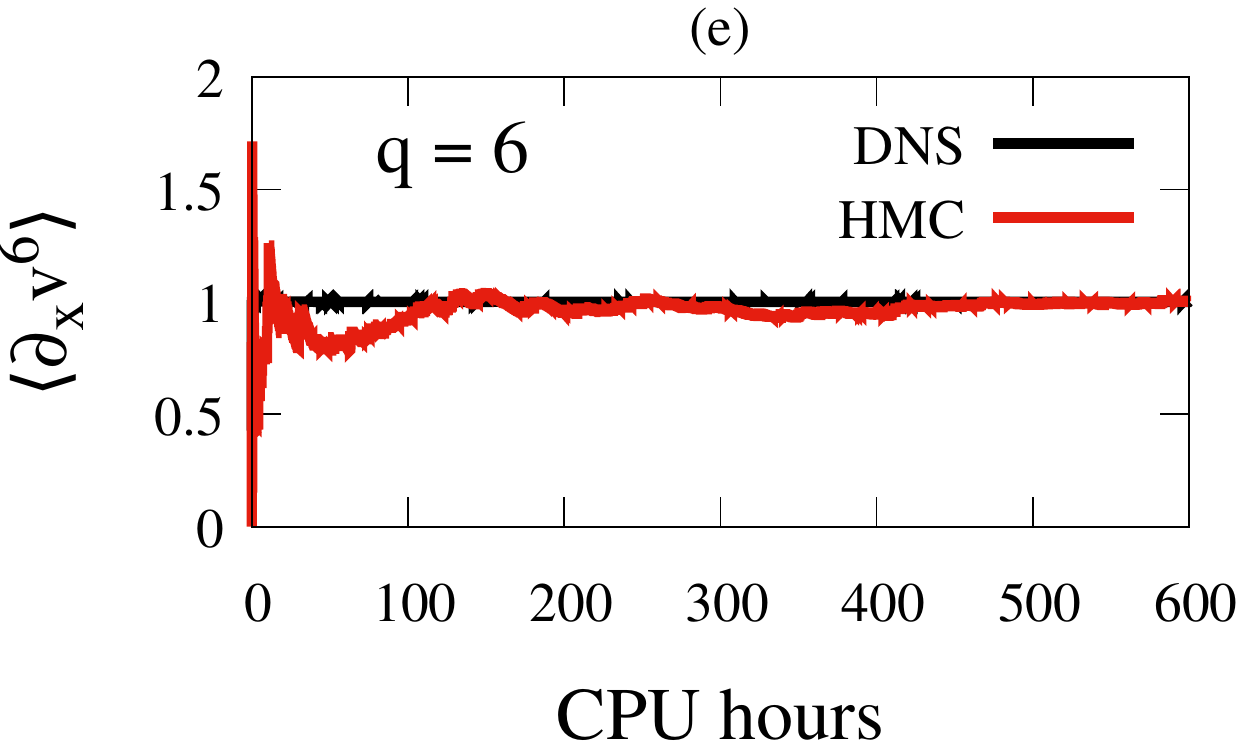} \quad
	\includegraphics[width=0.31\textwidth]{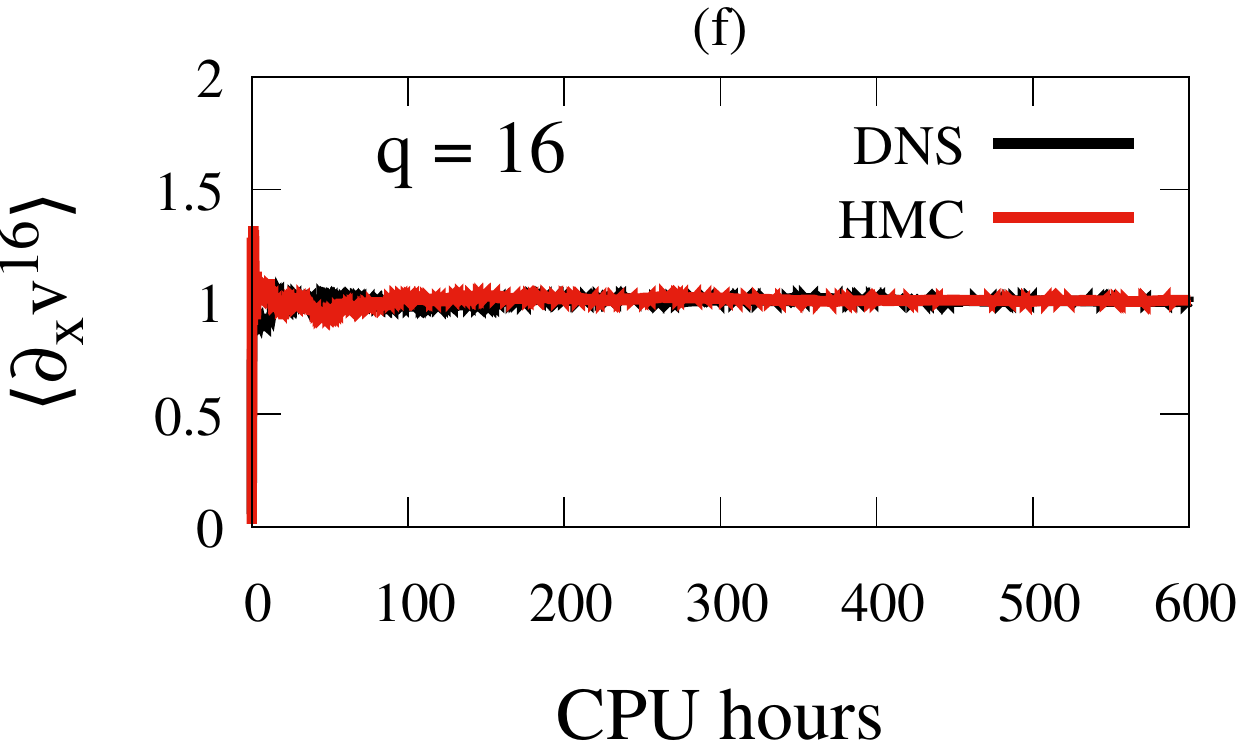} \quad
	\includegraphics[width=0.31\textwidth]{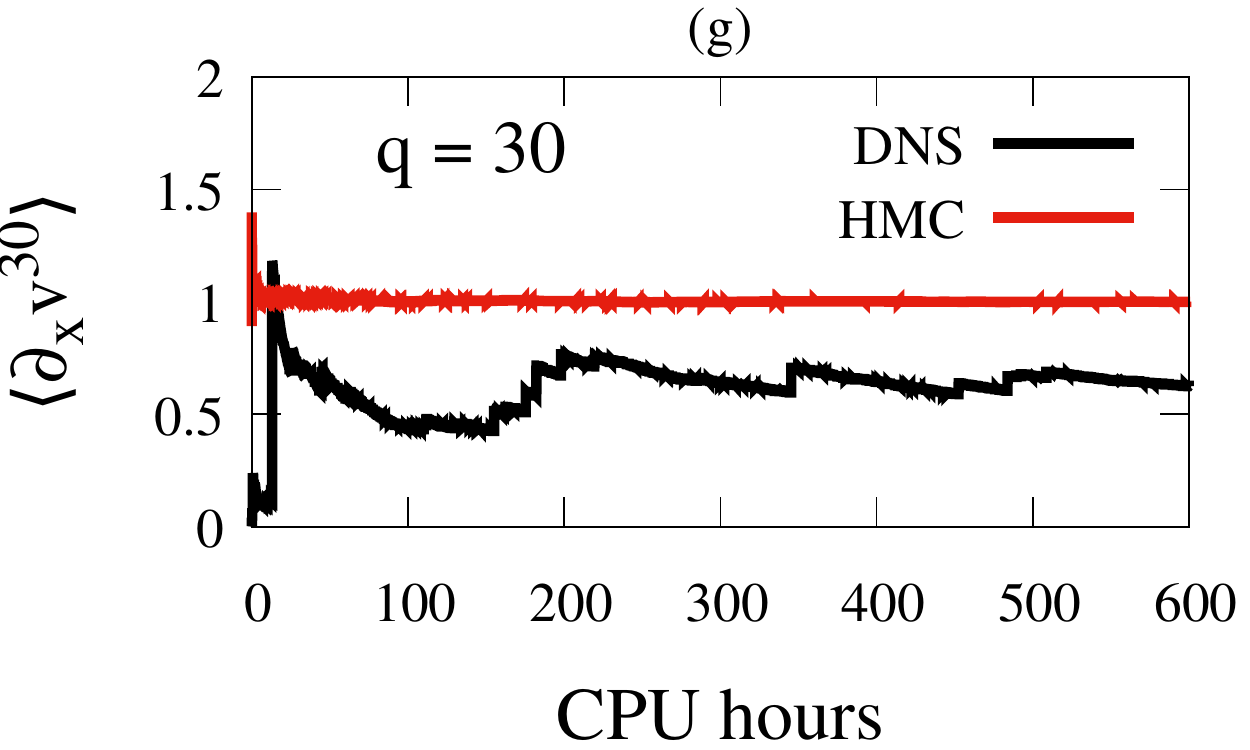}
	\caption{(a)--(c) Velocity gradient PDF multiplied by a moment $w^q$. The $P(w)$ of the HMC is reweighted, rescaled with $\kappa$, and we consider only the lattice point at which the constraint acts. Here we used $\Delta \S_1$ for $c_1=1.9$.\\\hspace{\textwidth}
		(d)--(f) Computational time to the stabilized running average of the velocity gradient moment $\langle (\partial_x v)^q \rangle$, divided with respect to the final stabilized value. Regarding DNS, any site belonging to the stationary regime is considered.}
	\label{fig:t_to_sol}
\end{figure*}

To further quantify the performance of the HMC for the purpose of systematically sampling very intense velocity gradients we provide Fig.\ \ref{fig:t_to_sol}. Figures \ref{fig:t_to_sol}(a)-- \ref{fig:t_to_sol}(c) show $P(w) w^q$, i.e., the reweighted and rescaled histogram of the velocity gradients multiplied by a moment $w^q$. The idea is that the higher the power $q$ is, the more we focus towards larger negative gradients. If the statistics of $P(w)$ are sufficient in the corresponding ``focused'' region, then $P(w) w^q$ has a clear peak and shape. Figures \ref{fig:t_to_sol}(d)-- \ref{fig:t_to_sol}(f) depict the computational cost that the ensemble running average of a moment of a velocity gradient $\langle (\partial_x v)^q \rangle$ requires in order to stabilize at a certain value and stop fluctuating. Here, for the HMC, we used the functional $\Delta \S_1$, for $c_1=1.9$, and we consider only the velocity gradient at the point ($x=0$, $t=t_f$). The data here are the same as the red and black data sets of Fig.\ \ref{fig:rescaling} for the HMC and the DNS, respectively. Also, the observable is reweighted according to Eq.\ \eqref{eq:reweighting} so that the comparison is equivalent. Finally, for visualization purposes, we normalize to one the observables by dividing them by the final value of the stabilized line (depending on $q$, this might be the line of either the HMC or DNS).

The plots in Fig.\ \ref{fig:t_to_sol} are complementary, as a specific power $q$ is chosen for each column. The plots in Figs. \ref{fig:t_to_sol}(a) and \ref{fig:t_to_sol}(d) are for a small $q=6$. In this region the DNS performs better as here the data of the HMC are only measured on the site on which the constraint acts, and therefore the appearance of relatively small negative gradients is suppressed. For $q=16$ [Figs. \ref{fig:t_to_sol}(b) and \ref{fig:t_to_sol}(e)] we see that both the HMC and the DNS are equivalent in terms of the computational cost and quality of the statistics. Finally, for $q=30$ [Figs. \ref{fig:t_to_sol}(c) and \ref{fig:t_to_sol}(f)] the HMC significantly outperforms the DNS, as it immediately converges to the expectation value, while for the DNS we would have to remarkably increase the computational cost to achieve comparable statistics. Note that the data for both the HMC and the DNS in Figs.\ \ref{fig:constr_HMC_pdf} and \ref{fig:t_to_sol} required the same computational cost to be produced, using the same processors. Overall, Fig.\ \ref{fig:t_to_sol} summarizes the ability of the HMC to consistently sample intense negative gradients that belong in the large-deviation regime and furthermore gives a qualitative measure of the computational performance gained over a standard DNS method.

\section{The relevance of instantons in extreme events}
\label{Sec:The relevance of instantons in extreme events}

The application of instantons in turbulent flows was first proposed in \cite{Gurarie:1995qc}, where the instanton contribution to the right tail of the velocity increment PDF was calculated for Burgers turbulence, while in a succeeding work \cite{Balkovsky:1997zz}, the left tail of the increment PDF was studied using the instanton approach. These works paved the way to other hydrodynamical models, such as the advection of a passive scalar by a turbulent velocity field \cite{Falkovich:1995fa,Chertkov:1996be}, shell models \cite{Daumont2000,Biferale1999}, geophysical flows \cite{Bouchet2011,Laurie2015,Bouchet2009}, and atmospheric and oceanic flows \cite{Weeks1997,Schmeits2001} (see also \cite{Grafke:2015yua} and references therein).

\subsection{Derivation of the instanton configuration}
\label{Sec:Derivation of the instanton configuration}

In order to calculate ensemble averages of observables $\langle\Ob_v\rangle$ as, e.g., the probability distribution of the gradient $P(\partial_x v = w) = \langle \delta(\partial_x v(x=0,t = t_f) - w)\rangle$, we utilize the path integral formulation introduced in Sec. \ref{Sec:Path integral for stochastic dynamics}: 
\begin{eqnarray}
  P(w) &\propto& \int \Dv \, \delta(\partial_x v(x=0,t=t_f)-w) e^{-\S} \nonumber \\
  &=& \int \Dv \, \int_{-i\infty}^{i\infty}  d\lambda \, e^{-S'(\lambda)} .
\end{eqnarray}
Here $\S' = \S'(\lambda)$ contains both the Onsager-Machlup action $\S$ [cf.\ Eq.\ \eqref{eq:action_S}] and the contribution of the observable $\delta(\partial_x v(x=0,t=t_f)-w)$:
\begin{eqnarray}
  &&\hspace{-20pt} \S' = \S + \lambda \left(\partial_x v(0,t_f) - w\right) \nonumber \\
  && \hspace{-10.5pt} = \int_{t_0}^{t_f} dt\, \left\{ \frac{1}{2} \left(F, \,\Gamma^{-1} \ast F\right) \right. \nonumber\\
  &&  + \lambda \left(\partial_x v(x,t) - w,\delta(x) \right) \delta(t-t_f) \Big\} - \ln \J
  \label{eq:tildeaction}
\end{eqnarray}

Instanton configurations are ``classical'' solutions that extremize the action and therefore dominate the path integral of the stochastic Burgers equation \eqref{eq:hami}. They can be computed by Laplace's method or alternatively, as in many applications, instantons are found by numerically minimizing the action directly (see, e.g., \cite{Bouchet2011}). Here, where the observable is evaluated only at the final time $t=t_f$, it is advantageous to switch to another equivalent formulation by applying a Hubbard-Stratonovich transformation \cite{hubbard:1959,stratonovich:1957}, which leads to the alternative representation of the partition sum
\begin{equation}
  \Z \propto \int \Dv \,\D\mu\, e^{\int dt \,\left\{ i (\mu,F) - \frac{1}{2} (\mu, \,\Gamma \ast \mu) \right\} + \ln \J} ,
  \label{eq:functionalint_msr}
\end{equation}
which prompts us to define
\begin{equation}
  \S_{{\rm MSRJD}} = -\int dt \,\left\{ i (\mu,F) - \frac{1}{2} (\mu, \,\Gamma \ast \mu) \right\} - \ln \J ,
\end{equation}
also known as Martin-Siggia-Rose -- Janssen -- de Dominicis (MSRJD) action \cite{Janssen1976,dedominicis:1976}. At the expense of an additional auxiliary field $\mu$, we have ``linearized'' the action with respect to the noise $\eta ~(=\!F)$. Furthermore, the force correlator $\Gamma$ now appears directly and not through its inverse $\Gamma^{-1}$. This allows for the implementation of more general types of forcing as the power-law forcing considered in this paper. Now the corresponding expression for the PDF of velocity gradients reads
\begin{equation}
  P(w) \propto \int \Dv\, \D\mu\, \int_{-i\infty}^{i\infty} d\lambda \, e^{-\S'_{{\rm MSRJD}}} ,
  \label{eq:prob_inst}
\end{equation}
with
\begin{eqnarray}
  \S'_{{\rm MSRJD}} &=& \S_{{\rm MSRJD}} + \lambda \left(\partial_x v(0,t_f) - w\right) .
\end{eqnarray}
Before we proceed, we note that attempting to compute path integrals of the form of Eq.\ \eqref{eq:functionalint_msr} is not straightforward and might be impossible for most cases. For instance, perturbative approaches might be helpful, depending on the problem. In the context of fluid dynamics, a diagrammatic approach (influenced by quantum field theory) was proposed by Wyld \cite{Wyld:1961}. Using perturbation theory to expand the exponential in Eq.\ \eqref{eq:functionalint_msr} in powers of the nonlinear term [see also Eq.\ \eqref{Eq:burgers}] proves insufficient in the turbulent limit $\nu \rightarrow 0$, since the path integral is dominated by the nonlinear term forming strong shocks. Therefore, perturbative approaches must be abandoned, as a large parameter is required \cite{Gurarie:1995qc}.

Nevertheless, the introduced Lagrange multiplier $\lambda$ in Eq.\ \eqref{eq:tildeaction} can be used as a large parameter. This allows the use of the saddle-point approximation, by which the variation of the integrand in Eq.\ \eqref{eq:prob_inst} is equal to zero. In the case of Burgers turbulence, we obtain the instanton equations (minimizer of the action $\S'_{{\rm MSRJD}}$)
\begin{subequations}
  \label{eq:instanton_burgers}
  \begin{eqnarray}
    \partial_t v + v\partial_x v - \nu \partial_x^2 v &=&  - i\, \Gamma \ast \mu , \label{eq:instanton_burgers_u}  \\
    \partial_t \mu + v\partial_x \mu + \nu \partial_x^2 \mu &=& i \lambda \delta'(x) \delta(t-t_f) ,\label{eq:instanton_burgers_mu}
  \end{eqnarray}
\end{subequations} 
where the term on the right-hand side. in Eq.\ \eqref{eq:instanton_burgers_mu} implements the boundary condition for $\mu$ at $t_f$ according to which $\mu(x,t_f) = i\lambda \delta'(x)$. Recall that in the case of Burgers equation $\J = {\rm const.}$ and therefore the Jacobian does not contribute to the saddle-point equations. In \cite{Chernykh:2001} an algorithm was proposed to numerically solve the above equations. In short, the sign in front of the viscous terms defines the temporal direction of the numerical integration, with $v$ being integrated forward in time and $\mu$ backward. Using $\mu(x,t_f) = i\lambda \delta'(x)$ as an initial condition  for some large value of $t_f$ and starting by setting $v(x,t)=0$, Eq.\ \eqref{eq:instanton_burgers_mu} is first integrated backward until $t_0$. Then the obtained $\mu(x,t)$ is used to integrate Eq.\ \eqref{eq:instanton_burgers_u} forward in time, with the whole procedure being iterated until convergence to the prescribed constraint $\partial_x v(0,t_f) = w$ is achieved. For more details see also \cite{Grafke:2013ska,Grafke2015,Grafke:2015yua}, where the aforementioned methodology is revisited.

\subsection{Numerical results}
\label{Sec:Numerical results}

Instantons, strong field-force fluctuations and extremal points of the action $\S'_{{\rm MSRJD}}$, may be considered as particular examples of extreme and rare events. Constraining the HMC to sample at large negative gradients, we observe that the generated configurations clearly resemble the classical instanton configurations determined via the saddle-point approximation. This will be checked directly via the averaged velocity-field profile and through the probability distribution function of velocity gradients.

\begin{figure}[t]
  \centering
  \includegraphics[width=0.45\textwidth]{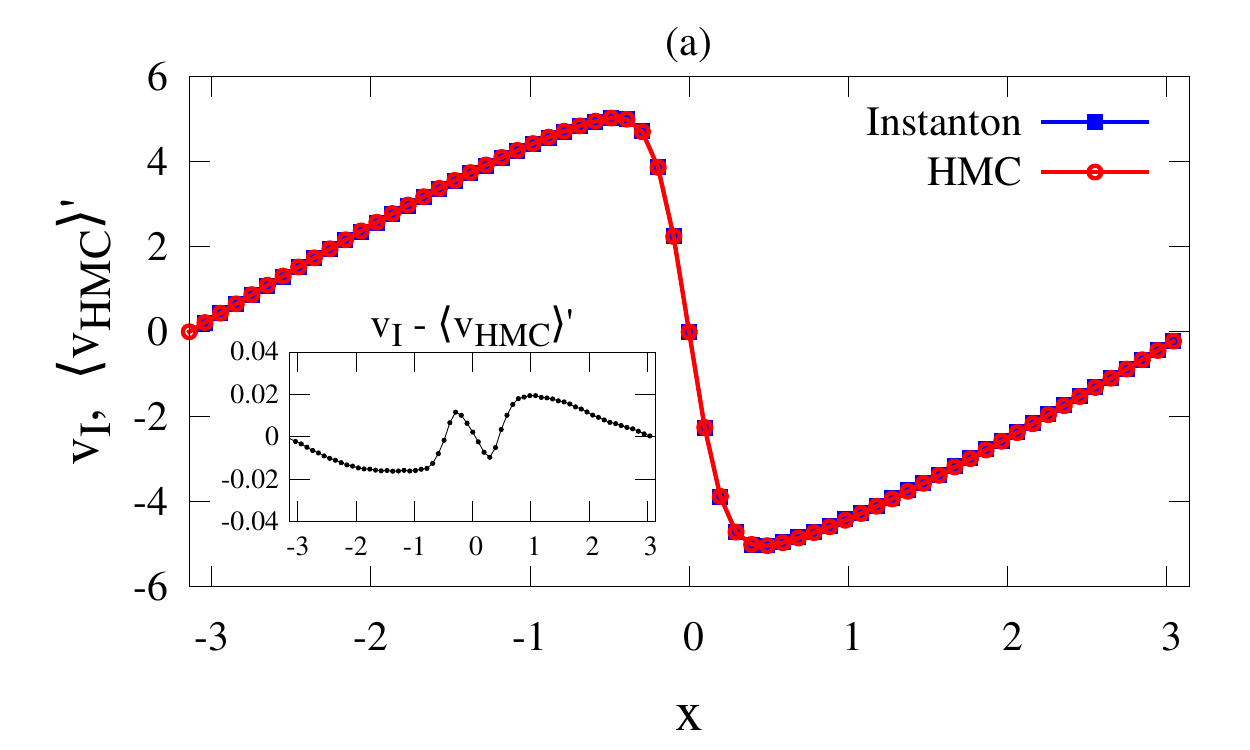} \quad
  \includegraphics[width=0.45\textwidth]{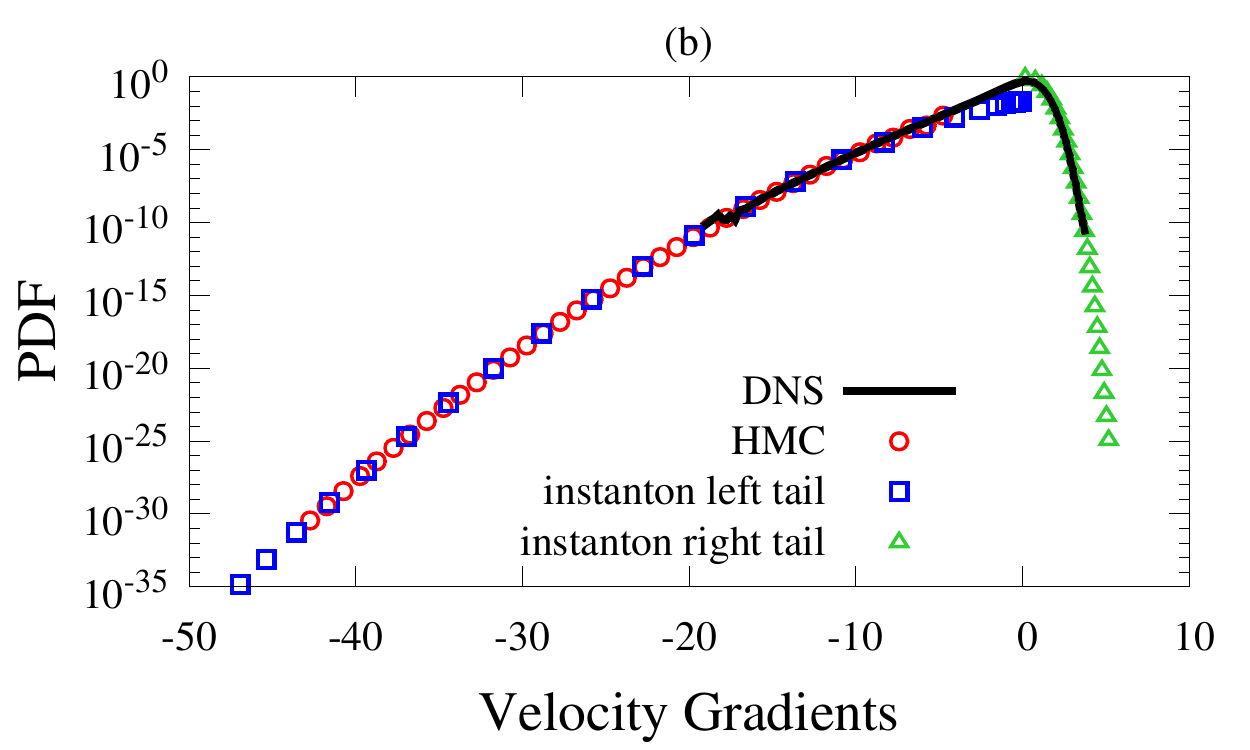}
  \caption{(a) Ensemble average of velocity configurations generated by the HMC using $\Delta \S_1$ with $c_1=1.9$ compared to the classical instanton velocity-field profile generated for $\lambda=-1.148$ and $w=-24.23$. (b) PDF of the velocity gradients for the classical instanton (for a range of values of $\lambda$ and $w$), HMC simulation, and DNS.}
  \label{fig:HMC_vs_inst}
\end{figure}

\begin{figure}[t]
  \centering
  \includegraphics[width=0.45\textwidth]{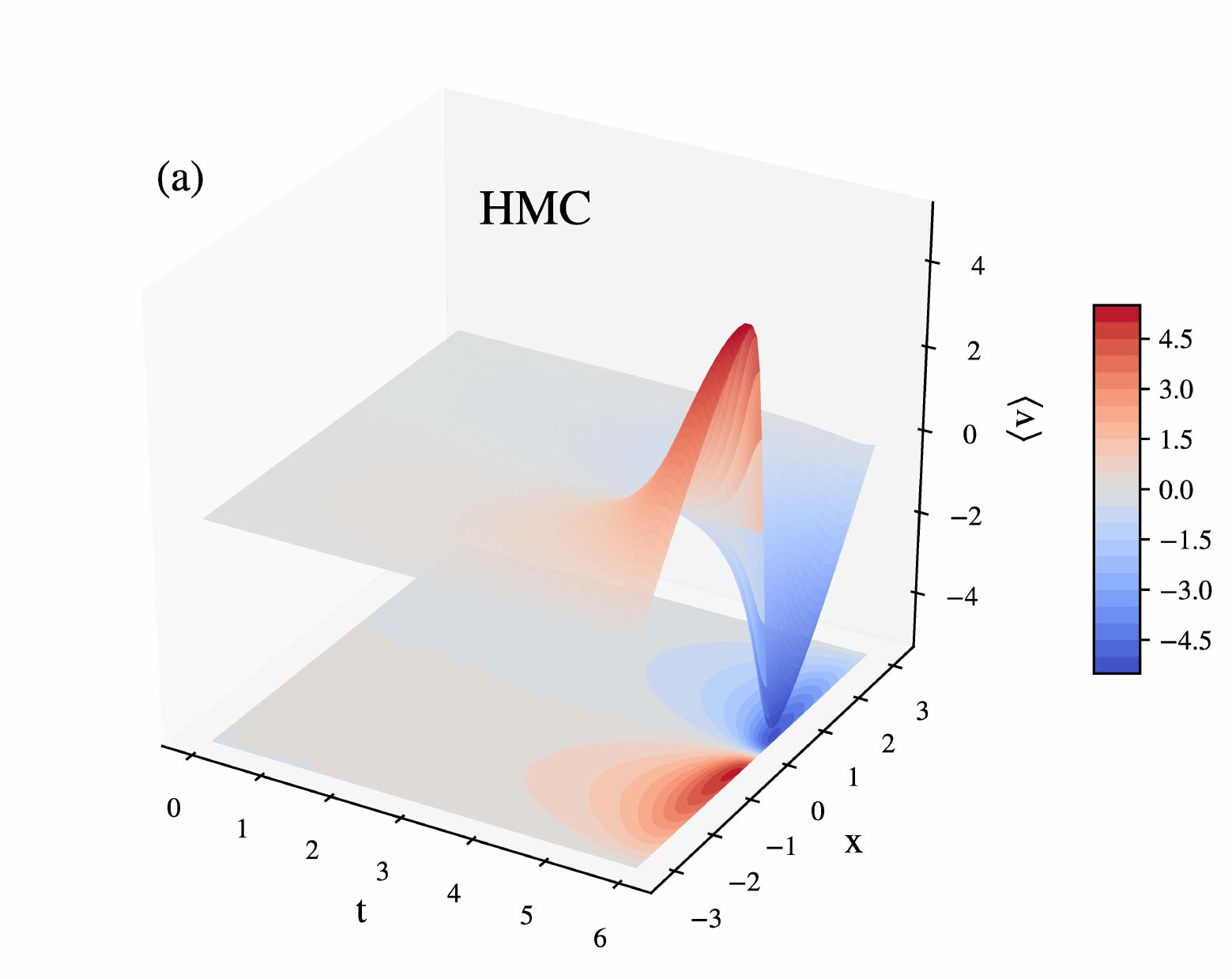} \quad
  \includegraphics[width=0.45\textwidth]{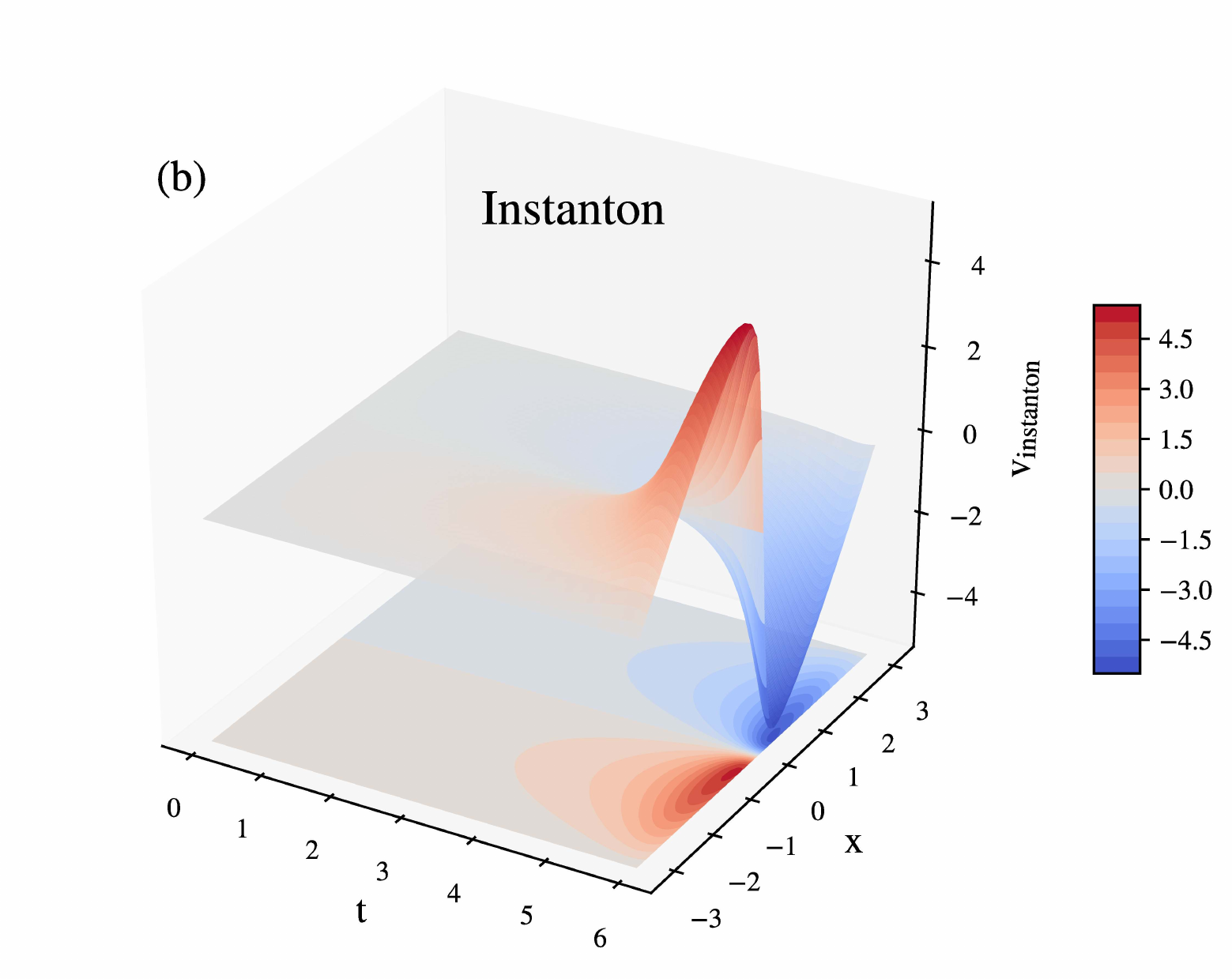} 
  \caption{(a) Ensemble average of velocity configurations generated by the HMC using $\Delta \S_1$ with $c_1=1.9$. (b) Instanton velocity field for $\lambda=-1.148$ and $w=-24.23$. By averaging over the HMC velocity-field ensemble the spatio temporal shape of the classical instanton is restored.}
  \label{fig:instanton}
\end{figure}

Figure \ref{fig:HMC_vs_inst}(a) compares the HMC ensemble average of the velocity field at the last time slice ($t=t_f$) with the velocity field obtained by performing the numerical integration of the instanton equations \eqref{eq:instanton_burgers}. The profile of the classical instanton at time $t=t_f$ s reproduced to a remarkable degree, implying that the ensemble average is equivalent to removing the fluctuations around the instanton. This confirms that instantons can be found in Burgers turbulence, as already shown in \cite{Grafke:2013ska} using a post production filtering protocol to consider only events with strong gradients generated using DNS. Furthermore, the inset depicts the difference of the two velocity fields, which are on the order of statistical error.  Similarly, in Fig.\ \ref{fig:instanton} we compare the whole averaged spatio temporal domain of the HMC with the instanton velocity field in space and time. For Figs.\ \ref{fig:HMC_vs_inst} and \ref{fig:instanton} a resolution of $N_t=576$ points in time was used for the HMC, the DNS, and the instanton, while the rest of the parameters are the same as in Table \ref{table3}.

Figure \ref{fig:HMC_vs_inst}(b) compares the PDF of the velocity gradients of the DNS, the HMC with constrained sampling, and the instanton. In the case of the instanton we plot $e^{-\S_{\textrm{inst}}}$, with $\S_{\textrm{inst}} = -\frac{1}{2}\int_{t_0}^{t_f} dt \, (\mu, \,\Gamma \ast \mu)$. We notice that the PDF predicted by the instanton follows the same trend as the HMC and the DNS, and the agreement is extraordinary. However, in order to correctly interpret this result one should note the following. On the one hand, the PDF prediction for positive gradients is valid independently of the Reynolds number $\Re$ and is actually valid for all positive values besides small corrections near $\partial _x v = 0$. This is a result already obtained by Feigel'man \cite{Feigelman:1980} in the context of charge density waves and also confirmed by the instanton formalism \cite{Gurarie:1995qc}. On the other hand, the PDF of negative gradients depends on the Reynolds number and for a given Reynolds number $\Re_\ell$ the instanton prediction is only valid for $|\partial _x v| > |\partial _x v^*(\Re_\ell)|$. A precise estimate for $\partial _x v^*$ is given in \cite{Grafke2015} [see Eq.\ (17) in the same reference]. For the Reynolds number $\Re_\ell = 1$ used in our simulation and depicted in Fig.\ \ref{fig:HMC_vs_inst}, this means that the instanton prediction is valid only for $\partial _x v < -10$. 

\section{Conclusion}
\label{Sec:Conclusions}

In this work we established how to apply Monte Carlo importance sampling for stochastic dynamics based on the Janssen--de Dominicis path integral, in order to address the statistics of large fluctuations in driven nonequilibrium systems. This approach allowed us to access the phase space of all possible field realizations of a stochastic system. Using reweighting techniques, we were able to systematically enhance the occurrence of extreme and rare events, by sampling in specific phase-space regions related to such events.

We have chosen to illustrate the HMC algorithm as an example of the random-noise-driven one-dimensional Burgers equation, which often is used as a model for benchmarking numerical methods in computational fluid dynamics. However, the HMC approach is generally applicable to any stochastic PDE and generally free from any modeling assumptions. Also, while the random forcing was chosen to be Gaussian, self-similar, and white in time, this is by no means necessary and other types of noises can be addressed within this approach. We thoroughly benchmarked our HMC implementation with a standard forward-time-integration pseudospectral method (see Figs.\ \ref{fig:dissip} and  \ref{fig:dv_pdf}). By constraining the sampling of the HMC to generate a strong negative velocity gradient at a specific site we increased the statistics of the left tail of the PDF of velocity gradients significantly, producing gradients as intense as 30 (and more) times the rms value (see Fig.\ \ref{fig:constr_HMC_pdf}). Although we restricted ourselves to the case of localized (in space and time) constraints, the technique can be easily extended to more general cases. Also, our constrained HMC sampling allowed us to decrease by order of magnitudes the time to the solution needed to collect sufficient statistics for high-order moments (up to order 30) if compared with DNS (see Fig.\ \ref{fig:t_to_sol}). We expect that the types of local constraints considered in this work might have an impact on similar studies in lattice gauge theories, where they may lead to new observables.

We demonstrated that instanton configurations can be found in Burgers turbulence. We have recovered the full shape of the classical instanton by averaging the generated ensemble of the constrained configurations, with the agreement of the HMC and the instanton being remarkable [see Figs.\ \ref{fig:HMC_vs_inst}(a) and \ref{fig:instanton}]. We further compared the PDF of the velocity gradients for a very large range of strong negative gradients and showed that, beyond a specific Reynolds-number-dependent threshold of applicability of the instanton method, both the HMC and the instanton produce the same left tail, which further ensures the relevance of instantons in Burgers turbulence [see Fig.\ \ref{fig:HMC_vs_inst}(b)]. Thus, we established a one-to-one correspondence between the biased realizations of the HMC and the fluctuations around instantons. The present study focused on low-Reynolds-number turbulence. However, the present method is not restricted to this case and actually opens the possibility to explore the role of fluctuations around instantons with unmatched precision. We are confident that the suggested approach can find suitable applications in the diverse field of stochastic PDEs and related studies on extreme and rare events.

\section*{Acknowledgments}

This research has benefited from high-performance computing resources provided by the J\"ulich Supercomputing Center (Germany) and the CINECA Supercomputing Center (Italy), under the $13^{\rm th}$ PRACE call. D.M.\ acknowledges support from the Swiss National Science Foundation, the HPC-Europa2 Transnational Access Program, which funded research visits at the University of Rome Tor Vergata and the University of Ferrara, as well as the German Academic Exchange Service (DAAD), which provided financial resources to present part of this work at the Institute of Pure and Applied Mathematics, Rio de Janeiro (Brazil). G.M.\ acknowledges funding from the European Union's Horizon 2020 Research and Innovation Programme under the Marie Sk\l{}odowska-Curie Grant Agreement No.\ 642069 (European Joint Doctorate Programme ``HPC-LEAP"). L.B.\ acknowledges funding from the European Research Council under the European Union's Seventh Framework Programme, ERC Grant Agreement No.\ 339032.

\appendix

\label{Appendix}

\section{Implementation of the leapfrog integrator and the Fourier acceleration}
\label{Sec_ap:Implementation of the leapfrog integrator and the Fourier acceleration}

The leapfrog integrator is a symplectic integrator that numerically integrates the Hamiltonian system of equations \eqref{eq:MD}. It advances the fields $(v_0,\pi_0)\mapsto (v_{\tau},\pi_{\tau})$, up to corrections $O(\Delta \tau^2)$, along a trajectory on a hypersurface ${\H}^{\textrm{eff}}(v_s,\pi_s) = const.$, where the effective Hamiltonian is defined by  ${\H}^{\textrm{eff}} = \frac{1}{2} \int dt (\pi_s, \Omega \ast \pi_s)+ \S[v_s]$. We also note that in our application we treat the fields in Fourier space, i.e., $\F(v_s(x,t)) \mapsto v_s(k,t)$, where the forward Fourier transform is denoted by $\F$, as in Fourier space many calculations of our interest simplify. For instance, the Fourier transform of the convolution between two fields $f$ and $g$ becomes a simple multiplication among the corresponding individually Fourier transformed fields, i.e., $\F( f\ast g) \sim \F( f) \cdot \F( g)$. Therefore, the effective Hamiltonian can be simplified to
\begin{equation}
{\H}^{\textrm{eff}} = \frac{1}{2} \int dt \int dk\, \Omega(k,t) \,\pi_s^2(k,t)+ \S[v_s]
\label{eq:new_eff_hami}
\end{equation}

 The numerical scheme of the leapfrog integrator starts with a half step $\Delta \tau/2$ evolution of the velocity, followed by a full step $\Delta \tau$ of the momenta, and finally another half-step $\Delta \tau/2$ of the velocity
{
\begin{eqnarray}
v_{_{\Delta \tau/2}}(k,t) &=& v_{_0}(k,t) + \Omega(k,t) \, \pi_{_0}(k,t) \, \frac{\Delta \tau}{2}, \label{eq1a} \\
\pi_{_{\Delta \tau}}(k,t) &=& \pi_{_0}(k,t) - \partial \S/\partial v(k,t)\big\rvert_{\Delta \tau/2}\, \Delta \tau , \label{eq1b} \\
v_{_{\Delta \tau}}(k,t) &=& v_{_{\Delta \tau/2}}(k,t) + \Omega(k,t) \, \pi_{_{\Delta \tau}}(k,t)\, \frac{\Delta \tau}{2}, \label{eq1c}
\end{eqnarray}}
where the derivative $\frac{\partial S}{\partial v}$ is evaluated at intermediate step $\Delta \tau/2$. The steps \eqref{eq1a} --\eqref{eq1c} are then repeated $N_{\Delta \tau}$ times until $\tau$ is reached.

 The Fourier acceleration effectively prescribes different trajectory lengths to different Fourier modes of the velocity in an effort to balance the scaling of the forces. This is implemented with an appropriate choice of $\Omega(k,t)$, where $k$ labels a particular wavenumber. Upon rescaling
\begin{eqnarray}
\pi_s(k,t) &\rightarrow& \xi_s(k,t) = \left[ \Delta t \,\Omega(k,t) \right]^{1/2} \pi_s(k,t) \\ \Delta \tau &\rightarrow& \Delta \tilde{\tau}(k,t) = \left( \frac{\Omega(k,t)}{\Delta t} \right)^{1/2} \Delta \tau ,
\end{eqnarray}
we note that the new stepsize $\Delta \tilde{\tau}$ carries a $k$ dependence. Then the equations of motion \eqref{eq1a} --\eqref{eq1c} transform as
\begin{eqnarray}
v_{_{\Delta \tilde{\tau}/2}}(k,t) &=& v_{_0}(k,t) + \xi_{_0}(k,t) \, \Delta \tilde{\tau}/2 , \label{eq1aMod} \\
\xi_{_{\Delta \tilde{\tau}/2}} (k,t) &=& \xi_{_0}(k,t) - \Delta t\, \partial \S/\partial v(k,t)\big\rvert_{\Delta \tilde{\tau}}\, \Delta \tilde{\tau} , \label{eq1bMod} \\
v_{_{\Delta \tilde{\tau}/2}}(k,t) &=& v_{_{\Delta \tilde{\tau}}}(k,t) + \xi_{_{\Delta \tilde{\tau}}} (k,t)\, \Delta \tilde{\tau}/2 . \label{eq1cMod}
\end{eqnarray}
We require that the fields should satisfy $\xi \sim O(1)$, with $\Delta \xi \sim O(\Delta \tau)$, i.e.,
\begin{eqnarray}
\Delta t \frac{\partial \S}{\partial v}(k,t)\, \Delta \tilde{\tau}(k,t) &=& \left( \Delta t \Omega(k,t) \right)^{\frac{1}{2}} \frac{\partial \S}{\partial v}(k,t)\, \Delta \tau \nonumber \\
 &\sim& O(\Delta \tau) .
\label{eq2}
\end{eqnarray}
Then, Eq. \eqref{eq2} gives a relation for the kernel $\Omega(k,t)$,
\begin{equation}
\Omega^{-1} (k,t) = \Delta t \left\langle \left|\frac{\partial \S}{\partial v}
(k,t) \right| \right\rangle^{2} .
\label{Eq:InvOmega}
\end{equation}

\section{Classical action for finite approximations of Burgers equation}
\label{Sec:Classical action for finite approximations of Burgers equation}

A numerical treatment of the path integral \eqref{eq:functionalint_mean} relies on a proper regularization of the functional integration measure and weight. For this purpose we employ finite approximations of the stochastic dynamics of Eq.\ \eqref{Eq:burgers} using a uniform grid in space and time. To make contact with standard approaches employed in the explicit-time integration of the Burgers equation, we adopt a discretization in Fourier space, where the velocity field $v(k,t)$ is defined on a finite set of wave numbers, $k = -N_x/2, -N_x/2 + 1, \ldots, N_x/2 - 1$ and a discrete set of points in time $t = t_0+ n\Delta t$, with $n = 0, 1, \ldots, N_t$, with $N_t\equiv M (t_f - t_0) = M \, T \in \mathbb{N}$ and $\Delta t \equiv T / N_t = 1/M$. Thus, we measure length in units of $L/(2\pi)$, time in units of $M \Delta t$, and velocity in units of $L/(2 \pi M \Delta t)$. In the following we will simply set $L = 2\pi$ and $M \Delta t = 1$.

We also employ the initial condition $v(k,t_0) = 0$ for all wavenumbers $k$ and restrict the time evolution to a finite time interval $t_0 < t \leq t_f$ of length $T=t_f-t_0$. Correspondingly, the functional measure is given by
\begin{equation}
  \int\, \Dv \equiv \prod_{k = -N_x/2}^{N_x/2-1} \prod_{n = 1}^{N_t} \int\, dv(k,t_0 + n\Delta t) .
\end{equation}

\subsection{Finite approximation of equation of motion}
\label{Sec:Finite approximation of equation of motion}

Passing from continuous space to a discrete finite number of Fourier modes, the original stochastic partial differential equation \eqref{Eq:burgers} becomes a high-dimensional set of coupled ordinary stochastic differential equations
\begin{subequations}
\begin{align}
  \frac{d}{dt} v(k,t) &= f^{(\nu)}(k,t) + \eta(k,t) , \label{eqm} \\
  f^{(\nu)}(k,t) &\equiv \frac{-i k}{2 (2\pi)}\sum_{l,m = -N_x/2}^{N_x/2 - 1} \bigg\{v(l,t) v(m,t) \delta_{k, l+m} \nonumber \\& \hspace{3cm}-\nu k^2 v(k,t) \bigg\}. \label{Eq:deterministic}
\end{align}
\end{subequations}
For later convenience, we have separated the equation of motion for each wave number into two parts that describe the deterministic and stochastic components of $dv/dt$, respectively. Note that $v(-k,t) \equiv v^{\ast}(k,t)$ and $\eta(-k,t) \equiv \eta^{\ast}(k,t)$, as well as $f^{(\nu)}(-k,t) \equiv [f^{(\nu)}(k,t)]^{\ast}$ for $k=1,2,\ldots,N_x/2-1$, while $v(k=0,t)$ and $v(k=N_x/2-1,t)$ are both real valued [and similarly $\eta(k=0,t)$ and $\eta(k=N_x/2-1,t)$, etc].

\subsection{De-aliasing}
\label{Sec:De-aliasing}

The nonlinear term leads to aliasing errors in the numerical integration of Eq.\ \eqref{eqm}, which can be avoided by applying the $2/3$ rule \cite{Orszag:1971}. Thus, to correct for these artifacts we introduce the projection operator $\mathsf{P}$ whose action is most conveniently defined in Fourier space
\begin{equation}
  \mathsf{P} (f(k)) = \left\{ \begin{array}{ll} f(k) , & |k|\leq N_x/3 , \\ 0, & |k| > N_x/3 , \end{array} \right.
  \label{eq:dealiasing}
\end{equation}
for any function $f(k)$. The dealiased interaction term is then defined as
\begin{equation}
\widehat{f}^{(\nu)}(k,t) \equiv \mathsf{P}(f^{(\nu = 0)}(k,t)) - \nu k^2 v(k,t) 
\end{equation}
and the corresponding equation of motion reads
\begin{equation}
  \label{Eq:burgers_discrete}
\frac{d}{dt} v(k,t) = \widehat{f}^{(\nu)}(k,t) + \eta(k,t) .\end{equation}

The representation \eqref{Eq:burgers_discrete} of the dynamics relies on standard approaches in the numerical treatment of partial differential equations, i.e., via spectral Galerkin or pseudospectral methods. We do not intend to advocate that these are in any way optimal in terms of their convergence properties (the interested reader is referred to \cite{Durran:2010}). We simply choose Eq.\ \eqref{Eq:burgers_discrete} as our starting point to benchmark the performance of the HMC algorithm.

\subsection{Discrete stochastic dynamics: Euler-Maruyama method}
\label{Sec:Discrete stochastic dynamics: Euler-Maruyama method}

To arrive at a discrete-time representation of the dynamics \eqref{eqm} we employ the stochastic Taylor expansion in time for $v(k)$,
\begin{align}
  v (k,t+\Delta t) 
  &= v(k,t) + \widehat{f}^{(\nu)}(k, t) \Delta t  \nonumber \\
  & + \bar{\eta}(k) \sqrt{\Delta t} + O(\Delta t) , \label{Eq:stochastic_taylor}
\end{align}
where we have used $\int dt\, \eta(k,t) = \bar{\eta}(k) \sqrt{\Delta t} + O(\Delta t)$. This corresponds to the weak first-order Euler scheme \cite{Kloeden:1992}. However, we may improve on the rate of convergence of the deterministic part $\widehat{f}^{(\nu)}(k, t) \Delta t$ by considering the variable transformation
\begin{equation}
  v'(k,t) = \Gnu(k,t_0 - t) v(k,t) ,
  \label{Eq:exact_viscous_term}
\end{equation}
where $\Gnu(k,t) = \exp ( -\nu k^2 t )$. Taking the time derivative on both sides of Eq.\ \eqref{Eq:exact_viscous_term}, we obtain
\begin{equation}
  \frac{d}{dt} v'(k,t) = \Gnu(k,t_0 - t) \left[ \widehat{f}^{(\nu = 0)}(k, t) + \bar{\eta}(k) \right] ,
\end{equation}
to which we may apply the stochastic Taylor expansion. Doing so we arrive at the  result
\begin{eqnarray}
  v(k,t+\Delta t) &=&  \Gnu(k,\Delta t) \big[ v(k,t) + \widehat{f}^{(\nu = 0)}(k, t) \Delta t \nonumber \\
  &+& \bar{\eta}(k) \sqrt{\Delta t} \big] + O(\Delta t) .
  \label{eq:eqm_EM}
\end{eqnarray}

The exact integration of the viscous term [cf.\ Eq.\ \eqref{Eq:exact_viscous_term}] significantly improves the convergence for large wavenumbers, provided the step size $\Delta t$ is sufficiently small.

\subsection{Finite approximation of the stochastic noise}
\label{Sec:Finite approximation of the stochastic noise}

Here we consider finite approximations of the stochastic noise, which is assumed to be centered, Gaussian, and white in time. On a finite set of Fourier modes $k = -N_x/2, -N_x/2 + 1, \ldots, N_x/2 -1$, the second moment takes the form
\begin{equation}
  \int\D\eta\, \P_\eta\, \eta(k,t)\eta(k',t') = \Gamma(k)  \delta(k+k') \delta (t-t'),
\end{equation}
for $t, t' > t_0$. In Eq.\ \eqref{eq:prob_distr_eta} we showed that $\P_\eta = {\mathrm{e}}^{-\frac{1}{2} \int dt\,  (\eta, \,\Gamma^{-1} \ast \eta)}$, by assuming that $\Gamma(k)\neq 0$. The finite approximation of this expression is given by
\begin{align}
  -\ln \P_\eta &= \sum_{n = 1}^{N_t} \Delta t \, \biggl\{ \Gamma^{-1}(0)\hspace{1pt} [\eta(0,t_n)]^2 \nonumber\\
  &+ \Gamma^{-1}(-N_x/2) \hspace{1pt} [\eta(-N_x/2,t_n)]^2  \nonumber\\
   &+\sum_{k = 1}^{N_x/2-1} \eta(k,t_n) \hspace{1pt}\Gamma^{-1}(k) \hspace{1pt} \eta(-k,t_n) \biggr\},
   \label{Eq:Finite_approx_distr}
\end{align}
where $t_n = t_0 + n \Delta t$, and for symmetry reasons we consider half of the modes, i.e., $\sum_{k = -N_x/2}^{N_x/2-1} = 2\, \sum_{k = 0}^{N_x/2-1}$. Note that $\eta(0,t)$ and $\eta(-N_x/2,t)$ are both real valued, while $\eta(k,t)$, $|k| = 1, 2, \ldots, N_x/2-1$ are generally complex. The expression in Eq.\ \eqref{Eq:Finite_approx_distr} can be further simplified through applying a UV cutoff by considering the 2/3 dealiasing rule of Eq.\ \eqref{eq:dealiasing}. Moreover, we consider $\eta(0,t)=0$, i.e., the zero mode $k=0$ is not forced. Finally, we note that we use a large-scale power-law forcing, i.e., $\Gamma(k) = \Gamma_0 |k|^\beta$, with $\beta=-3$ throughout all the simulations in this article. Altogether we have
\begin{equation}
-\ln \P_\eta = \frac{1}{\Gamma_0} \sum_{n = 1}^{N_t} \Delta t \sum_{k = 1}^{N_x/3} \eta(k,t_n) \hspace{1pt} \,k^{-\beta}\hspace{1pt} \eta(-k,t_n).
\label{Eq:Finite_approx_distr_2}
\end{equation}

This expression will be useful below, when we construct the finite-time-discretized approximations of the classical action $\mathcal{\S}$, based on the regularized continuous time stochastic dynamics.

\subsection{Classical action}
\label{Sec:Classical action}

From Eq.\ \eqref{eq:eqm_EM} we extract the deterministic part of the time-discrete representation of the equation of motion, i.e.,
\begin{eqnarray}
F(k,t) &=& \frac{v(k,t)}{\Delta t} - \frac{\Gnu (k,\Delta t)}{\Delta t} \big[ v(k,t - \Delta t) \nonumber\\ &+& \Delta t \widehat{f}^{(\nu = 0)}(k, t - \Delta t) \big] ,
\end{eqnarray}
which enters the (reparametrized) classical action following Eqs.\ \eqref{Eq:Finite_approx_distr_2} and \eqref{eq:general_action}
\begin{subequations}
\begin{align}
  \S &= - \ln \P_v + \textrm{const} \label{Eq:Finite_approx_action_a} \\
  &= \frac{1}{\Gamma_0} \sum_{n = 1}^{N_t} \Delta t \sum_{k = 1}^{N_x/3} F(k,t_n) \hspace{1pt} \Gnu (k,\Delta t)^{-2} k^{-\beta}\hspace{1pt} F(-k,t_n).
   \label{Eq:Finite_approx_action}
\end{align}
\end{subequations}
Notice that Eq.\ \eqref{Eq:Finite_approx_action} is the discretized version of Eq.\ \eqref{eq:action_S_burgers}. The const. in Eq.\ \eqref{Eq:Finite_approx_action_a} is related to the contribution of  $\ln J$ to $\S$, which is constant in the case of the explicit-time schemes as introduced in Sec. \ref{Sec:Discrete stochastic dynamics: Euler-Maruyama method} with fixed initial boundary conditions, e.g., here $v(x,t_0)= 0$, and open final conditions \cite{Nakazato:1990kk}.

In general, any constant contribution to the action can be removed, as the sampling is left unaffected, since the HMC considers the differences of the Hamiltonian. On the other hand, as stated in Sec.\ \ref{Sec:Time-periodic boundary conditions} in the case of periodic boundary conditions in time $v(x,t+T)= v(x,t)$, the contribution $\ln J$ to $\S$ is field dependent \cite{Nakazato:1990kk} and therefore, in principle, it has to be evaluated during the course of the simulation.

\bibliographystyle{apsrev4-1}
\bibliography{references}

\end{document}